\documentclass[11pt,aps,prd,nofootinbib,superscriptaddress,preprint,preprintnumbers
]{revtex4-2}
\usepackage{amssymb}
\usepackage{amsmath}
\usepackage{graphics}
\usepackage{epsfig}
\usepackage{multirow}
\usepackage{color}
\usepackage{mathrsfs}
\usepackage{graphicx}
\usepackage{slashed}
\usepackage{hyperref}

\begin{document}

\title{$e^+e^-\gamma$ production at photon-photon colliders at complete electroweak NLO accuracy}

\author{He-Yi Li}
\affiliation{State Key Laboratory of Particle Detection and Electronics, University of Science and Technology of China, Hefei 230026, Anhui, People's Republic of China}
\affiliation{Department of Modern Physics, University of Science and Technology of China, Hefei 230026, Anhui, People's Republic of China}

\author{Ren-You Zhang}
\email{Corresponding author: zhangry@ustc.edu.cn}
\affiliation{State Key Laboratory of Particle Detection and Electronics, University of Science and Technology of China, Hefei 230026, Anhui, People's Republic of China}
\affiliation{Department of Modern Physics, University of Science and Technology of China, Hefei 230026, Anhui, People's Republic of China}

\author{Wen-Gan Ma}
\affiliation{State Key Laboratory of Particle Detection and Electronics, University of Science and Technology of China, Hefei 230026, Anhui, People's Republic of China}
\affiliation{Department of Modern Physics, University of Science and Technology of China, Hefei 230026, Anhui, People's Republic of China}

\author{Yi Jiang}
\affiliation{State Key Laboratory of Particle Detection and Electronics, University of Science and Technology of China, Hefei 230026, Anhui, People's Republic of China}
\affiliation{Department of Modern Physics, University of Science and Technology of China, Hefei 230026, Anhui, People's Republic of China}

\author{Xiao-Zhou Li}
\affiliation{State Key Laboratory of Particle Detection and Electronics, University of Science and Technology of China, Hefei 230026, Anhui, People's Republic of China}
\affiliation{Department of Modern Physics, University of Science and Technology of China, Hefei 230026, Anhui, People's Republic of China}

\date{\today}

\begin{abstract}
We present the NLO electroweak radiative corrections to the $e^+e^-\gamma$ production in $\gamma\gamma$ collision, which is an ideal channel for calibrating the beam luminosity of a Photon Linear Collider. We analyze the dependence of the total cross section on the beam colliding energy, and then investigate the kinematic distributions of final particles at various initial photon beam polarizations at EW NLO accuracy. The numerical results indicate that the EW relative corrections to the total cross section are non-negligible and become increasingly significant as the increase of the beam colliding energy, even can exceed $-10\%$ in the $\text{J} = 2$ $\gamma\gamma$ collision at $\sqrt{\hat{s}}=1~ \text{TeV}$. Such EW corrections are very important and should be taken into consideration in precision theoretical and experimental studies at high-energy $\gamma\gamma$ colliders.
\end{abstract}

\maketitle

\section{Introduction}
\label{section-1}
\par
There is a general consensus that the next large high-energy project after the Large Hadron Collider shall be a linear collider. As a potential extension to the positron-electron ($e^+e^-$) collision mode, the photon-photon ($\gamma\gamma$) collision \cite{Telnov:1995hc,Asner:2013psa} at the International Linear Collider (ILC), i.e., the Photon Linear Collider (PLC), can help us to better understand the electroweak symmetry breaking and search for new physics beyond the standard model (BSM), such as supersymmetry, quantum gravity, extra dimensions, etc. Although the photon beam luminosity of PLC \cite{Ginzburg:1981vm,Ginzburg:1982yr,Zarnecki:2002qr} can reach only approximately $80\%$ of the electron beam luminosity at the ILC, the production rate of any charged particle pair in the $\gamma\gamma$ collision is typically about one order of magnitude higher than that in the $e^+e^-$ collision. Therefore, the PLC provides a promising platform to test the standard model and search for BSM physics at the terascale. \cite{Boos:2000gr,Velasco:2002vg,Niezurawski:2005cp}.

\par
According to the ``ILC Higgs White Paper'' \cite{Asner:2013psa}, the incoming photons at the high-energy PLC are  produced via the backward Compton scattering (BCS) of the laser light off the linear electron beams \cite{Badelek:2001xb,Telnov:2006cj}, which  benefit from the monochromatic characteristic (concentrated at high energies in a narrow energy spread) and adjustable polarization mode. The $\gamma\gamma$ collision is known to have two polarization configurations: $\text{J} = 0$ and $\text{J} = 2$, where $\text{J}$ denotes the total helicity of the two incoming photons. As is well known, $\gamma\gamma \rightarrow l^+l^-$ is the most promising channel for calibrating the photon beam luminosity of the the PLC \cite{Telnov:1995hc,Ginzburg:1982yr,Badelek:2001xb,Yasui:1992fh}, and the precision predictions for the light fermion-antifermion pair production via $\gamma\gamma$ scattering up to the electroweak (EW) next-to-leading order (NLO) are given in \cite{Denner:1998tb,Demirci:2021zwz}. However, such a reaction is incapable of determining the luminosity of $\text{J} = 0$ polarized incoming photon beams, cause the production rate of this channel is suppressed by a factor of $m_{l}^2/\hat{s}$ in the $\text{J} = 0$ polarization configuration \cite{Dittmaier:1998nn}. On the other hand, as the $\text{J} = 0$ polarization mode is preferred by Higgs physics due to the unique channel $\gamma\gamma \rightarrow H^0 \rightarrow X$ ($H^0$ is any neutral scalar/pseudoscalar particle) in searching for BSM signals \cite{Asner:2001ia}, the related high precision studies at $\gamma\gamma$ colliders are indispensable correspondingly. The $\gamma\gamma \rightarrow W^+W^-$ process is also proposed for measuring the photon luminosity of PLC \cite{Yasui:1992fh}, but it can only work for the circumstance with colliding energy exceeding the threshold of $W$-boson pair, and its cross section has to be measured precisely at first since such reaction could be accompanied by some BSM effects. Although the cross section for the $\gamma\gamma \rightarrow l^+l^-l^+l^-$ process is large and insensitive to the beam polarization, it is, unfortunately, rather small at large scattering angle where the lepton momenta can be measured precisely \cite{Ginzburg:1981vm,Pak:2003jq}.

\par
Compared to the $\gamma\gamma \rightarrow l^+l^-$ process, the lepton pair production in association with an extra photon via $\gamma\gamma$ scattering is suppressed by an additional fine structure constant, but free from the helicity suppression due to the radiated spin-1 photon in the final state. Consequently, the $\gamma\gamma \rightarrow l^+l^-\gamma$ process can be adopted to measure the photon luminosity of the $\text{J} = 0$ $\gamma\gamma$ collision mode, and the precision theoretical predictions for $\gamma\gamma \rightarrow l^+l^-\gamma$ are necessary. Since the incoming photon beams at the PLC are only partially polarized, the ratio of the cross section for the $l^+l^-\gamma$ production via the $\text{J} = 0$ $\gamma\gamma$ scattering to that via $\text{J} = 2$ $\gamma\gamma$ scattering should be sufficiently high to calibrate the $\text{J} = 0$ $\gamma\gamma$ collision precisely. The dependences of the cross section for $\gamma\gamma \rightarrow l^+l^-\gamma$ on the helicity and colliding energy of the two incoming photons, as well as the kinematic cuts on the final state have already been analyzed at the lowest order \cite{Makarenko:2003vx,Makarenko:2003xg}. Careful and detailed studies show that the $\gamma\gamma \rightarrow l^+l^-\gamma$ process is a unique channel in measuring the photon luminosity together with the $\gamma\gamma \rightarrow l^+l^-$ process. In addition, the $e^+e^- \rightarrow l^+l^-$ and $e^+e^- \rightarrow l^+l^-\gamma$ processes also garner attention for determining the luminosity of the positron-electron collision mode at the ILC, and the related investigations at the EW NLO accuracy have been accomplished in \cite{Khiem:2014dka,Bardin:2017mdd}.

\par
In this study, we calculate the complete NLO EW radiative corrections to the $e^+e^-\gamma$ production in $\gamma\gamma$ collision, and provide the integrated cross sections and some kinematic distributions of final particles for both $\text{J} = 0$ and $\text{J} = 2$ polarization configurations. In Sec.\ref{section-2} we describe in detail the analytical calculation strategy, and then present the numerical results and discussion for the integrated and differential cross sections in both inclusive and exclusive event selection schemes in Sec.\ref{section-3}. Finally, a short summary is given in Sec.\ref{section-4}.

\vskip 5mm

\section{Outline of calculations}
\label{section-2}
\subsection{LO calculation}
\label{subsection-2A}
\par
\begin{figure}[htbp]
\begin{center}
\includegraphics[scale=1.0]{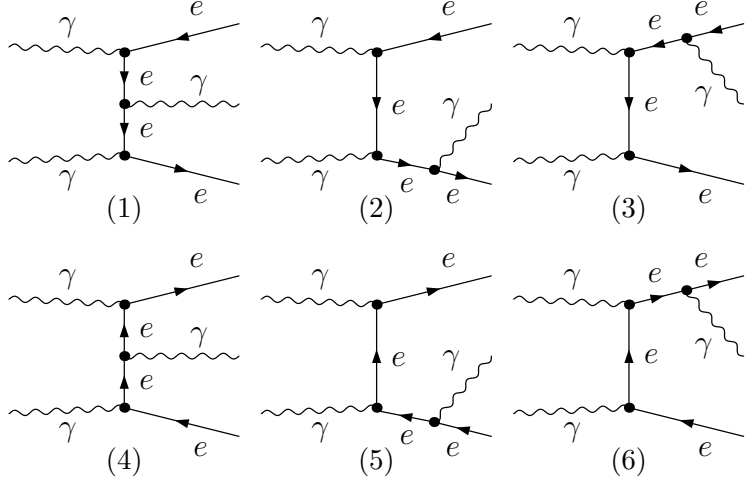}
\caption{Leading order Feynman diagrams for $\gamma\gamma \rightarrow e^+e^-\gamma$.}
\label{fig1}
\end{center}
\end{figure}

We consider the process
\begin{equation}
\label{rreer}
\gamma_{\lambda_1}(q_1) + \gamma_{\lambda_2}(q_2) \rightarrow e^+_{h_1}(p_1) + e^-_{h_2}(p_2) + \gamma_{\lambda_3}(q_3)\,,
\end{equation}
where $\lambda_i = \pm$ and $q_i = ( q_i^0 \equiv \vert \vec{q}_i \vert\,,~ \vec{q}_i )$ $(i = 1,\, 2,\, 3)$ are the helicities and four-momenta of the incoming and outgoing photons, while $h_j = \pm$ and $p_j = ( p_j^0 \equiv \sqrt{\vert \vec{p}_j \vert^2 + m_e^2}\,,~ \vec{p}_j )$ $(j = 1,\, 2)$ are the helicities and four-momenta of the two final-state fermions (positron and electron). Then, the total helicity of the two incoming photons is given by $\text{J} = \vert \lambda_1 - \lambda_2 \vert$. We denote the differential cross section in the center-of-mass (c.m.) frame of the initial-state $\gamma\gamma$ system as $d \hat{\sigma}^{\lambda_1 \lambda_2 h_1 h_2 \lambda_3}(\vec{q}_1,\, \vec{q}_2;\, \vec{p}_1,\, \vec{p}_2,\, \vec{q}_3)$. If the two incoming photon beams are partially polarized, the differential cross section for the unpolarized $e^+e^-\gamma$ production (i.e., the polarizations of the final-state $e^+$, $e^-$ and $\gamma$ are not measured) in $\gamma\gamma$ collision is given by \cite{MoortgatPick:2005cw}
\begin{equation}
\label{polariedCS}
d \hat{\sigma}({\cal P}_1,\, {\cal P}_2;\, \vec{q}_1,\, \vec{q}_2;\, \vec{p}_1,\, \vec{p}_2,\, \vec{q}_3)
=
\frac{1}{4}
\sum_{\lambda_{1,2} = \pm}
\left( 1 + \lambda_1 \mathcal{P}_1 \right) \left( 1 + \lambda_2 \mathcal{P}_2 \right)
d \hat{\sigma}^{\lambda_1 \lambda_2}(\vec{q}_1,\, \vec{q}_2;\, \vec{p}_1,\, \vec{p}_2,\, \vec{q}_3)\,,
\end{equation}
where $\mathcal{P}_1$ and $\mathcal{P}_2$ are the degrees of polarization of the two incoming photon beams\footnote{The degree of polarization of a photon beam is defined as ${\cal P} = \dfrac{N_+ - N_-}{N_+ + N_-}$, where $N_+$ and $N_-$ are the numbers of right- and left-handed photons, respectively.}, and
\begin{equation}
d \hat{\sigma}^{\lambda_1 \lambda_2}(\vec{q}_1,\, \vec{q}_2;\, \vec{p}_1,\, \vec{p}_2,\, \vec{q}_3)
=
\sum_{\lambda_3, h_{1,2} = \pm}
d \hat{\sigma}^{\lambda_1 \lambda_2 h_1 h_2 \lambda_3}(\vec{q}_1,\, \vec{q}_2;\, \vec{p}_1,\, \vec{p}_2,\, \vec{q}_3)\,.
\end{equation}
The tree-level Feynman diagrams for $\gamma\gamma \rightarrow e^+e^-\gamma$ are depicted in Fig.\ref{fig1}. By using $\mathcal{C}$, $\mathcal{P}$, $\mathcal{CP}$ and ${\it Bose}$ symmetries \cite{Makarenko:2003xg}, we obtain
\begin{equation}
\label{symmetries}
\begin{aligned}
& \mathcal{C}:
&&
d \hat{\sigma}^{\lambda_1 \lambda_2 h_1 h_2 \lambda_3}_{\text{LO}}(\vec{q}_1,\, \vec{q}_2;\, \vec{p}_1,\, \vec{p}_2,\, \vec{q}_3)
=
d \hat{\sigma}^{\lambda_1 \lambda_2 h_2 h_1 \lambda_3}_{\text{LO}}(\vec{q}_1,\, \vec{q}_2;\, \vec{p}_2,\, \vec{p}_1,\, \vec{q}_3)
\\
& \mathcal{P}:
&&
d \hat{\sigma}^{\lambda_1 \lambda_2 h_1 h_2 \lambda_3}_{\text{LO}}(\vec{q}_1,\, \vec{q}_2;\, \vec{p}_1,\, \vec{p}_2,\, \vec{q}_3)
=
d \hat{\sigma}^{-\lambda_1 -\lambda_2 -h_1 -h_2 -\lambda_3}_{\text{LO}}(-\vec{q}_1,\, -\vec{q}_2;\, -\vec{p}_1,\, -\vec{p}_2,\, -\vec{q}_3)
\\
& \mathcal{CP}:
&&
d \hat{\sigma}^{\lambda_1 \lambda_2 h_1 h_2 \lambda_3}(\vec{q}_1,\, \vec{q}_2;\, \vec{p}_1,\, \vec{p}_2,\, \vec{q}_3)
=
d \hat{\sigma}^{-\lambda_1 -\lambda_2 -h_2 -h_1 -\lambda_3}(-\vec{q}_1,\, -\vec{q}_2;\, -\vec{p}_2,\, -\vec{p}_1,\, -\vec{q}_3)
\\
& {\it Bose}:
&&
d \hat{\sigma}^{\lambda_1 \lambda_2 h_1 h_2 \lambda_3}(\vec{q}_1,\, \vec{q}_2;\, \vec{p}_1,\, \vec{p}_2,\, \vec{q}_3)
=
d \hat{\sigma}^{\lambda_2 \lambda_1 h_1 h_2 \lambda_3}(\vec{q}_2,\, \vec{q}_1;\, \vec{p}_1,\, \vec{p}_2,\, \vec{q}_3)
\end{aligned}
\end{equation}
It should be noted that $\mathcal{C}$ and $\mathcal{P}$ are simply the Born-level symmetries for the $\gamma\gamma \rightarrow e^+e^-\gamma$ process, because $\gamma\gamma \rightarrow e^+e^-\gamma$ is a pure QED scattering process at the lowest order, and the weak interaction is only involved in the high-order radiative corrections. From Eq.(\ref{symmetries}) we obtain
\begin{equation}
\label{CP-Bose}
\begin{aligned}
& \mathcal{CP}+{\it Bose}:
&&
\begin{aligned}
&
d \hat{\sigma}^{--}(\vec{q}_1,\, \vec{q}_2;\, \vec{p}_1,\, \vec{p}_2,\, \vec{q}_3)
=
d \hat{\sigma}^{++}(-\vec{q}_2,\, -\vec{q}_1;\, -\vec{p}_2,\, -\vec{p}_1,\, -\vec{q}_3)
\\
&
d \hat{\sigma}^{+-}(\vec{q}_1,\, \vec{q}_2;\, \vec{p}_1,\, \vec{p}_2,\, \vec{q}_3)
=
d \hat{\sigma}^{+-}(-\vec{q}_2,\, -\vec{q}_1;\, -\vec{p}_2,\, -\vec{p}_1,\, -\vec{q}_3)
\end{aligned}
\end{aligned}
\end{equation}
It clearly demonstrates that the differential distributions as well as the integrated cross section for $\gamma_-\gamma_-$ collisions can be obtained straightforwardly from the corresponding ones for $\gamma_+\gamma_+$ collisions. Therefore in the following discussion, we only consider the $\gamma_+\gamma_+ \rightarrow e^+e^-\gamma$ and $\gamma_+\gamma_- \rightarrow e^+e^-\gamma$ channels for $\text{J} = 0$ and $\text{J} = 2$ polarization configurations of the $\gamma\gamma$ system, respectively. Moreover, we can conclude that the final-state electron and positron in the $\text{J} = 2$ collision mode should have identical kinematic behaviors from Eq.(\ref{CP-Bose}), while there is no such coincidental feature in the $\text{J} = 0$ collision mode.

\par
The leading order (LO) differential cross section for $\gamma_{\lambda_1}(q_1) + \gamma_{\lambda_2}(q_2) \rightarrow e^+_{h_1}(p_1) + e^-_{h_2}(p_2) + \gamma_{\lambda_3}(q_3)$ can be expressed as
\begin{equation}
d \hat{\sigma}^{\lambda_1 \lambda_2 h_1 h_2 \lambda_3}_{\text{LO}}(\vec{q}_1,\, \vec{q}_2;\, \vec{p}_1,\, \vec{p}_2,\, \vec{q}_3)
=
\dfrac{1}{2 \hat{s}}
\left|
\mathcal{M}^{\lambda_1 \lambda_2 h_1 h_2 \lambda_3}_{{\rm LO}}(\vec{q}_1,\, \vec{q}_2;\, \vec{p}_1,\, \vec{p}_2,\, \vec{q}_3)
\right|^2
d\Phi_{3}(\vec{p}_1,\, \vec{p}_2,\, \vec{q}_3)\,,
\end{equation}
where $\hat{s} = (q_1 + q_2)^2$, $\mathcal{M}^{\lambda_1 \lambda_2 h_1 h_2 \lambda_3}_{\text{LO}}(\vec{q}_1,\, \vec{q}_2;\, \vec{p}_1,\, \vec{p}_2,\, \vec{q}_3)$ is the LO helicity amplitude that can be acquired by applying the Weyl-van-der-Waerden spinor technique \cite{Dittmaier:1998nn,DeCausmaecker:1981wzb,Berends:1981uq}, and $d\Phi_{3}(\vec{p}_1,\, \vec{p}_2,\, \vec{q}_3)$ represents the $e^+e^-\gamma$ final-state phase-space element, which is given by
\begin{equation}
d\Phi_{3}(\vec{p}_1,\, \vec{p}_2,\, \vec{q}_3) = (2 \pi)^4 \delta^{(4)}(p_1+p_2+q_3-q_1-q_2)
\dfrac{d^3\vec{p}_1}{(2 \pi)^3 2 p_1^0}
\dfrac{d^3\vec{p}_2}{(2 \pi)^3 2 p_2^0}
\dfrac{d^3\vec{q}_3}{(2 \pi)^3 2 q_3^0}\,.
\end{equation}
In the limit of $m_e \rightarrow 0$, $\mathcal{M}^{\lambda_1 \lambda_2 h_1 h_2 \lambda_3}_{\text{LO}} = 0$ if $h_1 = h_2$ or $\lambda_1 = \lambda_2 = -\lambda_3$. All the non-vanishing LO helicity amplitudes can be obtained from $\mathcal{M}_{\text{LO}}^{+--++}$ by using $\mathcal{C}$, $\mathcal{P}$ and ${\it Bose}$ symmetries in Eq.(\ref{symmetries}), and the crossing symmetry between the initial and final photons \cite{Makarenko:2003xg}. By adopting the helicity amplitude method \cite{DeCausmaecker:1981wzb}, we obtain
\begin{equation}
\big|\mathcal{M}^{+--++}_{\text{LO}}(\vec{q}_1,\, \vec{q}_2;\, \vec{p}_1,\, \vec{p}_2,\, \vec{q}_3)\big|^2
=
4 e^6
\dfrac{(p_1 \cdot p_2) (p_2 \cdot q_2)^2}{(p_1 \cdot q_1) (p_1 \cdot q_3) (p_2 \cdot q_1) (p_2 \cdot q_3)}\,.
\end{equation}

\subsection{NLO EW corrections}
\par
\begin{figure}[htb]
\begin{center}
\includegraphics[scale=1.0]{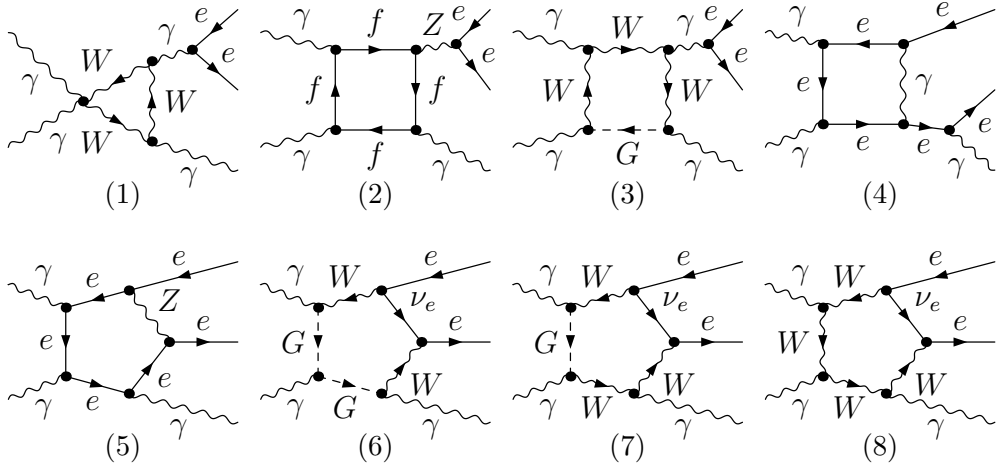}
\caption{Representative box and pentagon Feynman diagrams for $\gamma\gamma \rightarrow e^+e^-\gamma$. }
\label{fig2}
\end{center}
\end{figure}

We employ the modified {\sc FeynArts-3.7}+{\sc FormCalc-7.3}+{\sc LoopTools-2.8} packages \cite{Hahn:1998yk,vanOldenborgh:1990yc,Hahn:2000kx} to generate Feynman diagrams, simplify Feynman amplitudes, and perform loop and phase-space integrations. The one-loop EW virtual correction to $\gamma \gamma \rightarrow e^+e^-\gamma$ includes $960$ Feynman diagrams, which can be categorized into self-energy ($36$), triangle ($438$), box ($414$), pentagon ($42$) and counterterm ($30$) diagrams. Some representative box and pentagon Feynman diagrams for $\gamma\gamma \rightarrow e^+e^-\gamma$ are depicted in Fig.\ref{fig2}. We can see that the loop diagram in Fig.\ref{fig2} (2) may induce $Z$ resonance in the vicinity of $M_{e^+e^-} = m_Z$, where $M_{e^+e^-}$ is the invariant mass of the final-state $e^+e^-$ system, due to the possible on-shell $Z$ propagator. To avoid the numerical divergence in loop calculation, we replace the $Z$ propagator $1/(p^2 - m_Z^2)$ by $1/(p^2 -m_Z^2 - i m_Z \Gamma_Z)$, where the contribution from the imaginary part is beyond the EW NLO and thus can be ignored. We adopt the dimensional regularization (DR) scheme \cite{tHooft:1972tcz} to regularize the ultraviolet (UV) divergences. In the DR scheme, the dimensions of spinor and space-time manifolds are all extended to $D = 4 - 2 \epsilon$. The $5$-point loop integrals are decomposed into $4$-point loop integrals by adopting the Denner-Dittmaier method \cite{Denner:2002ii}, and all the $N$-point ($N \leqslant 4$) tensor integrals are reduced into scalar integrals recursively by adopting the Passarino-Veltman reduction algorithm \cite{Passarino:1978jh}. In the calculation of $4$-point scalar integrals, numerical instability would occur at some phase-space regions with small Gram determinants. Generally, this problem can be solved by adopting the quadruple precision arithmetic proposed in \cite{Nhung:2013jta}.

\par
The renormalized electric charge is defined by $e^{(B)} = ( 1 + \delta{Z_e} ) e$, where $e^{(B)}$ is the bare electric charge and $\delta Z_e$ is the corresponding renormalization constant. We adopt the $\alpha(0)$ scheme to renormalize the electric charge, in which the fine structure constant is set to its Thomson limit $\alpha(0)$ and the electric charge renormalization constant $\delta Z^{\alpha(0)}_e$ is given by \cite{Denner:1991kt}
\begin{equation}
\label{Counterm-Ze}
\delta Z^{\alpha(0)}_e
=
-\dfrac{1}{2} \delta Z_{AA} - \dfrac{1}{2} \tan\theta_W \delta Z_{ZA}
=
\dfrac{1}{2} \dfrac{\partial \sum^{AA}_T(p^2)}{\partial p^2}\bigg{|}_{p^2 = 0} - \tan\theta_W \dfrac{\sum^{AZ}_T(0)}{m^2_Z}\,,
\end{equation}
where $\theta_W$ is the weak mixing angle and $\sum^{ab}_T$ represents the transverse part of the unrenormalized self-energy of the $a \rightarrow b$ transition. For the field and mass renormalization, we employ the on-mass-shell scheme. The definitions and expressions of the relevant renormalization constants in the on-mass-shell scheme can be found in \cite{Denner:1991kt}. After performing the renormalization procedure, all the UV singularities are canceled and thus the virtual correction is UV finite.

\par
The virtual photon in loops can induce soft and quasi-collinear infrared (IR) divergences. We introduce an infinitesimal fictitious photon mass to regularize IR divergences. According to the Kinoshita-Lee-Nauenberg (KLN) theorem \cite{Lee:1964is,Kinoshita:1975bt}, we should consider the contribution from the real photon emission process $\gamma\gamma \rightarrow e^+e^-\gamma\gamma$ to get IR-safe observables at the EW NLO. The IR divergences of the virtual correction can be canceled exactly by those of the real photon emission correction. We extract the IR singularities from the real emission correction by employing the dipole substraction (DS) method \cite{Catani:1996jh,Catani:1996vz,Dittmaier:1999mb,Dittmaier:2008md} and combine them with the virtual correction. In the DS method, a parameter $\alpha \in (0, 1]$ is introduced to control the subtraction region \cite{Nagy:1998bb,Nagy:2003tz}, but the total cross section is independent of $\alpha$. We also employ the two cutoff phase space slicing (TCPSS) method \cite{Harris:2001sx} to deal with the real photon emission process for comparison, and find that the integrated cross sections obtained by using the DS and TCPSS methods are in good agreement with each other within the calculation errors.

\subsection{$ee \rightarrow \gamma\gamma \rightarrow e^+e^-\gamma$  at the PLC}
\par
Among all the methods of $e \rightarrow \gamma$ conversion, the best one is BCS of laser light on high energy electrons. The photons after BCS have energy close to that of the initial electrons and follow their directions with some small angular spread. Given that the photon beams at the PLC are generated via BCS mechanism, $\gamma\gamma \rightarrow e^+e^-\gamma$ can be regarded as the subprocess of $ee \rightarrow \gamma\gamma \rightarrow e^+e^-\gamma$. The production cross section for the parent process $ee \rightarrow \gamma\gamma \rightarrow e^+e^-\gamma$ can be obtained by folding the cross section for $\gamma\gamma \rightarrow e^+e^-\gamma$ with the photon luminosity, i.e.,
\begin{equation}
\sigma(ee \rightarrow \gamma\gamma \rightarrow e^+e^-\gamma;\, s)
=
\int_{2 m_e/\sqrt{s}}^{x_{\text{max}}}
dz \frac{d\mathcal{L}_{\gamma\gamma}}{dz}(z)
\hat{\sigma}(\gamma\gamma \rightarrow e^+e^-\gamma;\, \hat{s} = z^2 s)\,,
\end{equation}
where $\sqrt{s}$ and $\sqrt{\hat{s}}$ are the c.m. colliding energies of $ee$ and $\gamma\gamma$ systems, respectively. The photon luminosity is given by
\begin{equation}
\frac{d\mathcal{L}_{\gamma\gamma}}{dz}(z)
=
2 z \int_{z^2/x_{\text{max}}}^{x_{\text{max}}}
\frac{dx}{x}\phi_{\gamma/e}(x) \phi_{\gamma/e}(z^2/x)\,,
\end{equation}
where $\phi_{\gamma/e}(x)$ denotes the photon structure function, and $x$ is the fraction of the energy of the incident electron carried by the back-scattered photon. As is well known, the degree of polarization of the back-scattered photon $\mathcal{P}_{\gamma}$ is also a function of the energy fraction $x$. Both $\phi_{\gamma/e}$ and $\mathcal{P}_{\gamma}$ depend on the degrees of polarization of the incident electron and laser light. In the numerical treatment, the structure function $\phi_{\gamma/e}(x)$, polarization degree function $\mathcal{P}_{\gamma}(x)$, and maximum energy fraction $x_{\text{max}}$ of the back-scattered photon are all obtained from {\sc CompAZ} \cite{Zarnecki:2002qr}, whose parametrization is based on a realistic TESLA prototype \cite{Ginzburg:1999wz}.

\par
Finally, the NLO EW corrected cross section for $ee \rightarrow \gamma\gamma \rightarrow e^+e^-\gamma$ is given by
\begin{equation}
\sigma_{\text{NLO}}
=
\sigma_{\text{LO}} + \left( 1 + \delta_{\text{EW}} \right)\,,
\qquad
\delta_{\text{EW}} = \frac{\Delta \sigma_{\text{vir}} + \Delta \sigma_{\text{real}}}{\sigma_{\text{LO}}}\,,
\end{equation}
where the NLO EW relative correction $\delta_{\text{EW}}$ can be decomposed into QED and genuine weak relative corrections, i.e., $\delta_{\text{EW}} = \delta_{\text{QED}} + \delta_{\text{W}}$. The same definitions also hold for $\gamma\gamma \rightarrow e^+e^-\gamma$ by substituting $\sigma$ with $\hat{\sigma}$. In the following discussion on the $\gamma\gamma \rightarrow e^+e^-\gamma$ channel, the superscript ``$~\hat{ }~$'' in $\hat{\sigma}$ will be omitted only for convenience.

\vskip 5mm

\section{Numerical results and discussion}
\label{section-3}
\par
In this section, we provide the total cross sections and kinematic distributions of final particles for $\gamma\gamma \rightarrow e^+e^-\gamma$ up to the EW NLO for both $\text{J} = 0$ and $\text{J} = 2$ photon beam polarization configurations. Some kinematic cuts are imposed on the final state to exclude inevitable infrared divergences and fulfil the experimental requirement in the luminosity measurement.

\subsection{Event selection criteria}
\par
At the EW NLO, both $\gamma\gamma \rightarrow e^+e^-\gamma$ and $\gamma\gamma \rightarrow e^+e^-\gamma \gamma$ channels are involved in the production of $e^+e^-\gamma$ at the PLC. When generating $e^+e^-\gamma$ event samples and calculating cross sections at the LO, the following set of kinematic and geometric acceptance requirements are applied on the final-state electron, positron and photon as event selection cuts:
\begin{equation}
\label{cuts}
\begin{aligned}
& \text{energies:}
&&
E_{e^{\pm}},\, E_{\gamma} \, \geqslant \, 10~\text{GeV} \\
& \text{scattering angles:}
&\quad&
\theta_{e^{\pm}},\, \theta_{\gamma} \, \in \, \left[ 10^{\circ},\, 170^{\circ} \right] \\
& \text{opening angles:}
&&
\theta_{e^{\pm}\gamma} ,\, \theta_{e^+e^-} \, \geqslant \, 10^{\circ}
\end{aligned}
\end{equation}
where $\theta_{i}~ (i = e^{\pm},\, \gamma)$ denotes the scattering angle of the final-state particle $i$ with respect to the incoming ``$+$"-polarized photon beam direction, and $\theta_{ij}~ (ij = e^{\pm}\gamma,\, e^+e^-)$ represents the opening angle between $i$ and $j$ in the c.m. frame of the initial-state $\gamma\gamma$ system. This set of kinematic cuts can also guarantee the IR safety at the LO, and all the final-state particles are well separated.

\par
For the $\gamma\gamma \rightarrow e^+e^-\gamma\gamma$ channel, the two tracks of the $\gamma\gamma$- or $\gamma e^{\pm}$-pair will be recombined as a quasi-particle if they are sufficiently collinear ($\theta_{\gamma\gamma}~ \text{or}~ \theta_{e^{\pm}\gamma} \leqslant 10^{\circ}$), and the final state is regarded as an $e^+e^-\gamma$ event; otherwise, it is categorized as an $e^+e^-\gamma\gamma$ event. In this study, we adopt both inclusive and exclusive event selection schemes in the numerical calculations. In the inclusive event selection scheme (denoted by scheme-I), only the baseline event selection cuts in Eq.(\ref{cuts}) are applied on the $e^+e^-\gamma$ and $e^+e^-\gamma\gamma$ events. It should be noted that only one of the two final-state photons needs to satisfy the kinematic constraints in Eq.(\ref{cuts}) for an $e^+e^-\gamma\gamma$ event. Hence, the events with two energetic and well separated photons are accepted in the inclusive event selection scheme. In contrast, an $e^+e^-\gamma\gamma$ event will be rejected in the exclusive event selection scheme (denoted by scheme-II) if both final-state photons can pass the kinematic cuts in Eq.(\ref{cuts}).

\subsection{Input parameters}
\par
The SM input parameters used in this paper are taken as \cite{Zyla:2020zbs}
\begin{equation}
\label{input parameters}
\begin{aligned}
&  m_e = 0.5109989461~ \text{MeV}
&\quad& m_{\mu} = 105.6583745~ \text{MeV}
&\quad& m_{\tau} = 1776.86~ \text{MeV}
\\
&  m_u = 62~ \text{MeV}
&& m_c = 1.5~ \text{GeV}
&& m_t = 172.76~ \text{GeV}
\\
&  m_d = 83~ \text{GeV}
&& m_s = 215~ \text{MeV}
&& m_b = 4.7~ \text{GeV}
\\
&  m_W = 80.379~ \text{GeV}
&& m_Z = 91.1876~ \text{GeV}
&& \Gamma_Z = 2.4952~ \text{GeV}
\\
& \alpha(0) = 1/137.035999084
\end{aligned}
\end{equation}
where the masses of light quarks can reproduce the hadronic contribution to the photon vacuum polarization \cite{Jegerlehner:2001ca}, and $\alpha(0)$ is the fine structure constant in the Thomson limit.

\par
Normally, there exists mass-singular terms $\log(m_f^2/\mu^2)$ in both the electric charge renormalization constant and the photon wave-function renormalization constant. If the number of external photons equals that of the EW couplings in the tree-level amplitude, the full NLO EW correction is free of these unpleasant large logarithms because of the exact cancellation between the logarithms in the vertex counterterm and in the photon wave-function counterterm. Therefore, it is reasonable to adopt the $\alpha(0)$ scheme for all the EW couplings involved in the $\gamma\gamma \rightarrow e^+e^-\gamma$ process.

\subsection{Integrated cross sections}
\par
In order to verify the correctness of our numerical calculations for the integrated cross section, we perform the following checks:
\begin{itemize}
\item We calculate the LO cross section for $\gamma\gamma \rightarrow e^+e^-\gamma$ in $\text{J} = 0$ collision at $\sqrt{\hat{s}} = 500~ \text{GeV}$ by employing our developed {\sc FeynArts-3.7}+{\sc FormCalc-7.3}+{\sc LoopTools-2.8} and {\sc MadGraph5-2.3.3} \cite{Alwall:2014hca} packages, respectively, and obtain
    \begin{equation}
    \sigma_{\text{LO}}~[\text{pb}] = \left\lbrace
    \begin{aligned}
    & 0.042817(5) &\quad& (\text{{\it FeynArts}+{\it FormCalc}+{\it LoopTools}})
    \\
    & 0.04279(1)  &\quad& (\text{{\it MadGraph}})
    \end{aligned}
    \right.
    \end{equation}
    These two results are in good agreement with each other.
\item We verify numerically the independence of the full NLO EW corrected cross section on the fictitious photon mass $m_{\gamma}$ in the range of $10^{-15} \leqslant m_{\gamma}/\text{GeV} \leqslant 1$.
\item We calculate the NLO EW corrected cross section for $\gamma\gamma \rightarrow e^+e^-\gamma$ in $\text{J} = 0$ collision at $\sqrt{\hat{s}} = 250, 500, 1000~ \text{GeV}$ in scheme-I by adopting the DS (with $\alpha = 0.1$) and TCPSS (with $\delta_s = \delta_c = 0.001$) methods separately, and find that the numerical results (shown in  Table \ref{tab1}) are coincident with each other within the calculation errors.
    \begin{table}[!htbp]
    \renewcommand \tabcolsep{6.0pt}
    \renewcommand \arraystretch{1.1}
    \centering
    \begin{tabular}{|c|c|ccc|}
    \hline\hline
    \multicolumn{2}{|c|}{$\sqrt{\hat{s}}~\text{[GeV]}$} & $250$ & $500$ & $1000$ \\
    \hline
    \multirow{2}*{$\sigma_{\text{NLO}}^{\text{(I)}}~ \text{[pb]}$} &
    DS      & $~~0.15534(4)~~$ & $~~0.04261(2)~~$ & $~~0.01090(1)~~$  \\
    & TCPSS & $~~0.15525(9)~~$ & $~~0.04260(3)~~$ & $~~0.01092(2)~~$  \\
    \hline\hline
    \end{tabular}
    \caption{
    \label{tab1}
    NLO EW corrected cross sections for $\gamma\gamma \rightarrow e^+e^-\gamma$ in $\text{J} = 0$ collision at $\sqrt{\hat{s}} = 250, 500, 1000~ \text{GeV}$ in the inclusive event selection scheme obtained by using the DS and TCPSS methods separately.}
    \end{table}
\end{itemize}
In further numerical calculations, we adopt only the DS method with $\alpha = 0.1$, and fix the fictitious photon mass as $m_{\gamma} = 10^{-2}~ \text{GeV}$. For brevity's sake, the NLO EW corrected cross sections and the corresponding EW and pure QED relative corrections in the inclusive and exclusive event selection schemes are denoted by $\sigma_{\text{NLO}}^{\text{(I, II)}}$, $\delta_{\text{EW}}^{\text{(I, II)}}$ and $\delta_{\text{QED}}^{\text{(I, II)}}$, respectively. Since the colliding energy dependence of the NLO EW corrected integrated cross section and differential distributions in scheme-I are almost the same as the LO predictions, we only depict the integrated and differential cross sections in scheme-II in the following discussion.

\par
\begin{figure}[!htbp]
\begin{center}
\includegraphics[width=0.45\textwidth]{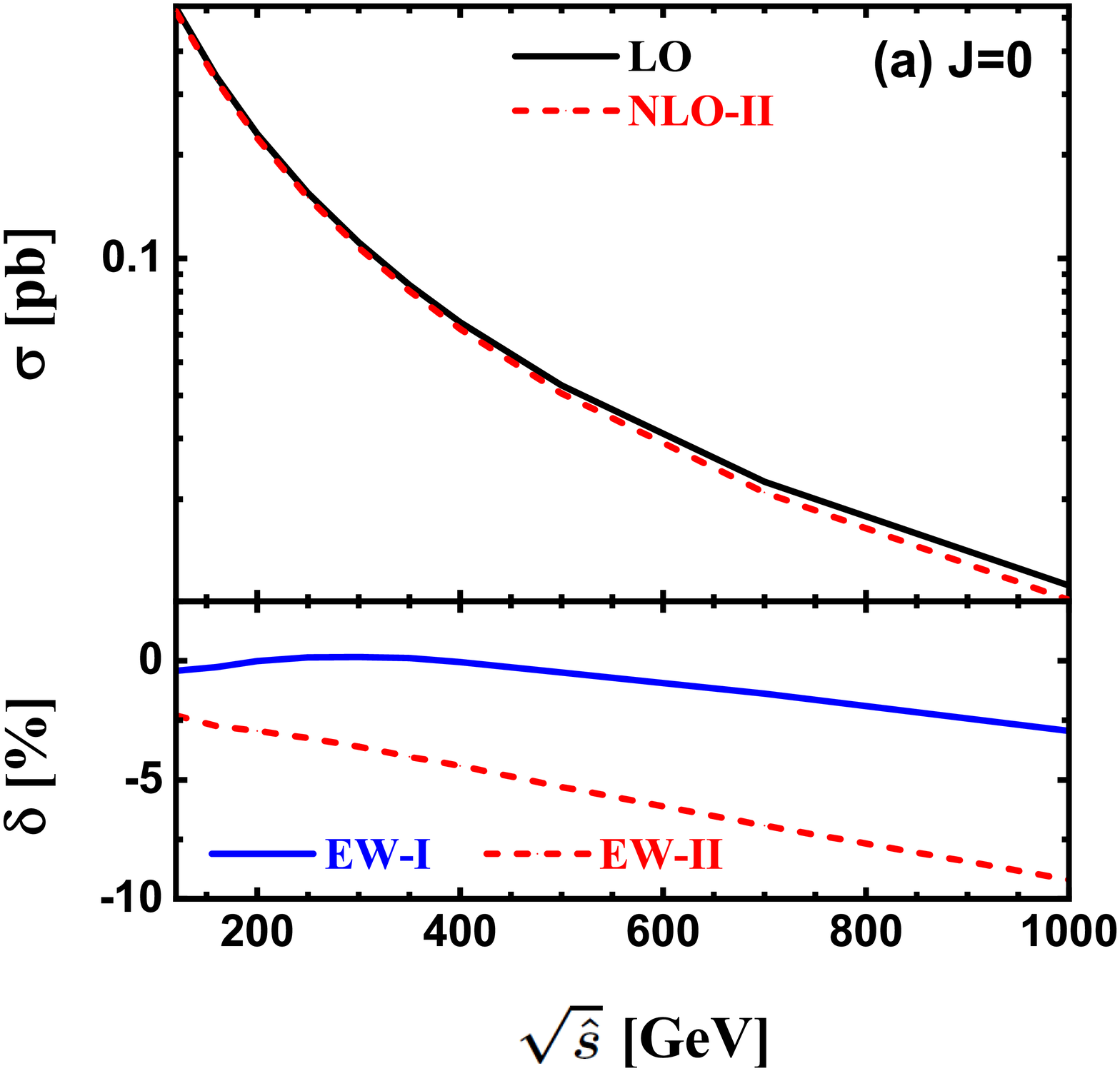}
\includegraphics[width=0.45\textwidth]{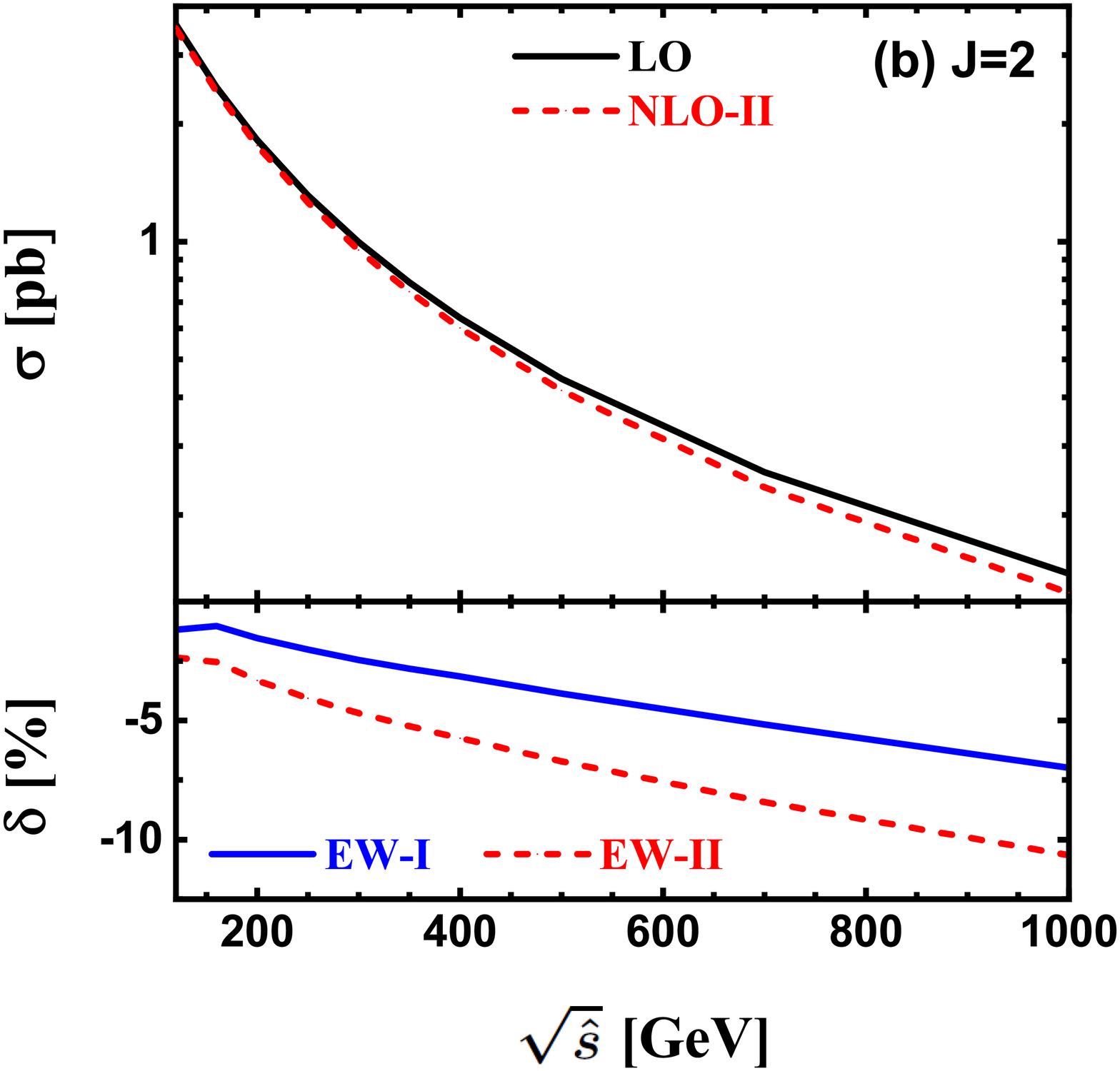}
\caption{LO, NLO EW corrected integrated cross sections in scheme-II and the EW relative corrections in both scheme-I and -II as functions of $\sqrt{\hat{s}}$ for the $e^+e^-\gamma$ production in (a) $\text{J} = 0$ and (b) $\text{J} = 2$ $\gamma\gamma$ collisions.}
\label{fig3}
\end{center}
\end{figure}

In Figs.\ref{fig3} (a) and (b), we present the integrated cross sections $\sigma_{\text{LO}}$, $\sigma^{\text{(II)}}_{\text{NLO}}$ (in upper panels) and the corresponding EW relative corrections $\delta_{\text{EW}}^{\text{(I, II)}}$ (in lower panels) as functions of the $\gamma\gamma$ c.m. colliding energy for the $e^+e^-\gamma$ production in $\text{J} = 0$ and $\text{J} = 2$ $\gamma\gamma$ collisions, respectively. As shown in this figure, the LO and NLO EW corrected cross sections for $\gamma\gamma \rightarrow e^+e^-\gamma$ in both $\text{J} = 0$ and $\text{J} = 2$ $\gamma\gamma$ collision modes decrease quickly as $\sqrt{\hat{s}}$ increases from $120~ \text{GeV}$ to $1~ \text{TeV}$. The EW relative correction is sensitive to the $\gamma\gamma$ colliding energy. In the inclusive event selection scheme, it increases in the low colliding energy region, reaches its maximum at $\sqrt{\hat{s}} \sim 300$ and $160~ \text{GeV}$ for the $\text{J} = 0$ and $\text{J} = 2$ collision modes, respectively, and then decreases gradually as the increase of $\sqrt{\hat{s}}$. In the exclusive event selection scheme, the EW relative correction is negative in the plotted $\sqrt{\hat{s}}$ region. It decreases monotonically from $-2.30\%$ to $-9.20\%$ and from $-2.36\%$ to $-10.65\%$ for $\text{J} = 0$ and $\text{J} = 2$, respectively, as $\sqrt{\hat{s}}$ varies from $120~ \text{GeV}$ to $1~ \text{TeV}$.

\par
\begin{figure}[!htbp]
\begin{center}
\includegraphics[width=0.45\textwidth]{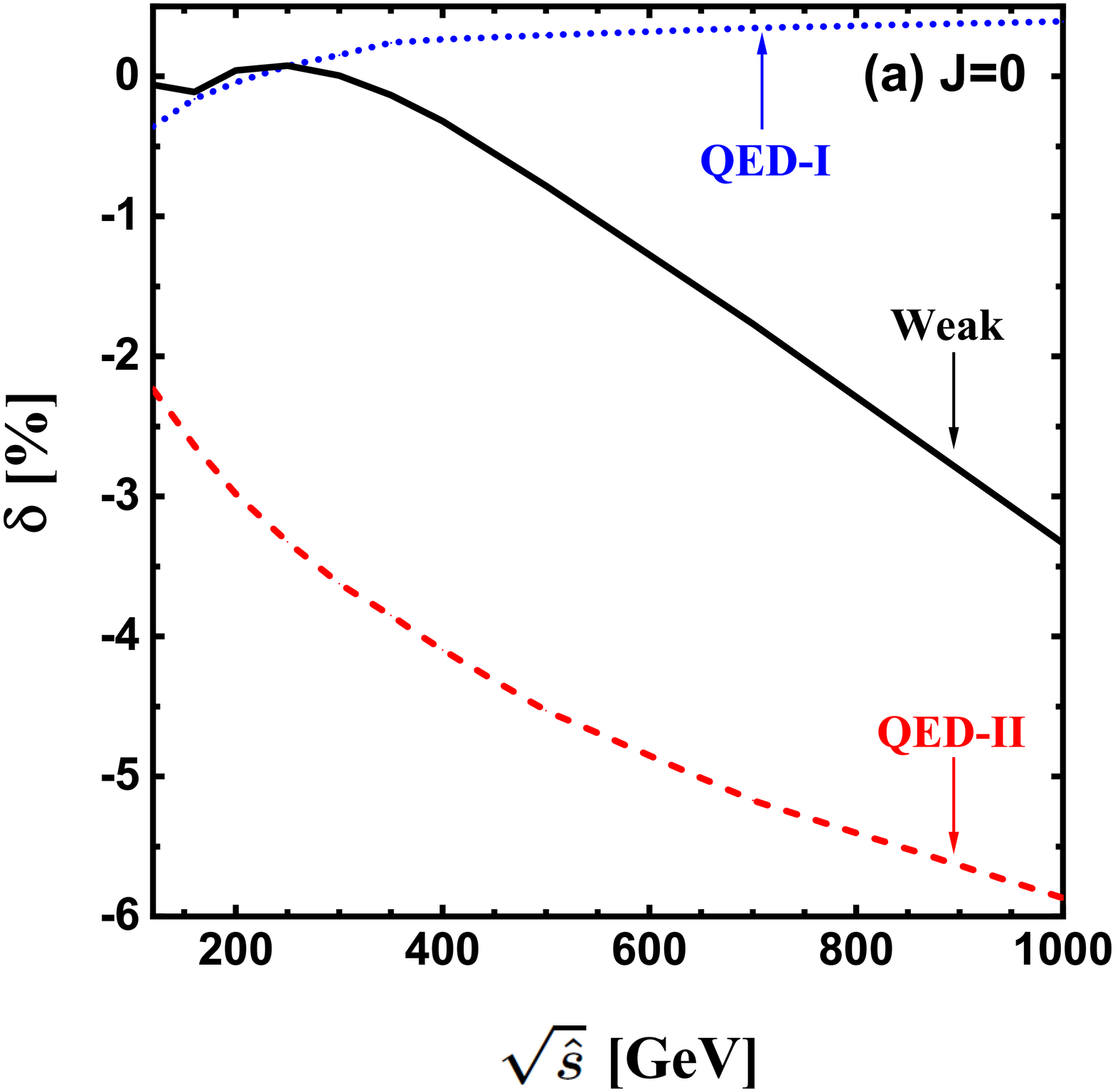}
\includegraphics[width=0.45\textwidth]{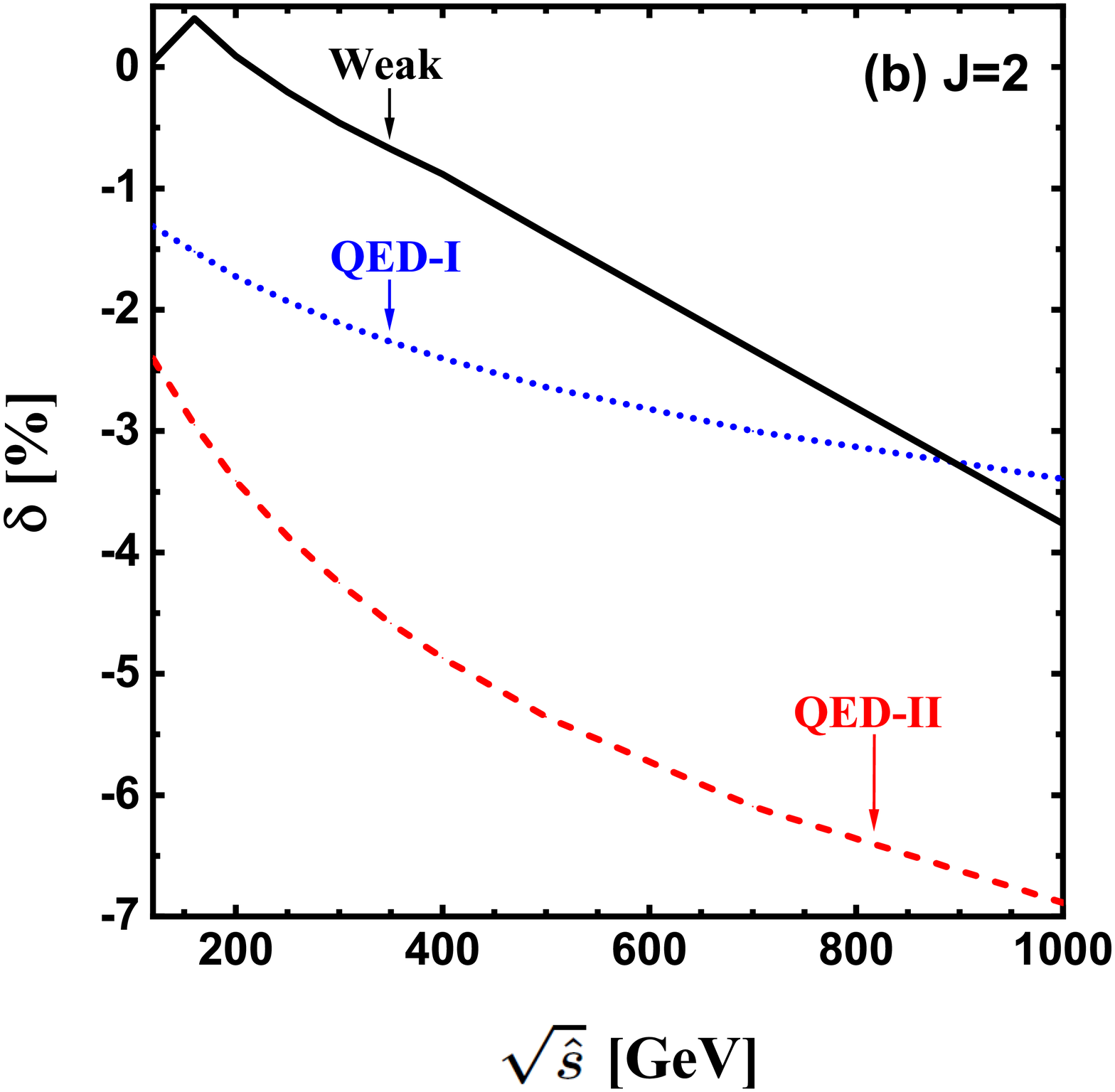}
\caption{QED and weak relative corrections in both inclusive and exclusive event selection schemes as functions of the $\gamma\gamma$ colliding energy for $e^+e^-\gamma$ production in (a) $\text{J} = 0$ and (b) $\text{J} = 2$ $\gamma\gamma$ collisions.}
\label{fig4}
\end{center}
\end{figure}

To analyze the constituents of the NLO EW correction more clearly, we depict the dependence of the pure QED and genuine weak relative corrections in both inclusive and exclusive event selection schemes on the $\gamma\gamma$ colliding energy for the $e^+e^-\gamma$ production in $\text{J} = 0$ and $\text{J} = 2$ $\gamma\gamma$ collisions in Figs.\ref{fig4} (a) and (b) separately\footnote{The weak relative correction in the exclusive scheme is the same as that in the inclusive scheme.}. In the exclusive event selection scheme, both QED and genuine weak relative corrections strongly depend on the $\gamma\gamma$ colliding energy, and the full NLO EW correction is dominated by the QED correction. As $\sqrt{\hat{s}}$ increases from $120~ \text{GeV}$ to $1~ \text{TeV}$, the pure QED relative correction decreases consistently from $-2.23\%$ to $-5.87\%$ for the $\text{J} = 0$ collision mode, and from $-2.40\%$ to $-6.89\%$ for the $\text{J = 2}$ collision mode. Compared to the QED relative correction, the weak relative correction is not a monotonically decreasing function of $\sqrt{\hat{s}}$. It can be seen from Fig.\ref{fig4} (b) that there is a small peak at $\sqrt{\hat{s}} \simeq 2 m_W \simeq 160~ \text{GeV}$ in the colliding energy distribution of the weak relative correction to the $e^+e^-\gamma$ production via $\text{J} = 2$ $\gamma\gamma$ collision, which corresponds to the $W$-pair resonance induced by the triangle loop in Fig.\ref{fig2} (1). For both the $\text{J} = 0$ and $\text{J} = 2$ collision modes, the weak relative correction is small ($\left| \delta_{\text{W}} \right| < 0.5\%$) when $\sqrt{\hat{s}} < 300~ \text{GeV}$, while it becomes relatively remarkable in the high energy region due to the Sudakov logarithms induced by the virtual exchange of soft or collinear massive weak gauge bosons \cite{Denner:2000jv,Denner:2001gw}. At $\sqrt{\hat{s}} = 1~ \text{TeV}$, $(\delta_{\text{W}},\, \delta_{\text{EW}}^{\text{(II)}}) = (-3.33\%,\, -9.20\%)$ and $(-3.76\%,\, -10.65\%)$ for $\text{J} = 0$ and $\text{J} = 2$, respectively. It clearly shows that the full EW relative correction to $e^+e^-\gamma$ production in the exclusive event selection scheme can reach and even exceed $-10\%$ at a TeV PLC. In the inclusive event selection scheme, the QED relative correction contributed by $e^+e^-\gamma\gamma$ events, i.e., $\delta_{\text{QED}}^{\text{(I)}} - \delta_{\text{QED}}^{\text{(II)}}$, is sizable, especially in the high energy region. It increases gradually from $1.88\%$ to $6.26\%$ and from $1.09\%$ to $3.49\%$ for $\text{J} = 0$ and $\text{J} = 2$, respectively, as $\sqrt{\hat{s}}$ increases from $120~ \text{GeV}$ to $1~ \text{TeV}$.

\subsection{Kinematic distributions}
\par
\begin{figure}[!htbp]
\begin{center}
\includegraphics[width=0.45\textwidth]{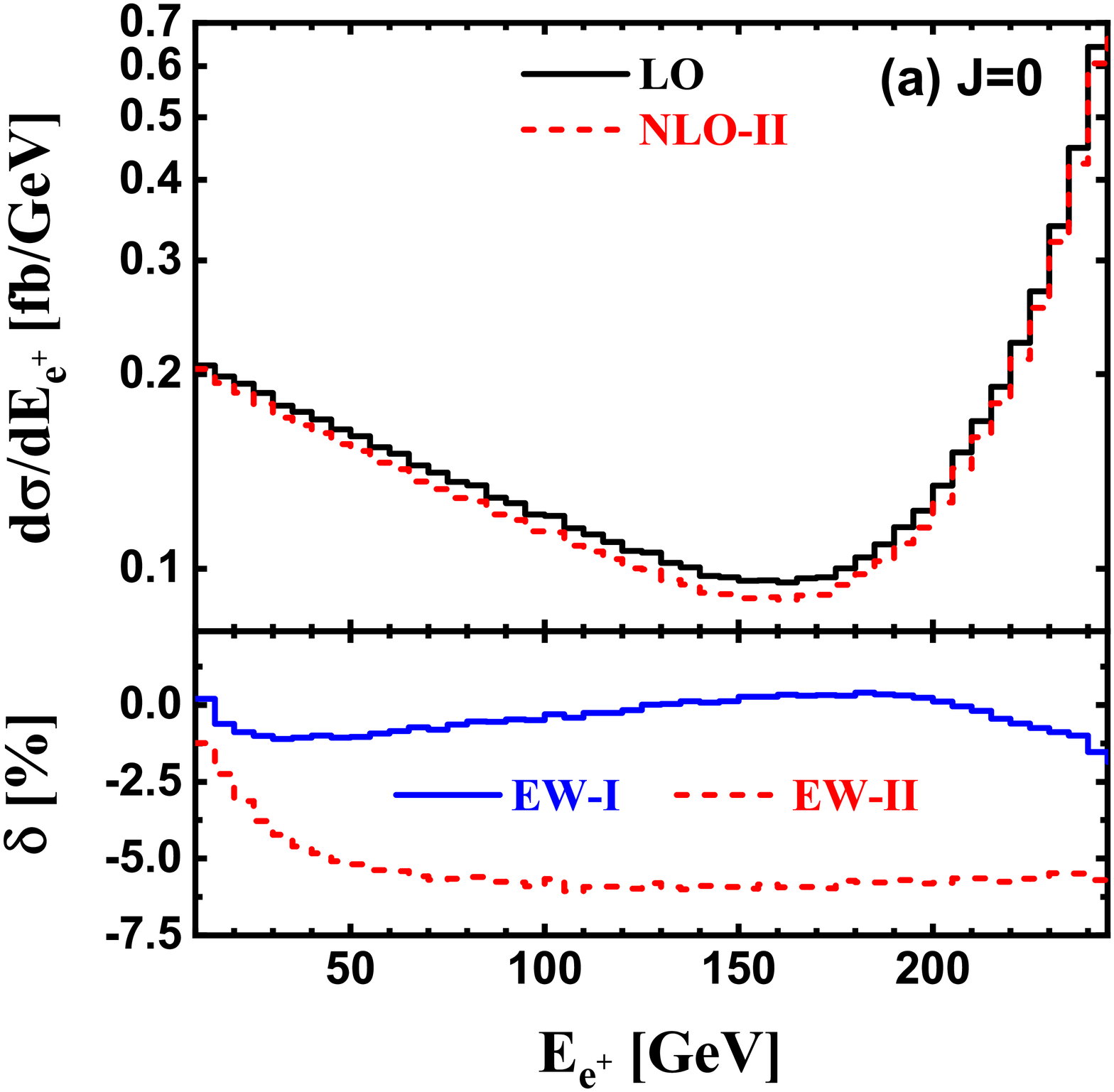}
\includegraphics[width=0.45\textwidth]{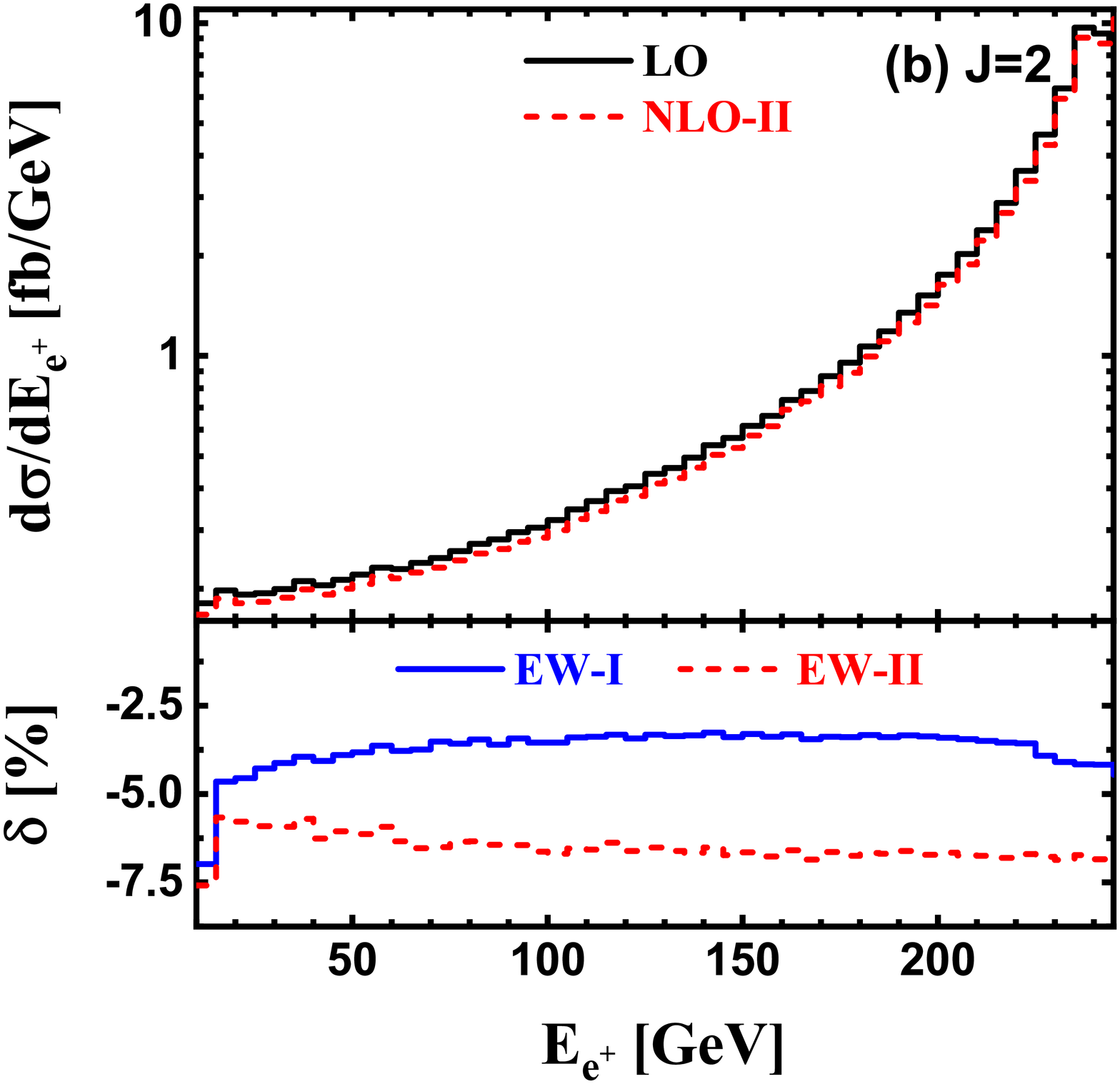}
\includegraphics[width=0.45\textwidth]{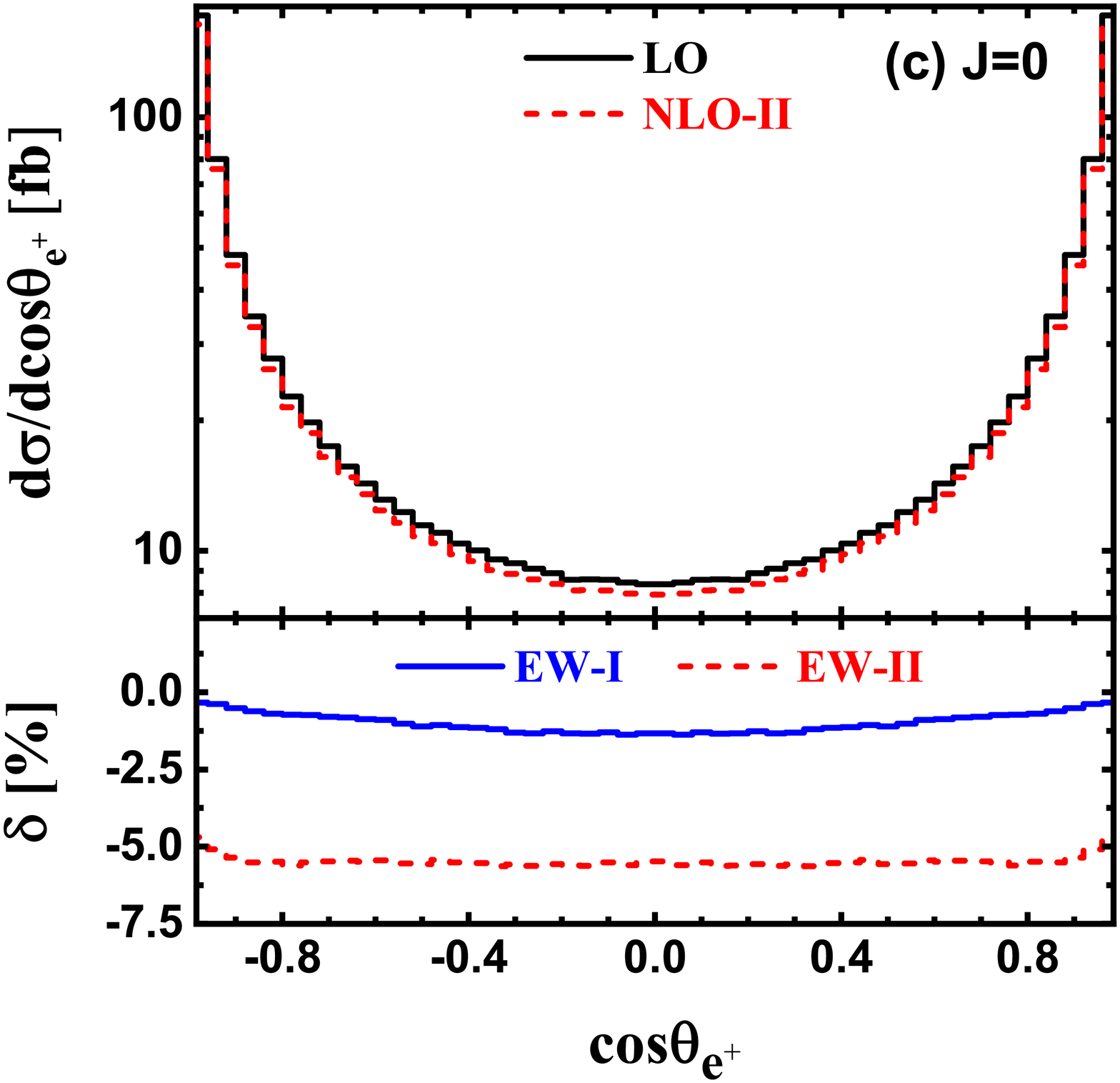}
\includegraphics[width=0.45\textwidth]{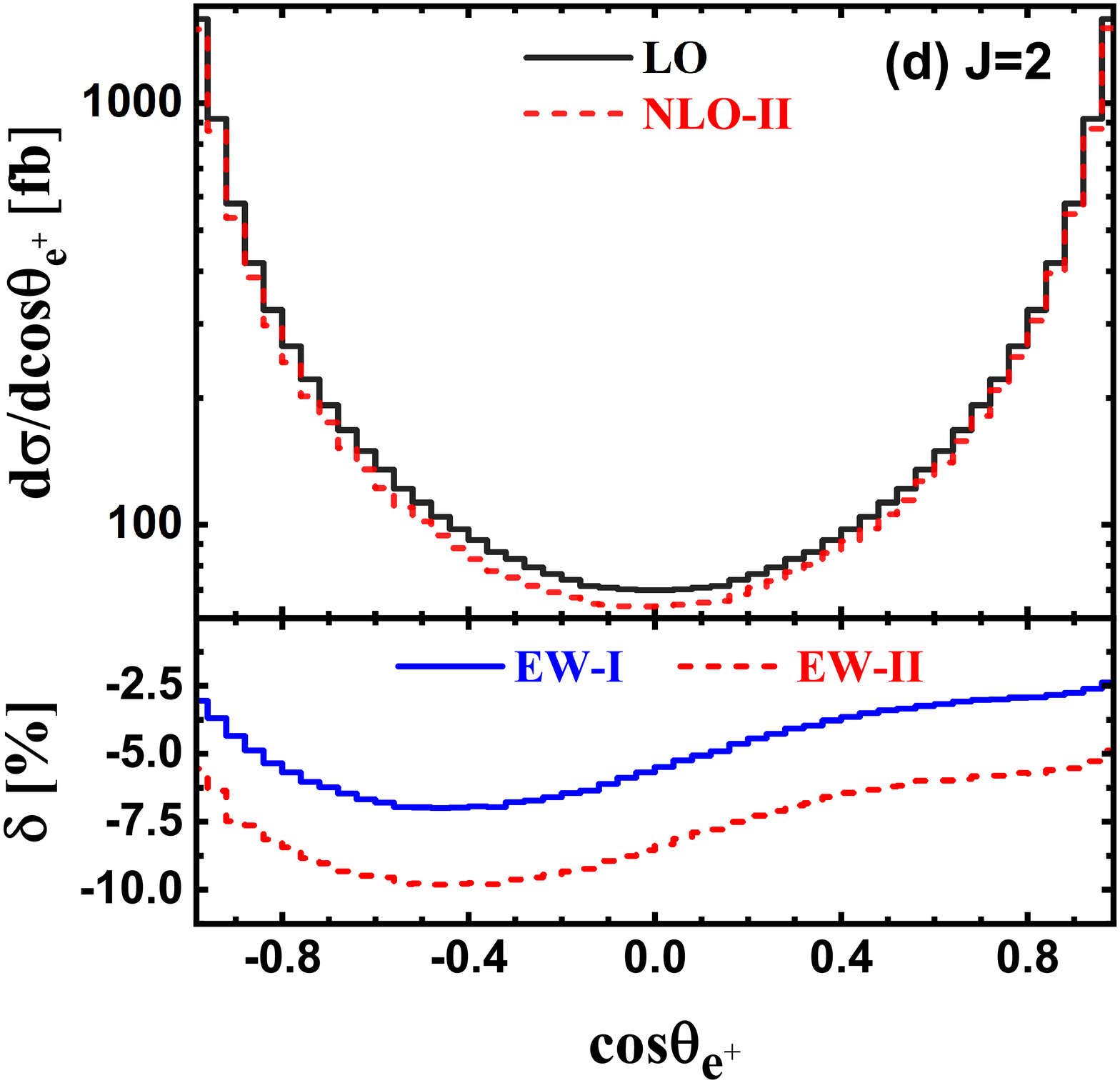}
\caption{LO, NLO EW corrected differential distributions of the final-state positron and the corresponding EW relative corrections for the $e^+e^-\gamma$ production at $\sqrt{\hat{s}} = 500~ \text{GeV}$ PLC. (a), (b), (c) and (d) are for the energy and scattering angle distributions in $\text{J} = 0$ and $\text{J} = 2$ collision modes, respectively.}
\label{fig5}
\end{center}
\end{figure}

The LO, NLO EW corrected energy distributions of the final-state positron in scheme-II and the corresponding EW relative corrections in both scheme-I and -II for the $e^+e^-\gamma$ production in $\text{J} = 0$ and $\text{J} = 2$ collisions at $\sqrt{\hat{s}} = 500~ \text{GeV}$ PLC are presented in Figs.\ref{fig5} (a) and (b), respectively. As $\sqrt{\hat{s}}$ increases, the LO and NLO EW corrected $E_{e^+}$ distributions for $\text{J} = 0$ $\gamma\gamma$ collision mode decrease gradually, reach their minima at $\sqrt{\hat{s}} \sim 160~ \text{GeV}$, and then increase rapidly to their maxima at $E_{e^+} \sim \sqrt{\hat{s}}/2 - E_{\gamma, \text{min}} = 240~ \text{GeV}$, which corresponds to the minimum-energy photon emitted from positron, while the $E_{e^+}$ distributions for $\text{J} = 2$ collision mode increases monotonically before reaching their maxima at $\sqrt{\hat{s}} \sim 240~ \text{GeV}$. The LO, NLO EW corrected scattering angle distributions of the final-state positron and the corresponding EW relative corrections are depicted in Figs.\ref{fig5} (c) and (d). For the $\text{J} = 0$ collision mode, both $d\sigma_{\text{LO}}/d\cos\theta_{e^+}$ and $d\sigma_{\text{NLO}}^{\text{(I, II)}}/d\cos\theta_{e^+}$ are symmetric with respect to $\cos\theta_{e^+} = 0$\footnote{The forward-backward symmetry of $d\sigma_{\text{NLO}}^{\text{(I)}}/d\cos\theta_{e^+}$ can be read off from the lower panel of Fig.\ref{fig5} (c).}. The corresponding EW relative correction increases from $-1.3\%$ to $-0.3\%$ as $\cos\theta_{e^+}$ varies from $0$ to $\cos 10^{\circ} = 0.985$ in the inclusive event selection scheme, while it is steady at around $-5.5\%$ for $\left| \cos\theta_{e^+} \right| < 0.9$ in the exclusive event selection scheme. For the $\text{J} = 2$ collision mode, although the LO $\cos\theta_{e^+}$ distribution is also symmetric with respect to $\cos\theta_{e^+} = 0$, the NLO EW correction breaks this forward-backward symmetry. In the exclusive event selection scheme, the EW relative correction is negative and notable ($| \delta_{\text{EW}}^{\text{(II)}} | > 5\%$), and can reach about $-10\%$ at $\cos\theta_{e^+} \sim 0.5$. The lineshape of the EW relative correction in the inclusive event selection scheme is the same as the exclusive event selection scheme. The QED relative correction contributed by $e^+e^-\gamma\gamma$ events is steady at around $2.8\%$.

\par
\begin{figure}[!htbp]
\begin{center}
\includegraphics[width=0.45\textwidth]{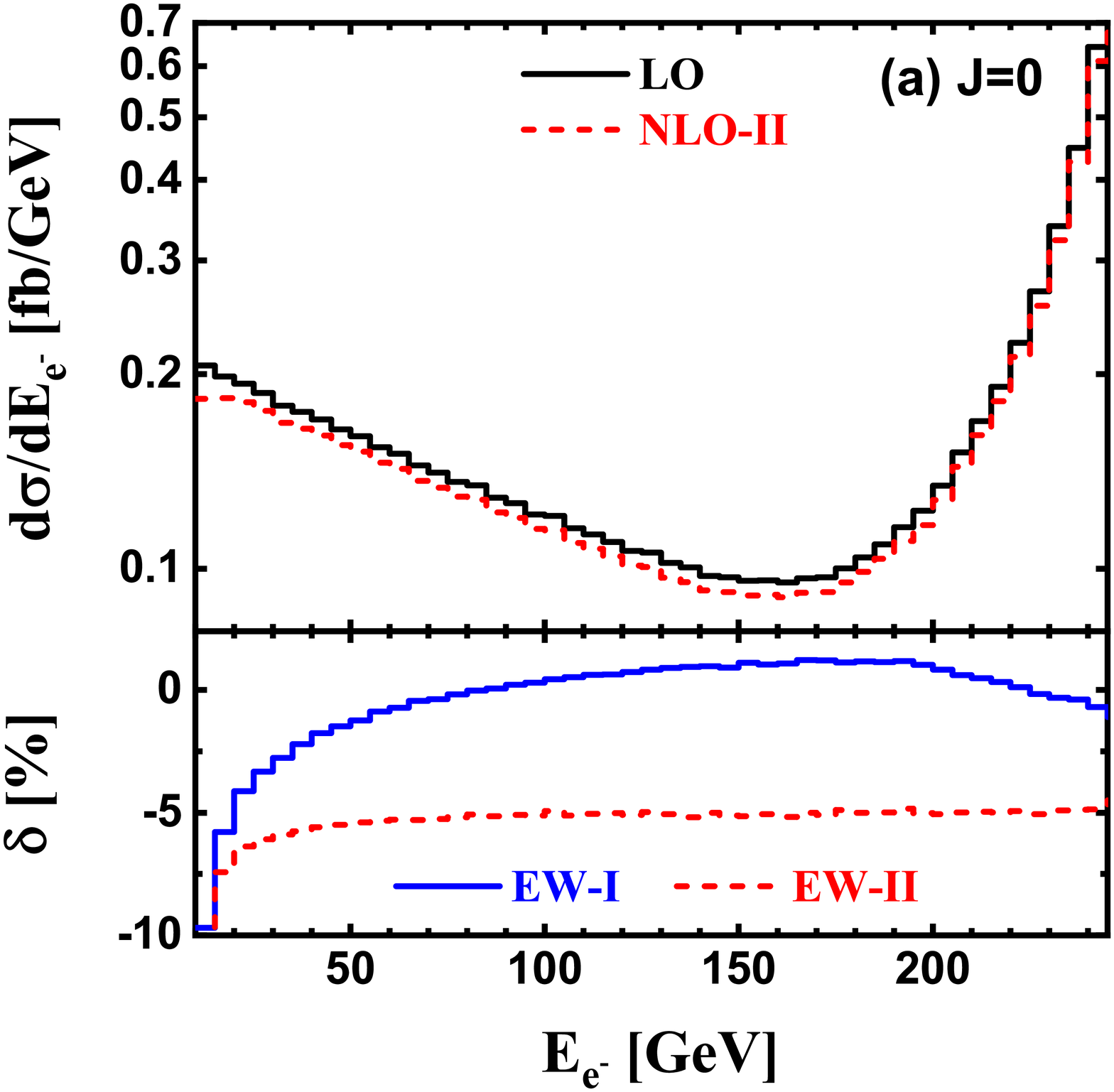}
\includegraphics[width=0.45\textwidth]{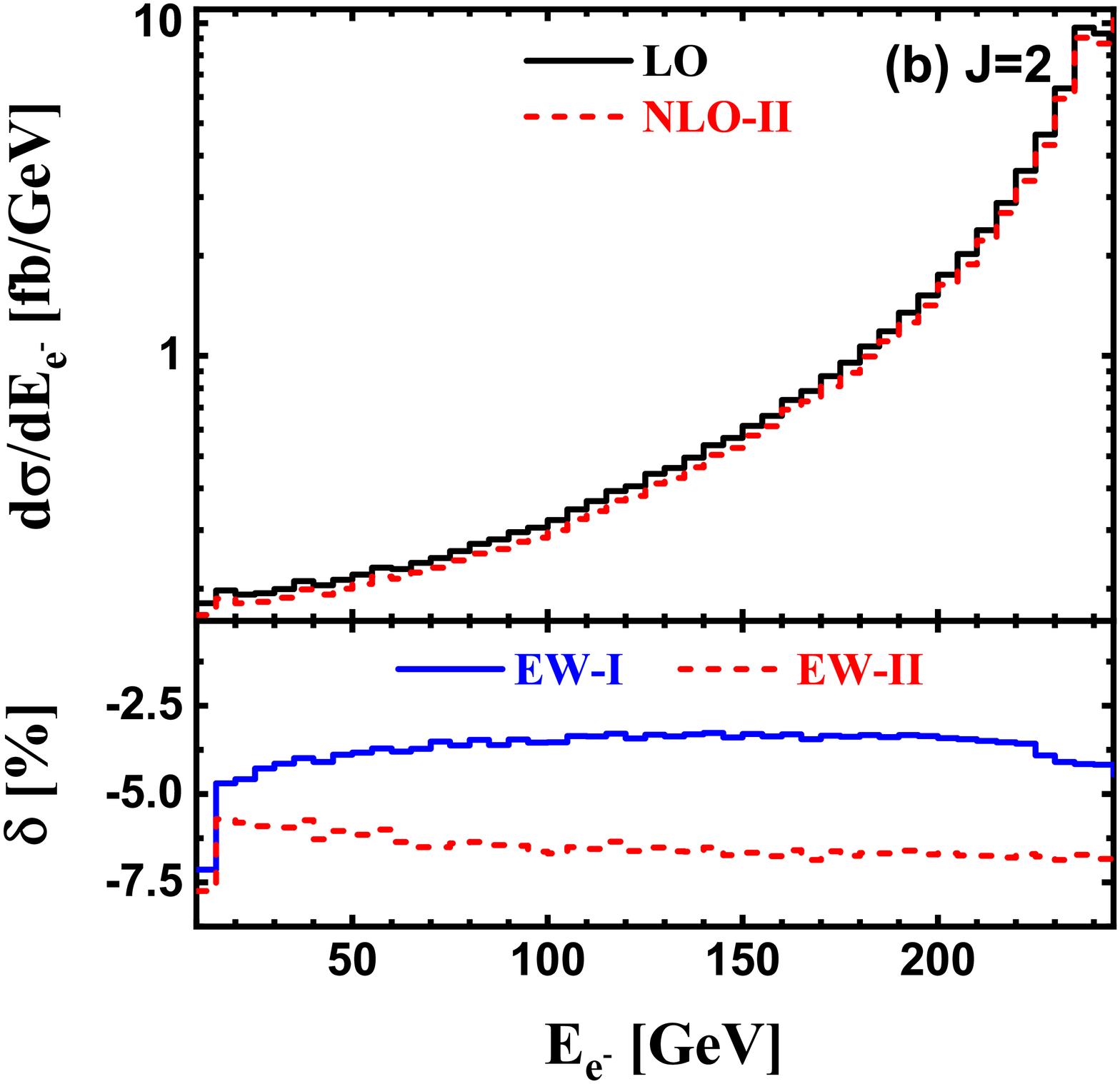}
\includegraphics[width=0.45\textwidth]{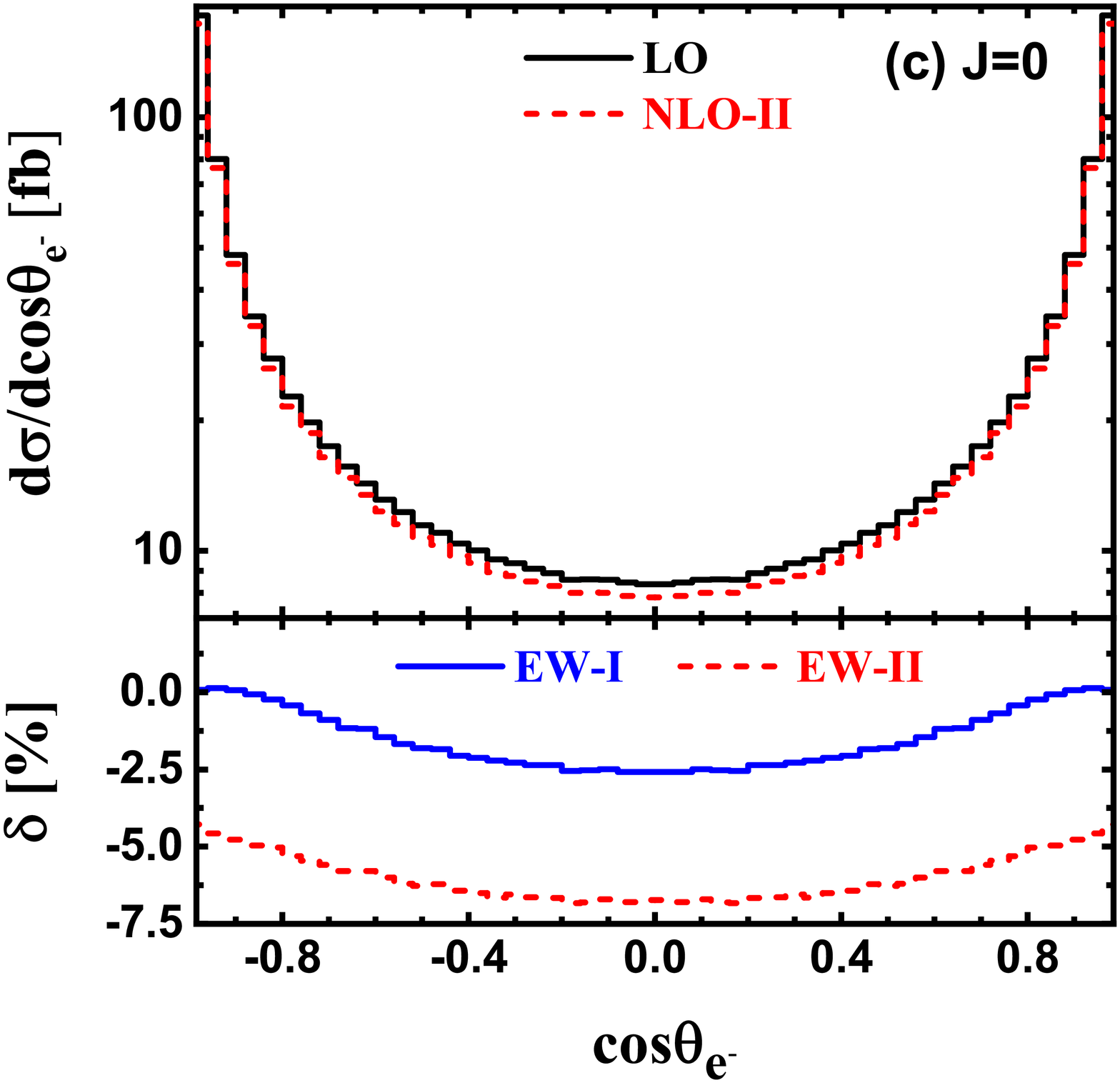}
\includegraphics[width=0.45\textwidth]{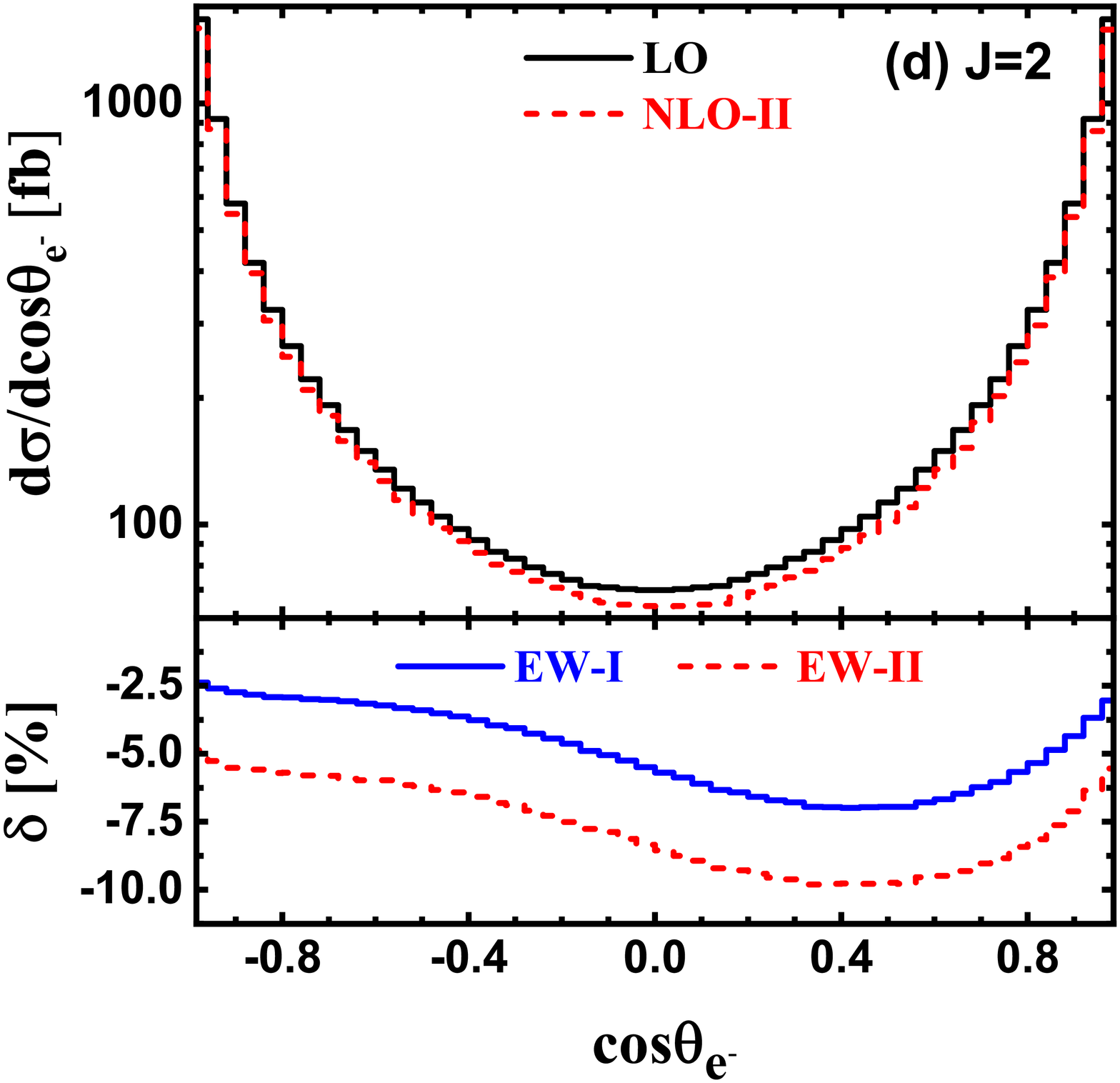}
\caption{Same as Fig.\ref{fig5}, but for the final-state electron.}
\label{fig6}
\end{center}
\end{figure}

The LO, NLO EW corrected energy and scattering angle distributions of the final-state electron as well as the corresponding EW relative corrections for the $e^+e^-\gamma$ production in $\text{J} = 0$ and $\text{J} = 2$ collisions at $\sqrt{\hat{s}} = 500~ \text{GeV}$ PLC are plotted in Figs.\ref{fig6} (a)-(d). As expected, all the kinematic distributions of the final-state electron are the same as the corresponding ones of the final-state positron for both $\gamma_+\gamma_+ \rightarrow e^+e^-\gamma$ ($\text{J}$ = 0) and $\gamma_+\gamma_- \rightarrow e^+e^-\gamma$ ($\text{J}$ = 2) channels at the LO due to the charge symmetry. However, it should be noted that the charge conjugation is only a Born-level symmetry operation for the $e^+e^-\gamma$ production at the PLC since $\gamma\gamma \rightarrow e^+e^-\gamma$ is a pure QED process at the lowest order, and the charge symmetry will be broken at the EW NLO by weak correction. Consequently, the EW relative corrections to the kinematic distributions of electron differ from the corresponding ones of the positron for the $e^+e^-\gamma$ production in the $\text{J} = 0$ $\gamma\gamma$ collision, as shown in the lower panels of Figs.\ref{fig5} (a, c) and Figs.\ref{fig6} (a, c). As stated in Sec.\ref{subsection-2A}, the kinematic behaviors of the final-state electron should be the same as positron for $\text{J} = 2$ $\gamma\gamma$ collision mode due to the $\mathcal{CP}\text{+}{\it Bose}$ symmetry. It can be concluded from Eq.(\ref{CP-Bose}) that
\begin{equation}
\label{CPB-symmetry}
\left.
\frac{d \sigma}{dE_{e^-}} = \frac{d \sigma}{dE_{e^+}}
\right|_{E_{e^+} \rightarrow E_{e^-}}
\qquad
\text{and}
\qquad
\frac{d \sigma}{d\cos\theta_{e^-}}
=
\left.
\frac{d \sigma}{d\cos\theta_{e^+}}
\right|_{\theta_{e^+} \rightarrow \pi - \theta_{e^-}}
\end{equation}
at both LO and EW NLO. By comparing Figs.\ref{fig6} (b, d) with Figs.\ref{fig5} (b, d) we can see that the numerical results for $E_{e^{\pm}}$ and $\cos\theta_{e^{\pm}}$ distributions of the $\gamma_+\gamma_- \rightarrow e^+e^-\gamma$ production channel satisfy Eq.(\ref{CPB-symmetry}) exactly. The consistency of our numerical results with the $\mathcal{CP}\text{+}{\it Bose}$ symmetry for $e^+e^-\gamma$ production in $\text{J} = 2$ $\gamma\gamma$ collision also verifies the correctness of our calculations.

\par
\begin{figure}[!htbp]
\begin{center}
\includegraphics[width=0.45\textwidth]{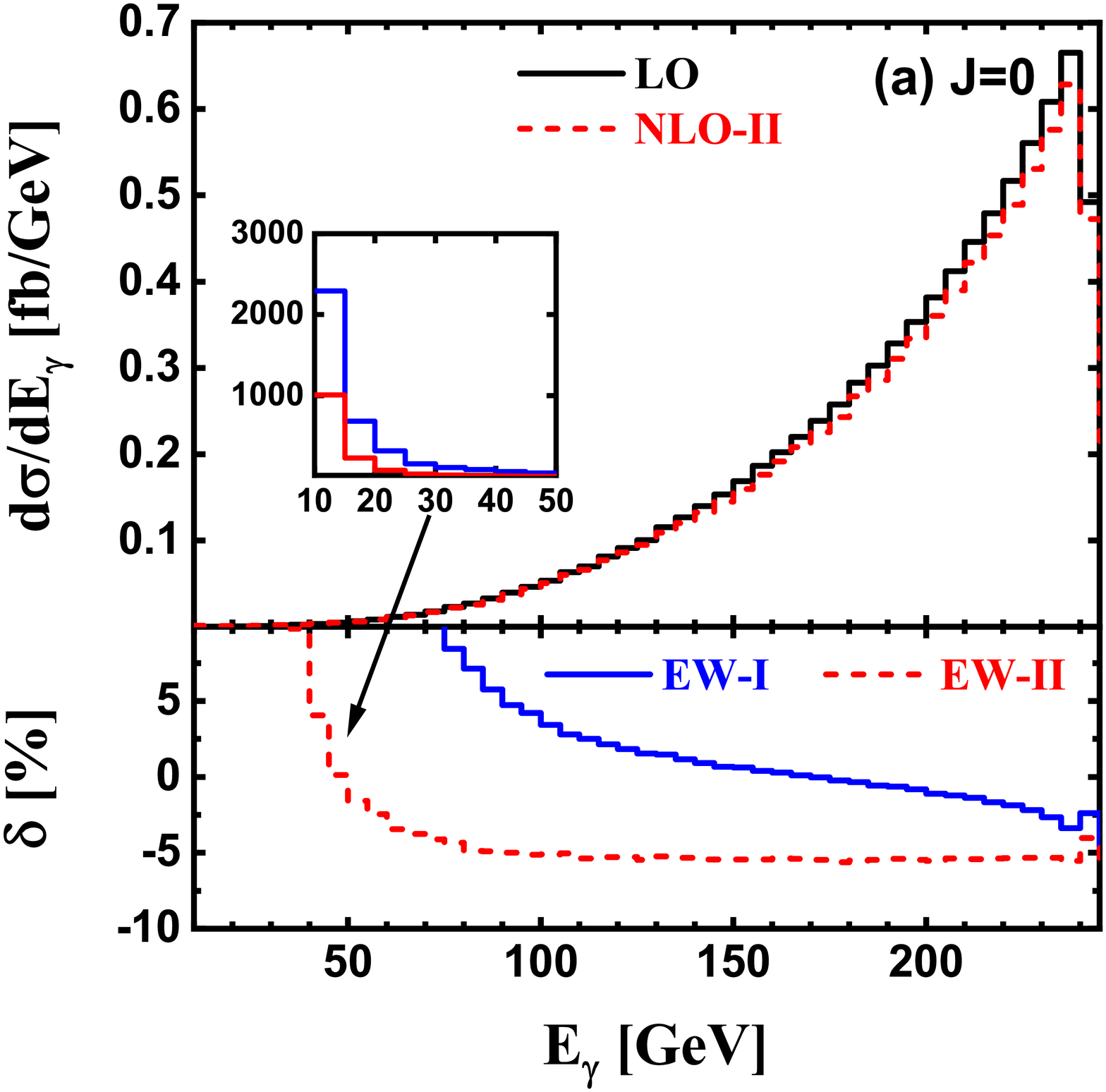}
\includegraphics[width=0.45\textwidth]{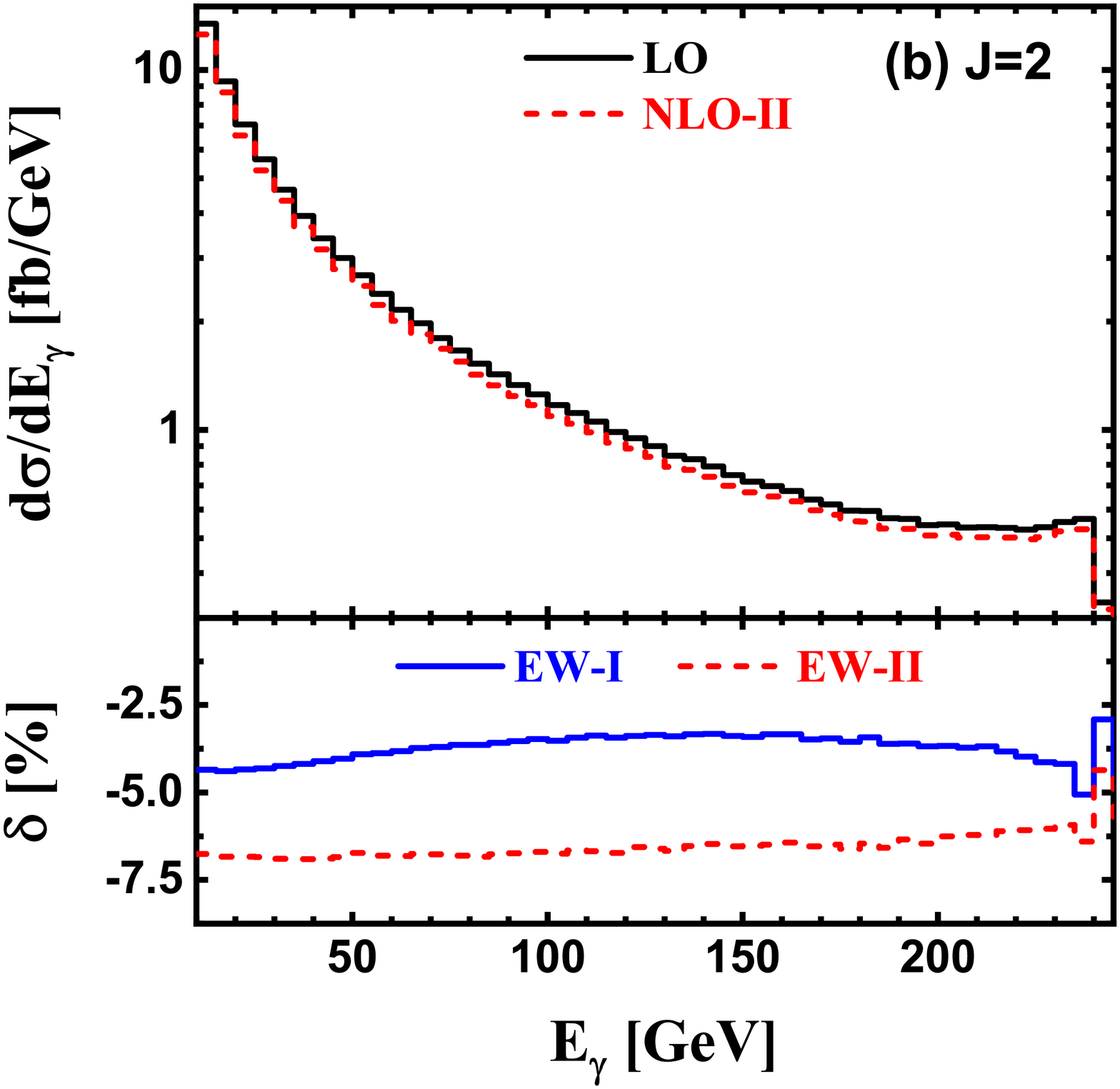}
\includegraphics[width=0.45\textwidth]{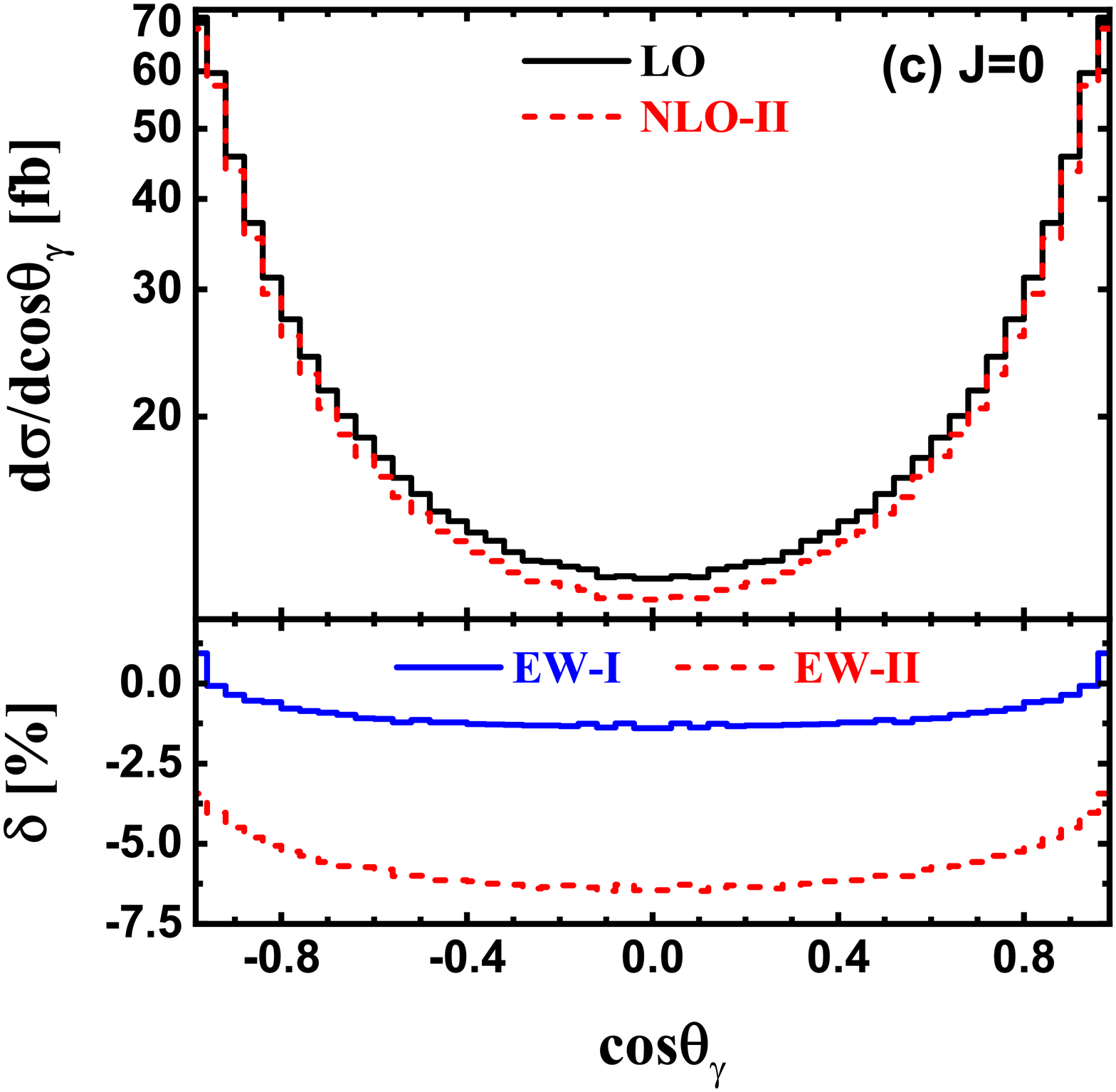}
\includegraphics[width=0.45\textwidth]{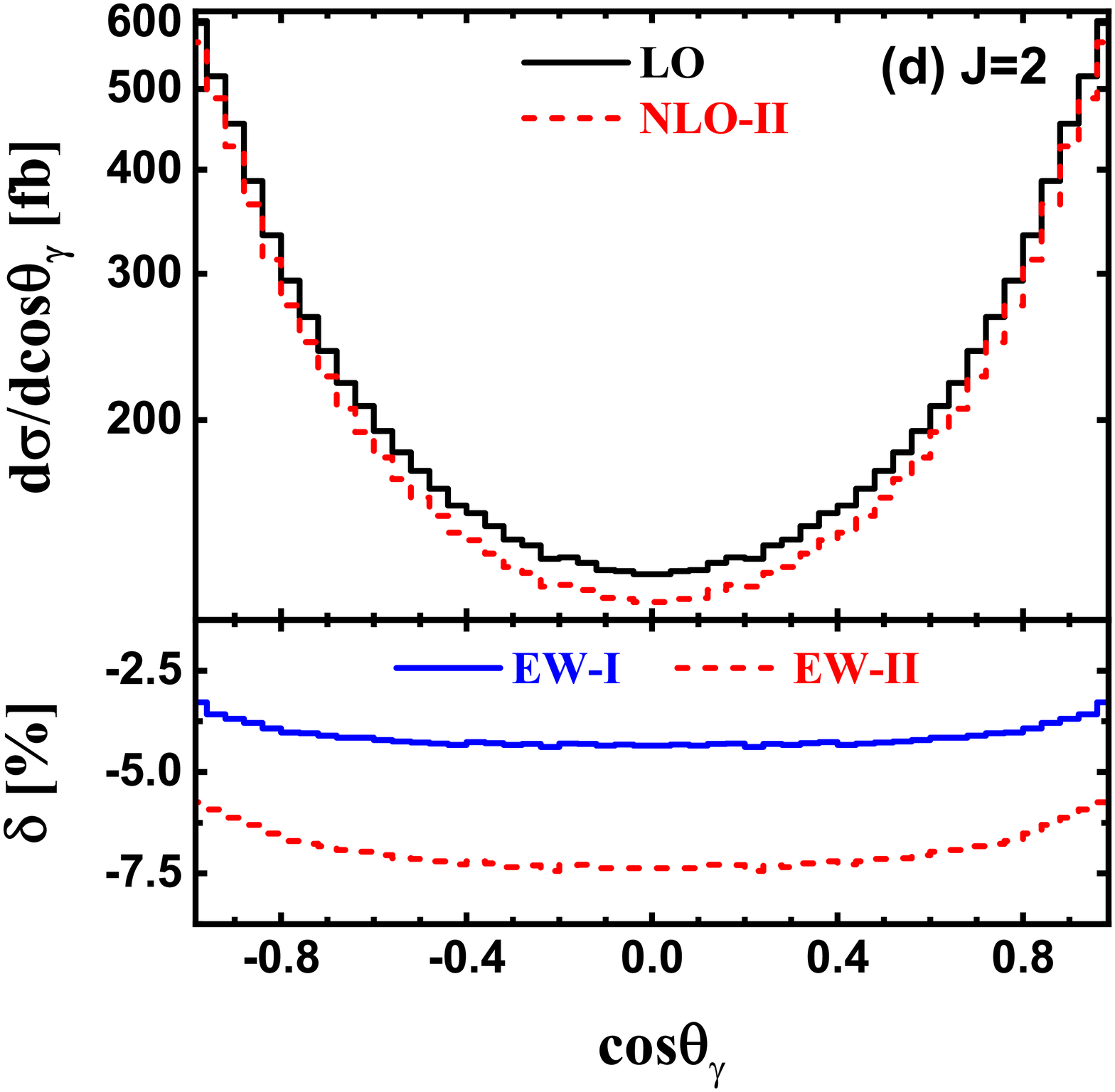}
\caption{Same as Fig.\ref{fig5}, but for the final-state leading photon.}
\label{fig7}
\end{center}
\end{figure}

For an $e^+e^-\gamma\gamma$ event collected in the inclusive event selection scheme, the two photons are called leading and sub-leading photons, respectively, according to their energies in decreasing order (The photon of an $e^+e^-\gamma$ event can also be named as leading photon.). In the upper panels of Figs.\ref{fig7} (a)-(d), we depict the energy and scattering angle distributions of the leading photon for $e^+e^-\gamma$ production in $\text{J} = 0$ and $\text{J} = 2$ $\gamma\gamma$ collisions at $\sqrt{\hat{s}} = 500~ \text{GeV}$ PLC at both LO and EW NLO. The corresponding EW relative corrections are provided in the lower panels. As shown in Figs.\ref{fig7} (a) and (b), the LO and NLO EW corrected energy distributions of the leading photon in scheme-II increase consistently as $\sqrt{\hat{s}}$ increases from $10$ to $240~ \text{GeV}$ for the $\text{J} = 0$ polarization configuration of the incoming photon beams, while decrease gradually as the increase of $\sqrt{\hat{s}}$ in most plotted $E_{\gamma}$ region for $\text{J} = 2$ $\gamma\gamma$ collision mode. The peak structure at $\sqrt{\hat{s}} \sim 240~ \text{GeV}$ in the EW relative corrections in both inclusive and exclusive event selection schemes can be attributed to the $Z$ resonance effect induced by the loop diagram in Fig.\ref{fig2} (2). At the $Z$ resonance, $M_{e^+e^-} = m_Z$, and thus $E_{\gamma} = (\hat{s} - m_Z^2)/(2 \sqrt{\hat{s}}) \sim 240~ \text{GeV}$ which corresponds to a photon recoiling against an on-shell $Z$ boson. From Figs.\ref{fig7} (c) and (d) we can see that both the LO $\cos\theta_{\gamma}$ distribution and the NLO EW correction are symmetric with respect to $\cos\theta_{\gamma} = 0$, and the differential distribution increases rapidly as the increase of $\left| \cos\theta_{\gamma} \right|$. It implies that the leading photon prefers to be produced along the incoming photon beam directions. In the exclusive event selection scheme, the EW relative correction is negative and sizable. It can exceed $-6\%$ for $\text{J} = 0$ and $-7\%$ for $\text{J} = 2$, respectively, when the final-state leading photon is produced centrally ($\left| \cos\theta_{\gamma} \right| < 0.5$).

\par
\begin{figure}[!htbp]
\begin{center}
\includegraphics[width=0.45\textwidth]{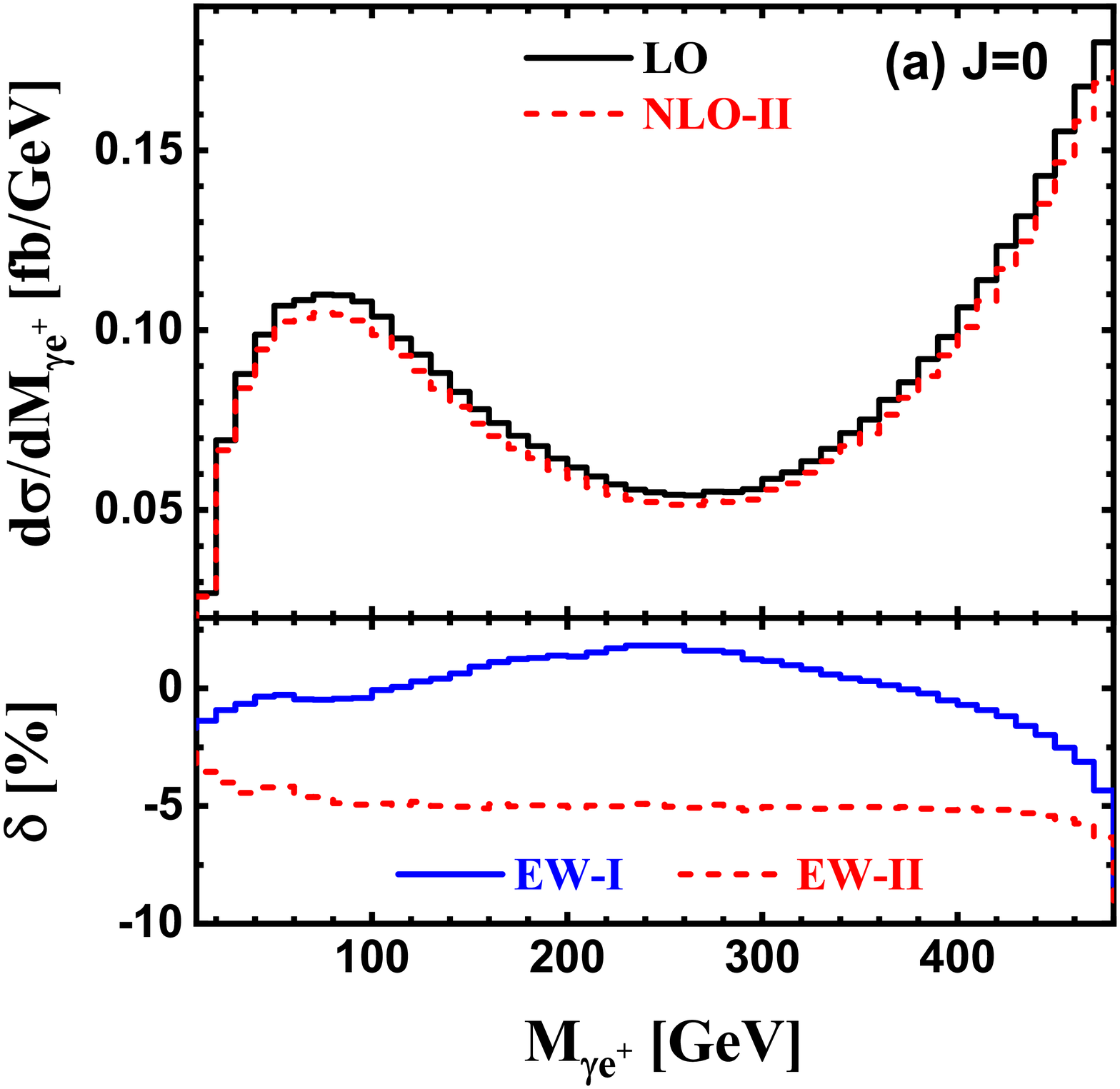}
\includegraphics[width=0.45\textwidth]{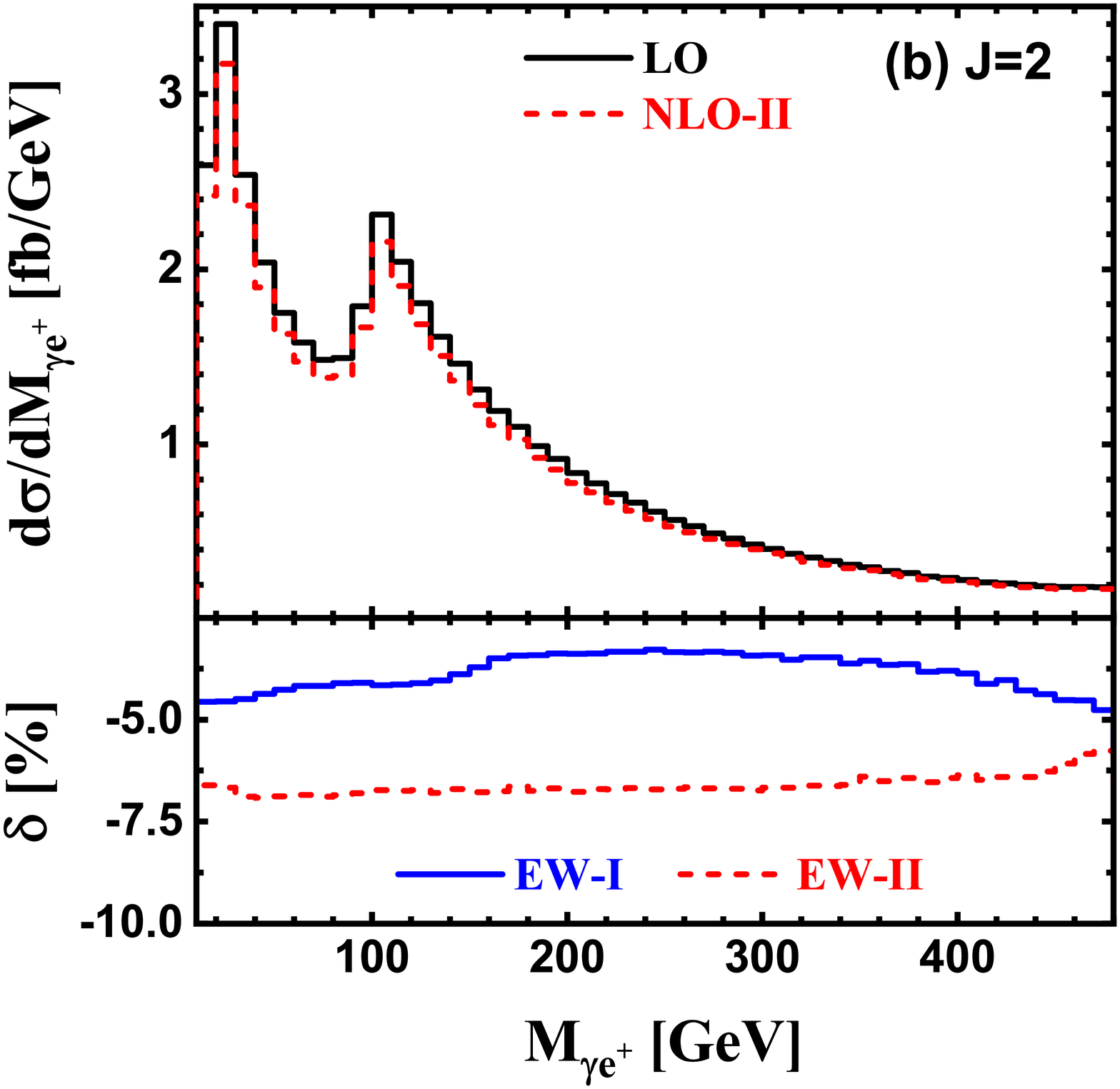}
\includegraphics[width=0.45\textwidth]{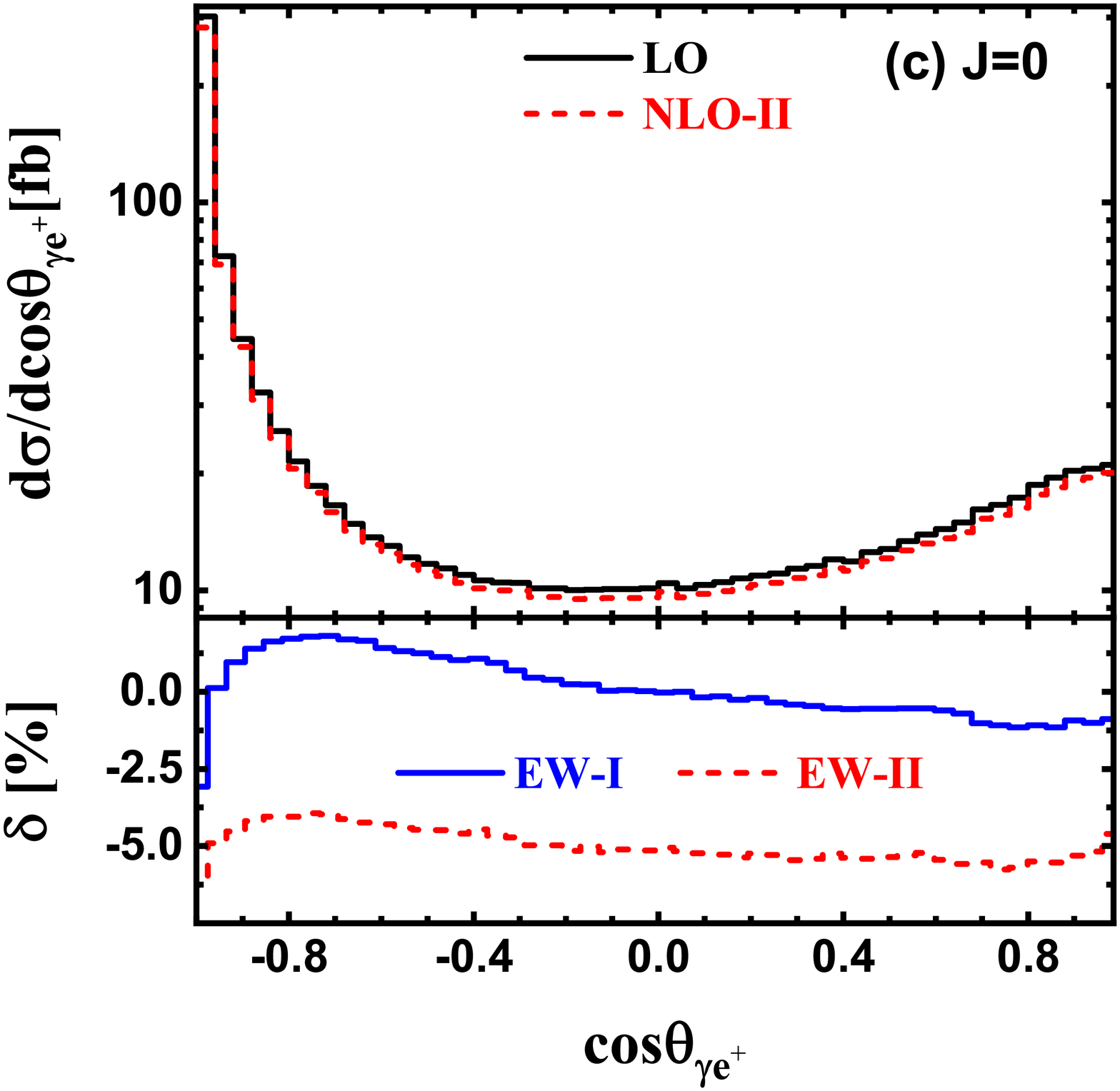}
\includegraphics[width=0.45\textwidth]{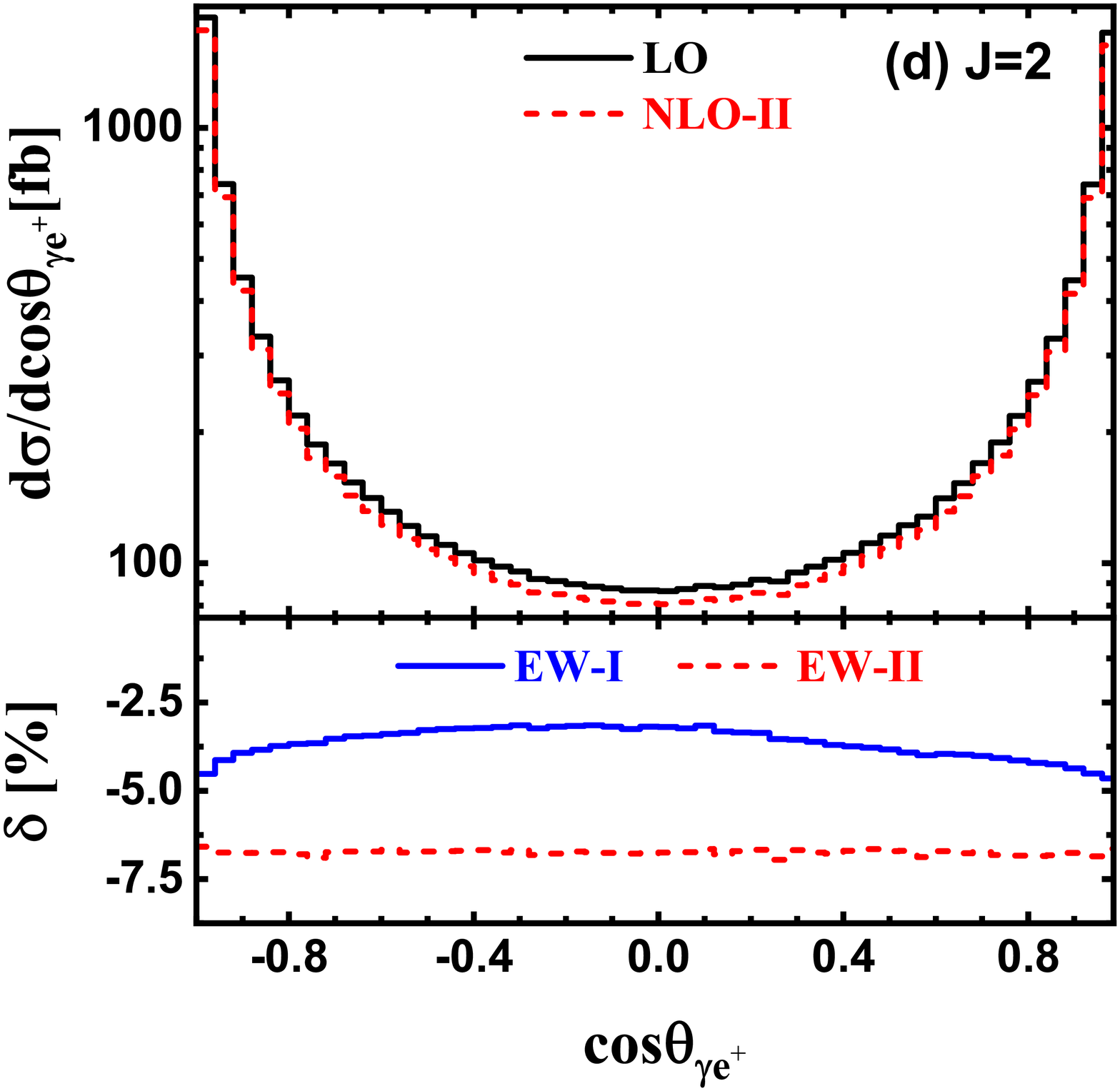}
\caption{LO, NLO EW corrected invariant mass and opening angle distributions of the final-state $\gamma e^+$ system and the corresponding EW relative corrections for the $e^+e^-\gamma$ production in $\text{J} = 0$ and $\text{J} = 2$ $\gamma\gamma$ collisions at $\sqrt{\hat{s}} = 500~ \text{GeV}$ PLC.}
\label{fig8}
\end{center}
\end{figure}

The LO, NLO EW corrected invariant mass and opening angle distributions of the final-state $\gamma e^+$ system (or more strictly speaking, the final-state leading photon and positron) and the corresponding EW relative corrections for $e^+e^-\gamma$ production via $\gamma_+\gamma_+$ and $\gamma_+\gamma_-$ collisions at $\sqrt{\hat{s}} = 500~ \text{GeV}$ PLC are presented in Figs.\ref{fig8} (a)-(b) separately. For $\text{J} = 0$ (i.e., $\gamma_+\gamma_+$) collision mode, the invariant mass distribution of the $\gamma e^+$ system increases sharply in the low $M_{\gamma e^+}$ region, reaches its maximum at $\sqrt{\hat{s}} \sim 70~ \text{GeV}$, and then decreases gradually as $M_{\gamma e^+}$ increases to around $260~ \text{GeV}$. Subsequently, the invariant mass distribution increases rapidly as the increase of $M_{\gamma e^+}$. The EW relative correction in the exclusive event selection scheme is steady at around $-5\%$ in most of the $M_{\gamma e^+}$ region. For $\text{J} = 2$ (i.e., $\gamma_+\gamma_-$) collision mode, there are two sharp peaks located at $M_{\gamma e^+} \sim 20~ \text{GeV}$ and $M_{\gamma e^+} \sim 100~ \text{GeV}$ in the $\gamma e^+$ invariant mass distribution. As we know, the invariant mass of the final-state $\gamma e^+$ system is given by
\begin{equation}
\label{Minv}
M_{\gamma e^+}
\approx
M_{\gamma e^+}^{\text{(LO)}}
=
\sqrt{\hat{s} + m_e^2 - 2 E_{e^-} \sqrt{\hat{s}}}\,,
\end{equation}
where $M_{\gamma e^+}^{\text{(LO)}}$ is the lowest-order approximation of $M_{\gamma e^+}$. From Eq.(\ref{Minv}) we can see that $M_{\gamma e^+} \sim 20~ \text{GeV}$ and $M_{\gamma e^+} \sim 100~ {\text{GeV}}$ are equivalent to $E_{e^-} \sim 250~ \text{GeV}$ and $E_{e^-} \sim 240~ \text{GeV}$, respectively. It is evident that those two peaks in the $\gamma e^+$ invariant mass distribution can be attributed to the highest-energy electron ($E_{e^-} \sim \sqrt{\hat{s}}/2$) and a minimum-energy photon emitted from electron ($E_{e^-} \sim \sqrt{\hat{s}}/2 - E_{\gamma, \text{min}}$), respectively. Moreover, we can see from the upper plots of Figs.\ref{fig8} (c) and (d) that the angular distributions of the final-state positron and leading photon are highly correlated. The positron and leading photon prefer to be produced back-to-back in the $\text{J} = 0$ $\gamma\gamma$ collision, while tend to be produced in the same direction or back-to-back in the $\text{J} = 2$ $\gamma\gamma$ collision. It is worth mentioning that the EW relative correction in the exclusive event selection scheme is relatively stable in the entire $\theta_{\gamma e^+}$ region, especially for the $\text{J} = 2$ collision mode.

\par
\begin{figure}[!htbp]
\begin{center}
\includegraphics[width=0.45\textwidth]{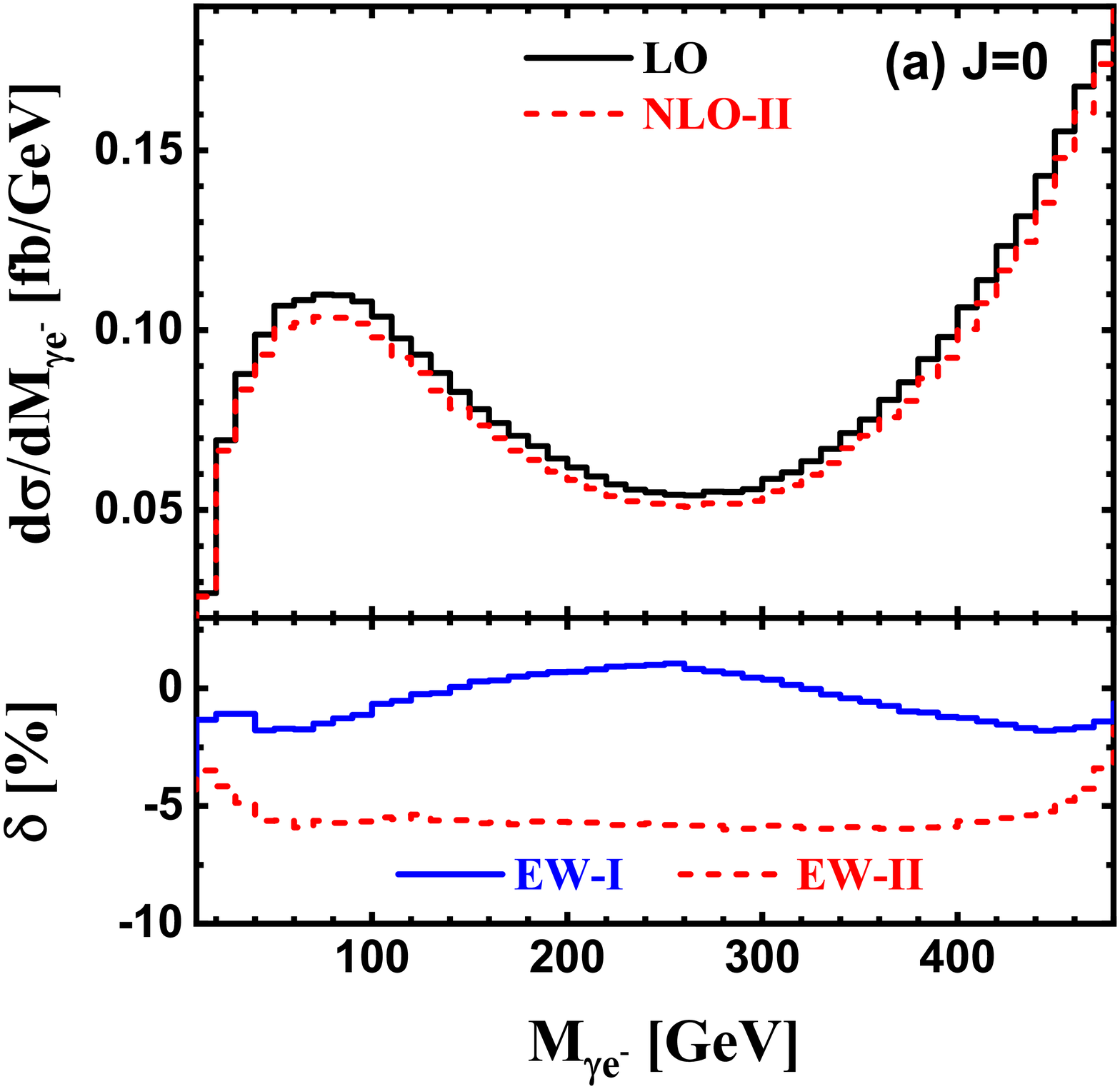}
\includegraphics[width=0.45\textwidth]{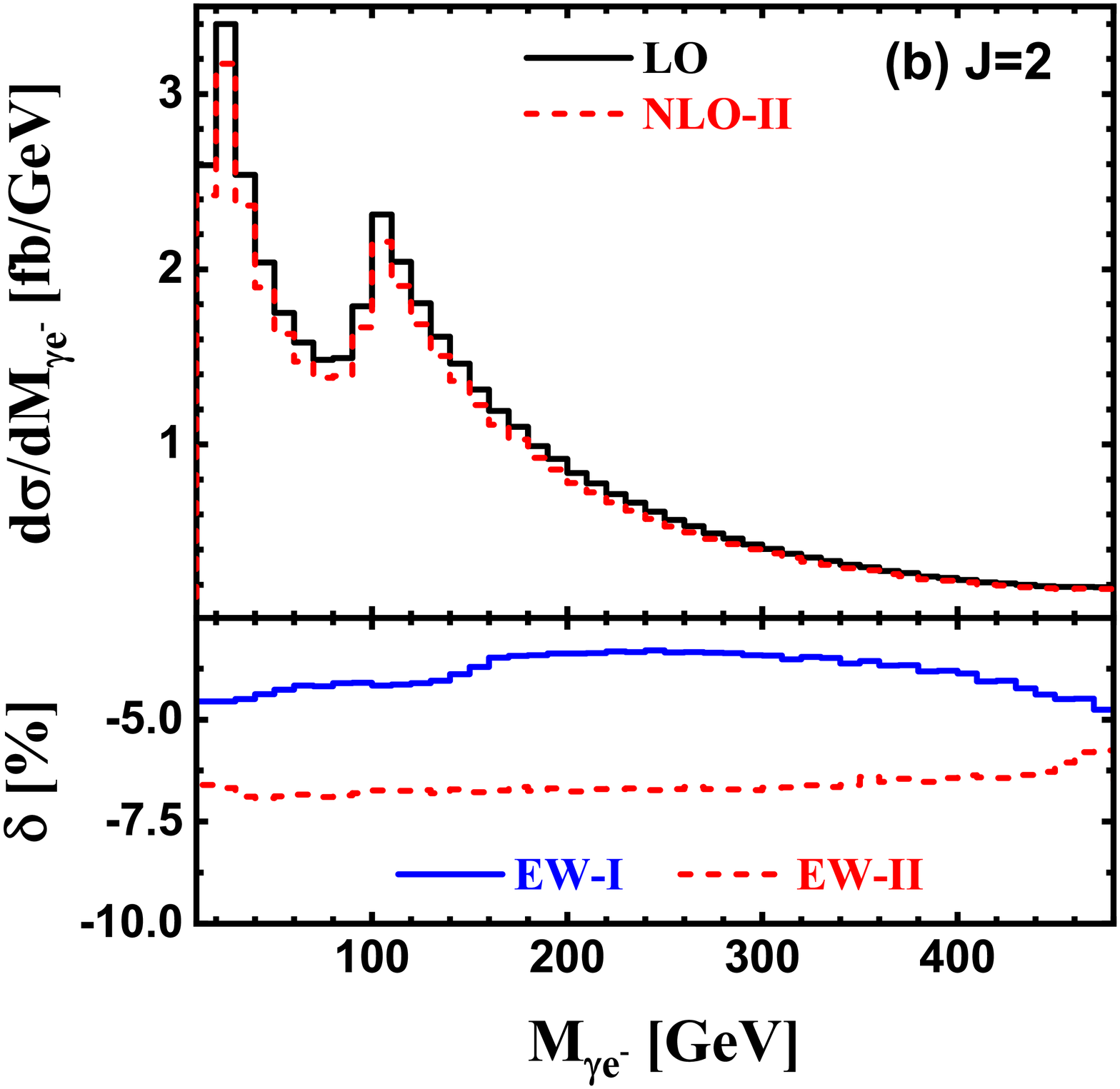}
\includegraphics[width=0.45\textwidth]{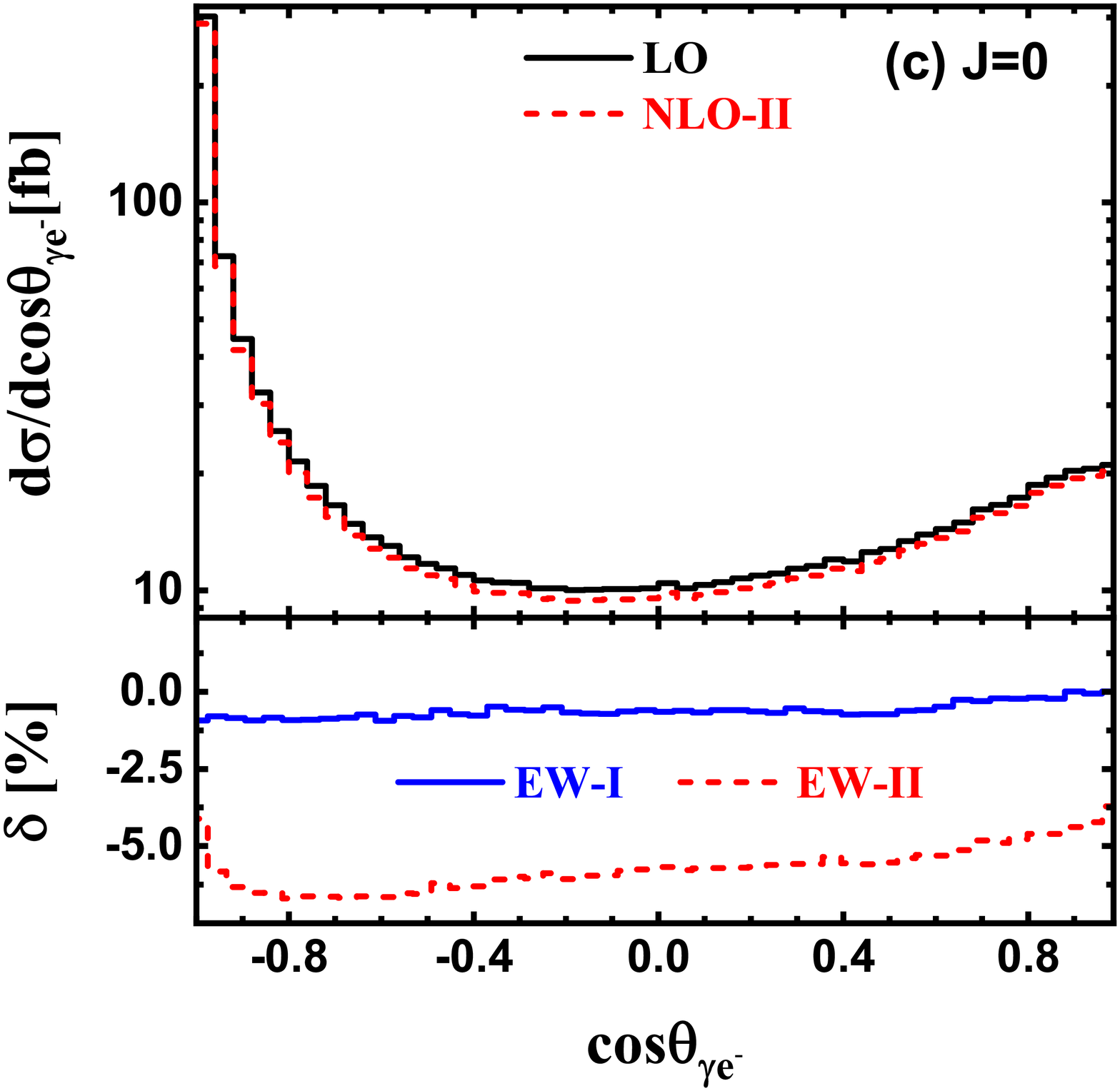}
\includegraphics[width=0.45\textwidth]{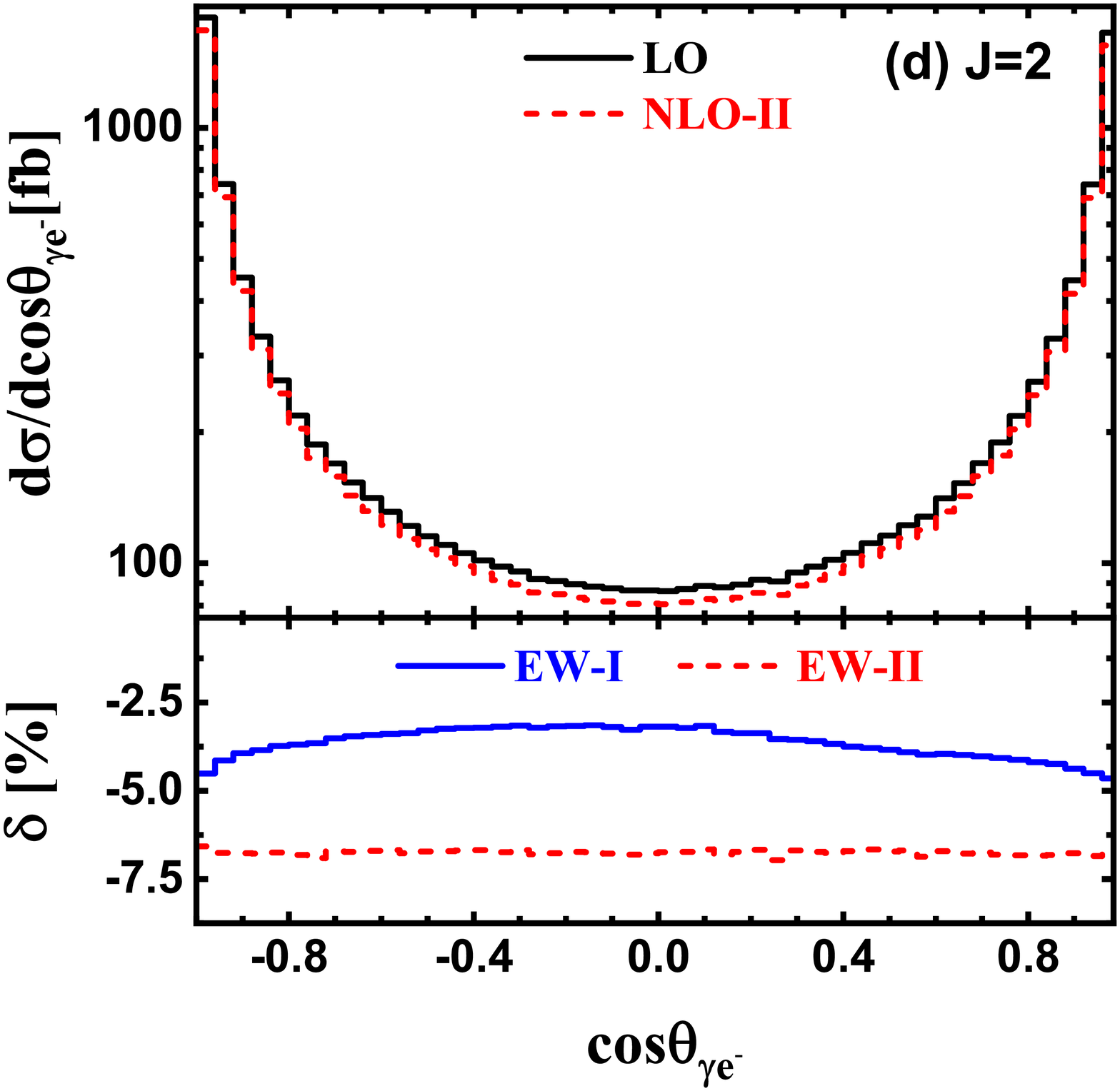}
\caption{Same as Fig.\ref{fig8}, but for the final-state $\gamma e^-$ system.}
\label{fig9}
\end{center}
\end{figure}

The differential cross sections with respect to the invariant mass and opening angle of the final-state $\gamma e^-$ system as well as the corresponding EW relative corrections are also provided in Figs.\ref{fig9} (a)-(d) for comparison. As expected, the invariant mass and opening angle distributions of the final-state $\gamma e^-$ system are exactly the same as the corresponding ones of $\gamma e^+$ system at the LO due to the charge conservation in the electromagnetic interaction. For the $\gamma_+\gamma_+ \rightarrow e^+e^-\gamma$ (i.e., $\text{J} = 0$) production channel, the EW relative corrections to $M_{\gamma e^-}$ and $\cos\theta_{\gamma e^-}$ distributions are significantly differ from the EW relative corrections to $M_{\gamma e^+}$ and $\cos\theta_{\gamma e^+}$ distributions, as shown in the lower panels of Figs.\ref{fig8} (a, c) and Figs.\ref{fig9} (a, c), due to the charge symmetry violation in weak interaction. In contrast, the EW relative corrections to the kinematic distributions of $\gamma e^-$ and $\gamma e^+$ systems are exactly the same for $\gamma_+\gamma_- \rightarrow e^+e^-\gamma$ (i.e., $\text{J} = 2$) production channel, as shown in the lower panels of Figs.\ref{fig8} (b, d) and Figs.\ref{fig9} (b, d), because of the $\mathcal{CP}\text{+}{\it Bose}$ symmetry. The numerical consistency between the kinematic distributions of the final-state $\gamma e^-$ and $\gamma e^+$ systems in the $\gamma_+\gamma_-$ collision reconfirms the conclusion declared in Sec.\ref{subsection-2A}.

\subsection{Parent process $ee \rightarrow \gamma\gamma \rightarrow e^+e^-\gamma$}
\par
Now we turn to the discussion of the parent process $ee \rightarrow \gamma\gamma \rightarrow e^+e^-\gamma$. The differential cross section for $ee \rightarrow \gamma\gamma \rightarrow e^+e^-\gamma$ with respect to a kinematic variable $\zeta$ can be expressed as
\begin{equation}
\frac{d \sigma}{d \zeta}
=
\frac{d \sigma^\text{(0)}}{d \zeta}
+
\frac{d \sigma^\text{(2)}}{d \zeta}\,,
\end{equation}
where $d \sigma^\text{(0)}/d \zeta$ and $d \sigma^\text{(2)}/d \zeta$ represent the contributions from $\text{J} = 0$ and $\text{J} = 2$ $\gamma\gamma$ collision modes, respectively. Considering that the two back-scattered photon beams are partially polarized, $d \sigma^\text{(0)}/d \zeta$ and $d \sigma^\text{(2)}/d \zeta$ are calculated by
\begin{equation}
\begin{aligned}
\frac{d \sigma^\text{(J)}}{d \zeta}(s)
=
\sum_{\lambda_1 = (-1)^{\frac{\text{J}}{2}} \lambda_2}
&
\int_{2 m_e/\sqrt{s}}^{x_{\text{max}}}
\frac{z dz}{2}
\int_{z^2/x_{\text{max}}}^{x_{\text{max}}}
\frac{dx}{x} \phi_{\gamma/e}(x) \phi_{\gamma/e}(z^2/x)
\\
&
\Big[ 1 + \lambda_1 \mathcal{P}_{\gamma}(x) \Big]
\Big[ 1 + \lambda_2 \mathcal{P}_{\gamma}(z^2/x) \Big]
\frac{d \hat{\sigma}^{\lambda_1\lambda_2}}{d \zeta}(\hat{s} = z^2 s)\,,
\qquad(\text{J} = 0,\, 2)
\end{aligned}
\end{equation}
When discussing the angular distributions of final particles in the c.m. frame of the initial $ee$ system, the reference direction can be chosen as either of the two electron beam directions due to the forward-backward symmetry.

\par
\begin{figure}[!htbp]
\begin{center}
\includegraphics[width=0.45\textwidth]{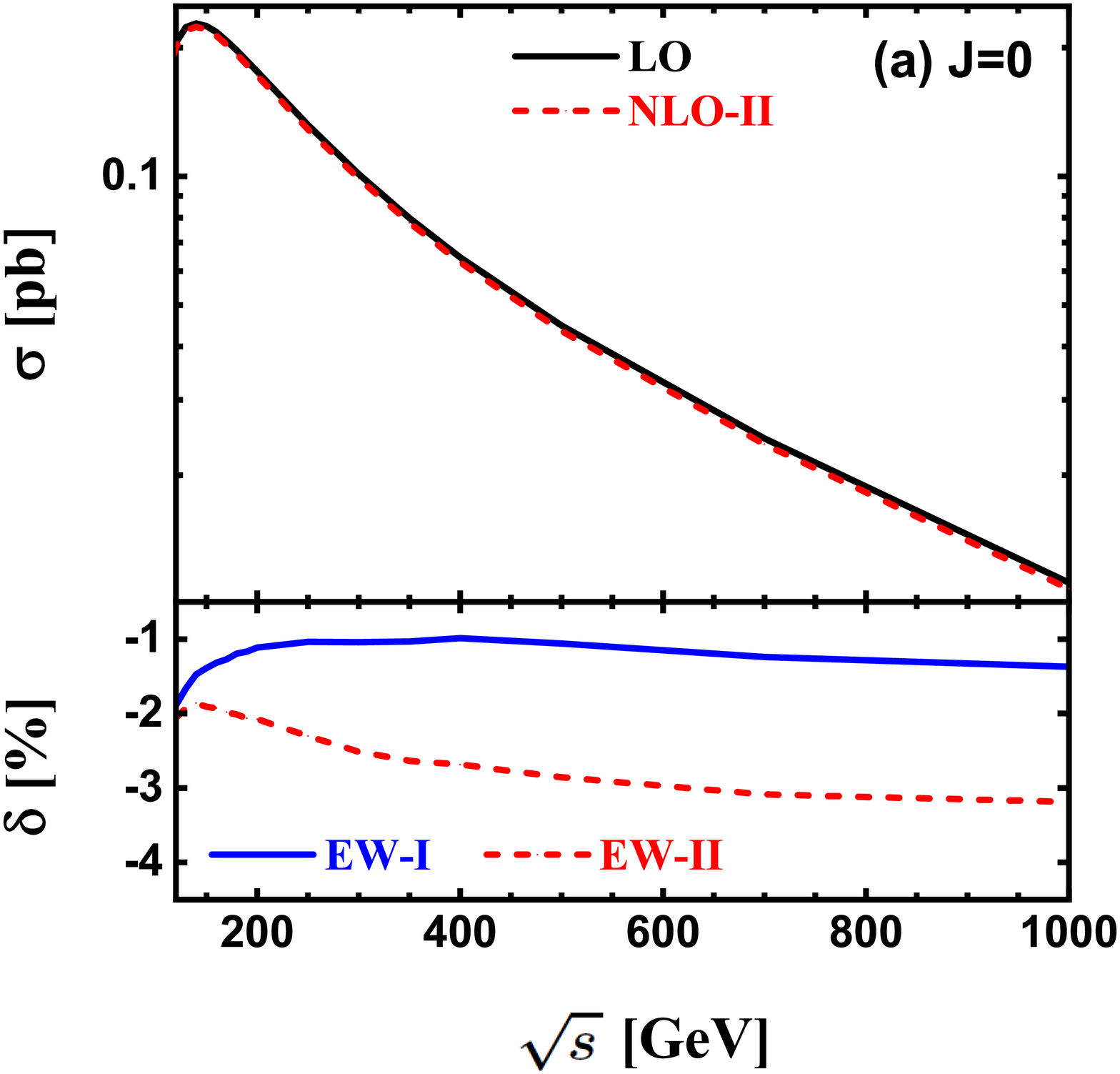}
\includegraphics[width=0.45\textwidth]{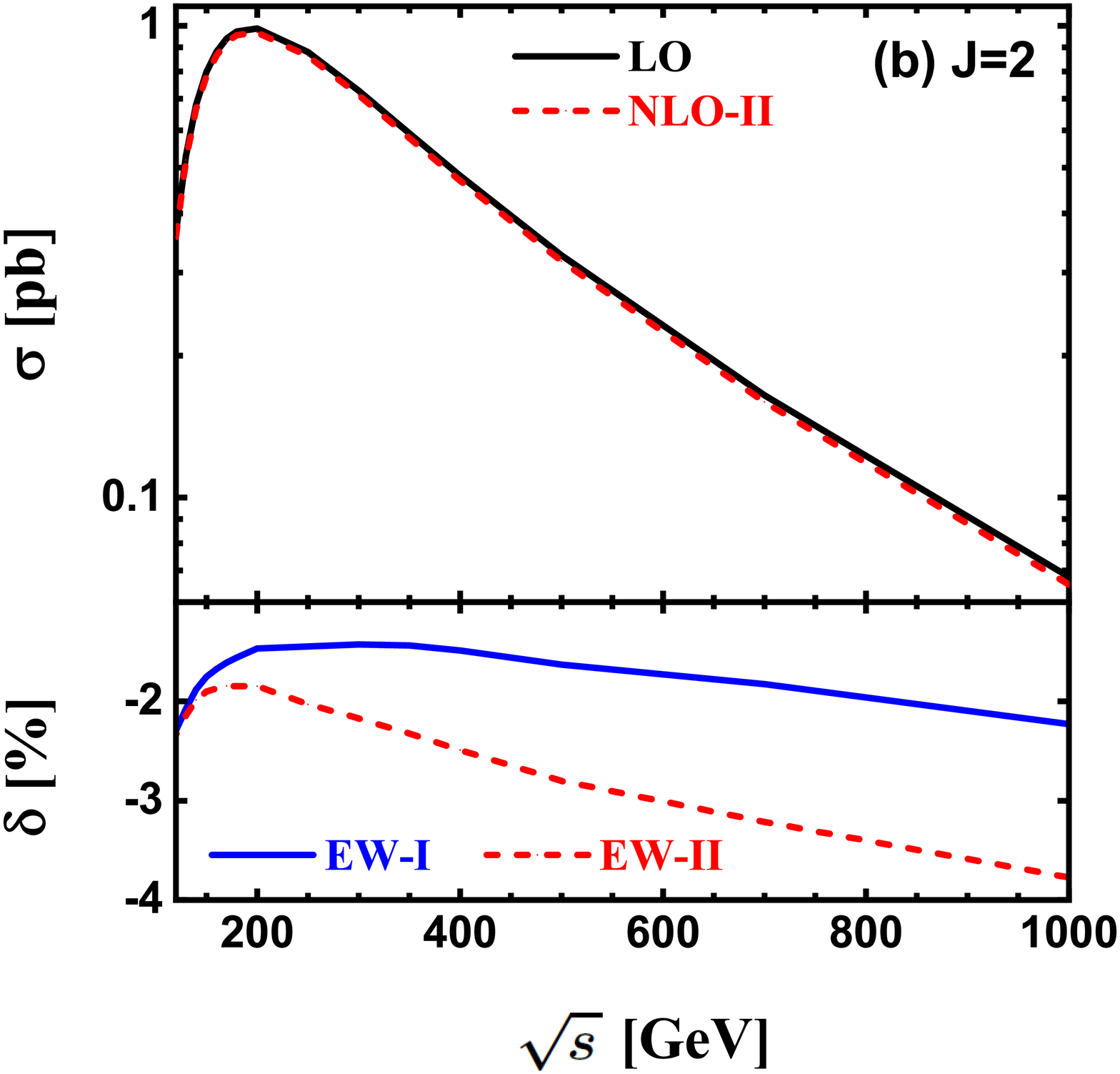}
\caption{LO, NLO EW corrected integrated cross sections from (a) $\text{J} = 0$ and (b) $\text{J} = 2$ $\gamma\gamma$ collisions in scheme-II and the corresponding EW relative corrections in both scheme-I and -II as functions of $\sqrt{s}$ for $ee \rightarrow \gamma\gamma \rightarrow e^+e^-\gamma$.}
\label{fig10}
\end{center}
\end{figure}
\begin{figure}[!htbp]
\begin{center}
\includegraphics[width=0.45\textwidth]{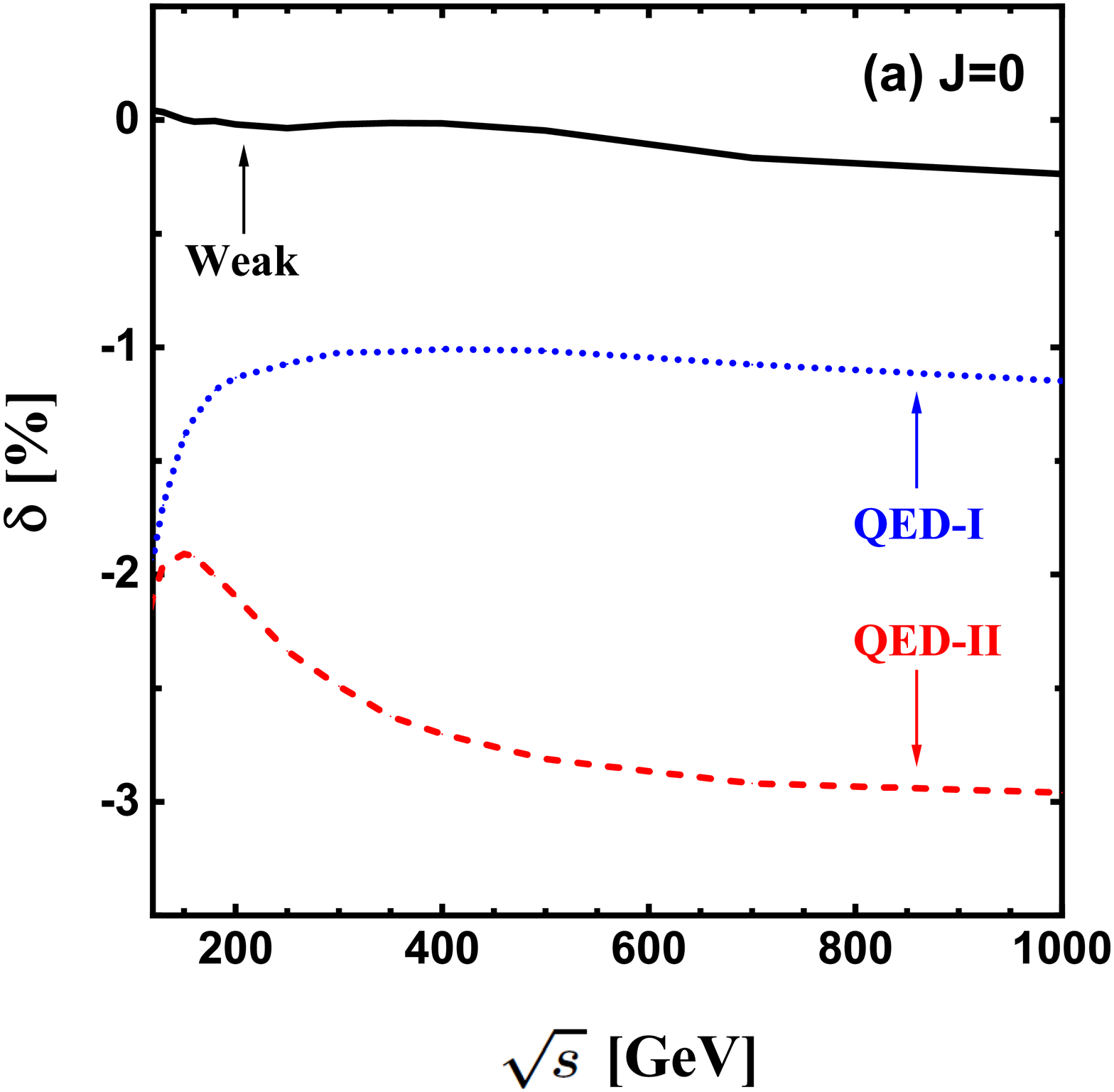}
\includegraphics[width=0.45\textwidth]{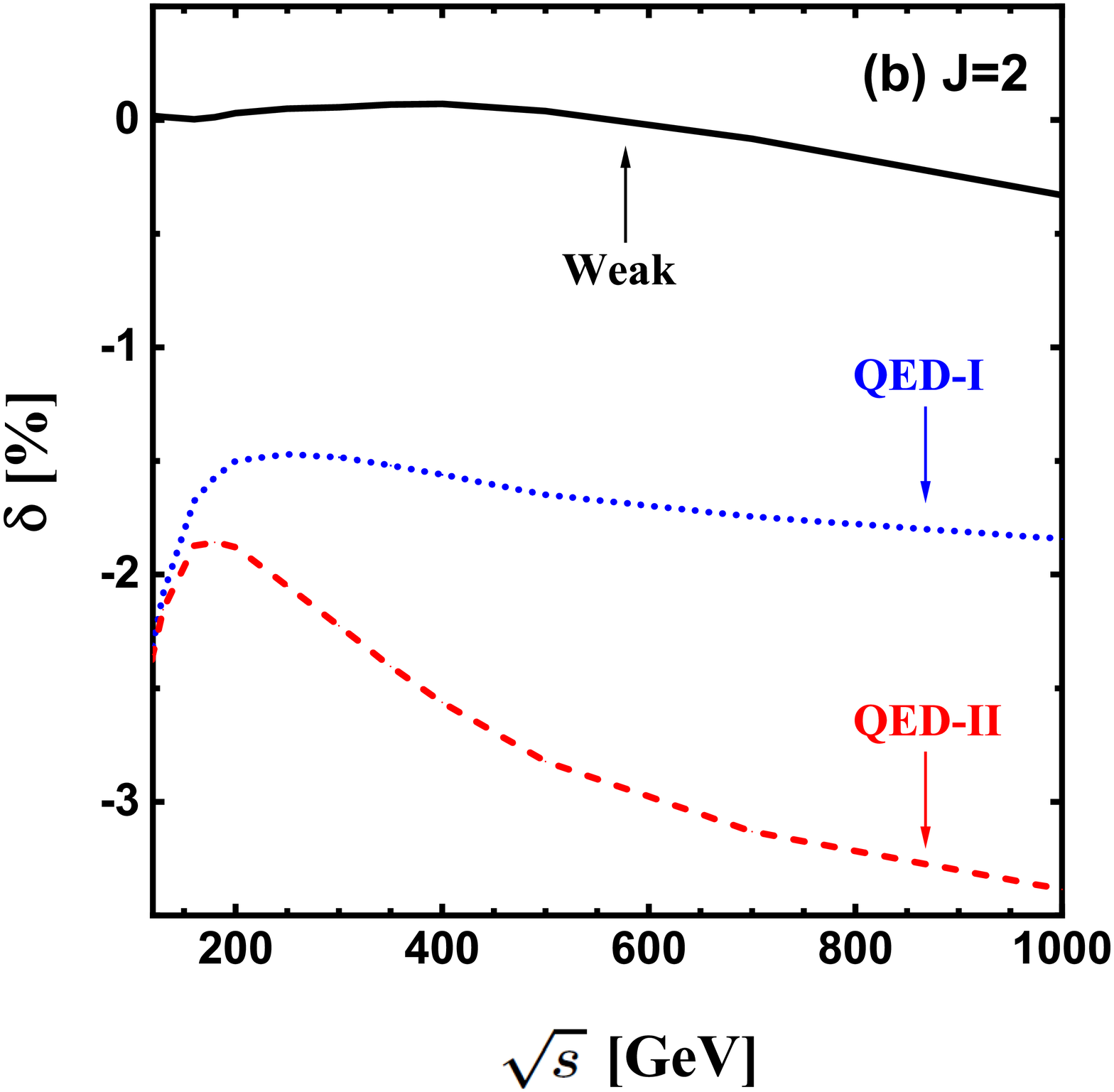}
\caption{QED and weak relative corrections to the integrated cross sections from (a) $\text{J} = 0$ and (b) $\text{J} = 2$ $\gamma\gamma$ collisions in both scheme-I and -II as functions of $\sqrt{s}$ for $ee \rightarrow \gamma\gamma \rightarrow e^+e^-\gamma$.}
\label{fig11}
\end{center}
\end{figure}

In Figs.\ref{fig10} (a) and (b), we depict the LO, NLO EW corrected integrated cross sections from the $\text{J} = 0$ and $\text{J} = 2$ $\gamma\gamma$ collisions in the exclusive event selection scheme as well as the corresponding EW relative corrections in both inclusive and exclusive event selection schemes as functions of the $ee$ c.m. colliding energy for $ee \rightarrow \gamma\gamma \rightarrow e^+e^-\gamma$. The integrated contributions from $\text{J} = 0$ and $\text{J} = 2$ $\gamma\gamma$ collisions increase at first, reach their maxima at $\sqrt{s} \sim 140$ and $200~ \text{GeV}$, respectively, and then decrease rapidly as $\sqrt{s}$ increases. In the exclusive event selection scheme, the EW relative correction to the integrated cross section from the $\text{J} = 0$ $\gamma\gamma$ collision is relatively stable as $\sqrt{s} \in [400,\, 1000]~ \text{GeV}$, varying in the range of $[-3.2\%,\, -2.7\%]$. Compared to $\text{J} = 0$, the EW relative correction to the integrated cross section from $\text{J} = 2$ $\gamma\gamma$ collision is more sensitive to the $ee$ c.m. colliding energy. It decreases approximately linearly from about $-2.5\%$ to about $-3.8\%$ as $\sqrt{s}$ increases from $400~ \text{GeV}$ to $1~ \text{TeV}$. Similar to the discussion on $\gamma\gamma \rightarrow e^+e^-\gamma$, we also separately provide weak and QED relative corrections to the parent process $ee \rightarrow \gamma\gamma \rightarrow e^+e^-\gamma$ for both $\text{J} = 0$ and $\text{J}=2$ polarization configurations of the Compton back-scattered photons. As shown in Figs.\ref{fig11} (a) and (b), the weak relative correction is small ($\left| \delta_{\text{W}} \right| < 0.5\%$) and the full NLO EW correction is dominated by the negative QED correction for both polarization modes of the back-scattered photons in the entire plotted $\sqrt{s}$ region ($120~ \text{GeV} < \sqrt{s} < 1000~ \text{GeV}$). In the exclusive event selection scheme, the QED relative correction reaches its maximum of about $-1.9\%$ at $\sqrt{s} \sim 150~ \text{GeV}$ and is roughly steady at $-2.9\%$ as $\sqrt{s} \in [500,\, 1000]~ \text{GeV}$ for $\text{J} = 0$ polarization of the back-scattered photons, while it decreases gradually from its maximum of about $-1.9\%$ to approximately $-3.4\%$ as $\sqrt{s}$ increases from $180~ \text{GeV}$ to $1~ \text{TeV}$ for the $\text{J} = 2$ $\gamma\gamma$ polarization mode. As is well known, the maximum energy fraction of the back-scattered photon is given by \cite{Zarnecki:2002qr}
\begin{equation}
x_{\text{max}}
=
\frac{4E_e E_0}{4E_e E_0 + m_e^2}\,,
\end{equation}
where $E_e$ represents the electron beam energy and $E_0$ denotes the energy of the laser photon. In this study, we take $E_0 = 1.17~ \text{eV}$ (i.e., the laser wave length $\lambda = 1.06~ \mu\text{m}$). At $\sqrt{s} = 120~ \text{GeV}$, the maximum c.m. colliding energy of the back-scattered photon beams is only approximately $60~ \text{GeV}$, thus the NLO QED contribution of the $e^+e^-\gamma\gamma$ events is too tiny to be ignored due to the small phase space of the $e^+e^-\gamma\gamma$ four-body final state. As $\sqrt{s}$ increases, the QED relative correction contributed by $e^+e^-\gamma\gamma$ events (i.e., $\delta_{\text{QED}}^{\text{(I)}} - \delta_{\text{QED}}^{\text{(II)}}$) becomes more and more notable and compensates the negative EW correction from $e^+e^-\gamma$ events. As $\sqrt{s} \in [600,\, 1000]~ \text{GeV}$, the QED relative correction in the inclusive event selection scheme is steady at around $-1.1\%$ and $-1.8\%$ for $\text{J} = 0$ and $\text{J} = 2$, respectively. Finally, in Table \ref{tab2} and Table \ref{tab3}, we present in detail the production cross sections and the corresponding NLO relative corrections for $ee \rightarrow \gamma\gamma \rightarrow e^+e^-\gamma$ at $\sqrt{s} = 250$, $500$ and $1000~ \text{GeV}$, which correspond to the three stages of the ILC \cite{Asner:2013psa}. For more details on the kinematic distributions of final products of $ee \rightarrow \gamma\gamma \rightarrow e^+e^-\gamma$, refer to the Appendix section of this article.
\begin{table}[!htbp]
\renewcommand \tabcolsep{6.0pt}
\renewcommand \arraystretch{1.1}
\centering
\begin{tabular}{|c|cccccccc|}
\hline\hline
  $\sqrt{s}~ \text{[GeV]}$
& $\sigma_{\text{LO}}~ \text{[fb]}$
& $\sigma_{\text{NLO}}^{\text{(I)}}~ \text{[fb]}$
& $\sigma_{\text{NLO}}^{\text{(II)}}~ \text{[fb]}$
& $\delta_{\text{EW}}^{\text{(I)}}~ \text{[\%]}$
& $\delta_{\text{EW}}^{\text{(II)}}~ \text{[\%]}$
& $\delta_{\text{QED}}^{\text{(I)}}~ \text{[\%]}$
& $\delta_{\text{QED}}^{\text{(II)}}~ \text{[\%]}$
& $\delta_{\text{W}}~ \text{[\%]}$
\\
\hline
  $250$
& $131.93$ & $130.46$ & $128.80$ & $-1.11$ & $-2.37$ & $-1.05$ & $-2.31$ & $-0.06$
\\
  $500$
& $44.729$ & $44.260$ & $43.452$ & $-1.05$ & $-2.85$ & $-1.01$ & $-2.81$ & $-0.04$
\\
  $1000$
& $11.195$ & $11.041$ & $10.838$ & $-1.38$ & $-3.19$ & $-1.14$ & $-2.95$ & $-0.24$
\\
\hline\hline
\end{tabular}
\caption{
\label{tab2}
LO, NLO EW corrected cross sections and the corresponding NLO relative corrections for $ee \rightarrow \gamma\gamma \rightarrow e^+e^-\gamma$ via $\text{J} = 0$ collision of Compton back-scattered photons at $\sqrt{s} = 250$, $500$ and $1000~ \text{GeV}$.}
\end{table}
\begin{table}[!htbp]
\renewcommand \tabcolsep{6.0pt}
\renewcommand \arraystretch{1.1}
\centering
\begin{tabular}{|c|cccccccc|}
\hline\hline
  $\sqrt{s}~ \text{[GeV]}$
& $\sigma_{\text{LO}}~ \text{[fb]}$
& $\sigma_{\text{NLO}}^{\text{(I)}}~ \text{[fb]}$
& $\sigma_{\text{NLO}}^{\text{(II)}}~ \text{[fb]}$
& $\delta_{\text{EW}}^{\text{(I)}}~ \text{[\%]}$
& $\delta_{\text{EW}}^{\text{(II)}}~ \text{[\%]}$
& $\delta_{\text{QED}}^{\text{(I)}}~ \text{[\%]}$
& $\delta_{\text{QED}}^{\text{(II)}}~ \text{[\%]}$
& $\delta_{\text{W}}~ \text{[\%]}$
\\
\hline
  $250$
& $880.36$ & $867.61$ & $862.51$ & $-1.45$ & $-2.03$ & $-1.48$ & $-2.06$ & $+0.03$
\\
  $500$
& $326.15$ & $320.84$ & $317.01$ & $-1.63$ & $-2.80$ & $-1.63$ & $-2.80$ & $+0.00$
\\
  $1000$
& $67.879$ & $66.366$ & $65.319$ & $-2.23$ & $-3.77$ & $-1.86$ & $-3.40$ & $-0.37$
\\
\hline\hline
\end{tabular}
\caption{
\label{tab3}
Same as Table \ref{tab2}, but for $\text{J} = 2$ collision mode of back-scattered photons.}
\end{table}

\vskip 5mm

\section{Summary}
\label{section-4}
\par
$\gamma\gamma \rightarrow l^+l^-\gamma$ as well as $\gamma\gamma \rightarrow l^+l^-$ is an ideal channel for calibrating the beam luminosity of the Photon Linear Collider, especially for the $\text{J} = 0$ polarization of the incident photon beams. In this paper, we present the full $\mathcal{O}(\alpha)$ EW corrected integrated cross sections and some kinematic distributions of final products for the $e^+e^-\gamma$ production in $\gamma\gamma$ collision. The production rate of $e^+e^-\gamma$ in $\text{J} = 2$ $\gamma\gamma$ collision is signigicantly larger than that in $\text{J} = 0$ collision mode. In the exclusive event selection scheme, the NLO EW correction is dominated by the QED contribution; the entire EW relative correction is sensitive to the $\gamma\gamma$ c.m. colliding energy and can exceed $-10\%$ at a TeV PLC for both $\text{J} = 0$ and $\text{J} = 2$ polarization configurations of photon beams. The kinematic behaviors of the final products in $\text{J} = 0$ $\gamma\gamma$ collision are quite different from those in $\text{J} = 2$ collision. At $\sqrt{\hat{s}} = 500~ \text{GeV}$, the EW relative correction is about $-7\% \sim -5\%$ in most of the final-state phase space and can even reach around $-10\%$ in some specific phase-space regions. We can conclude that the NLO EW correction exerts important impact on both integrated and differential cross sections, and thus is significant in the precise determination of incoming photon beam luminosity at PLC.

\vskip 5mm

\noindent{\large\bf Acknowledgments:}

This work is supported in part by the National Natural Science Foundation of China (Grants No. 11775211 and No. 12061141005) and the CAS Center for Excellence in Particle Physics (CCEPP).

\vskip 5mm

\nocite{*}
\bibliography{ref}

\vskip 20mm

\noindent{\bf APPENDIX}
\vskip 5mm
\par
\begin{figure}[!htbp]
\begin{center}
\includegraphics[width=0.45\textwidth]{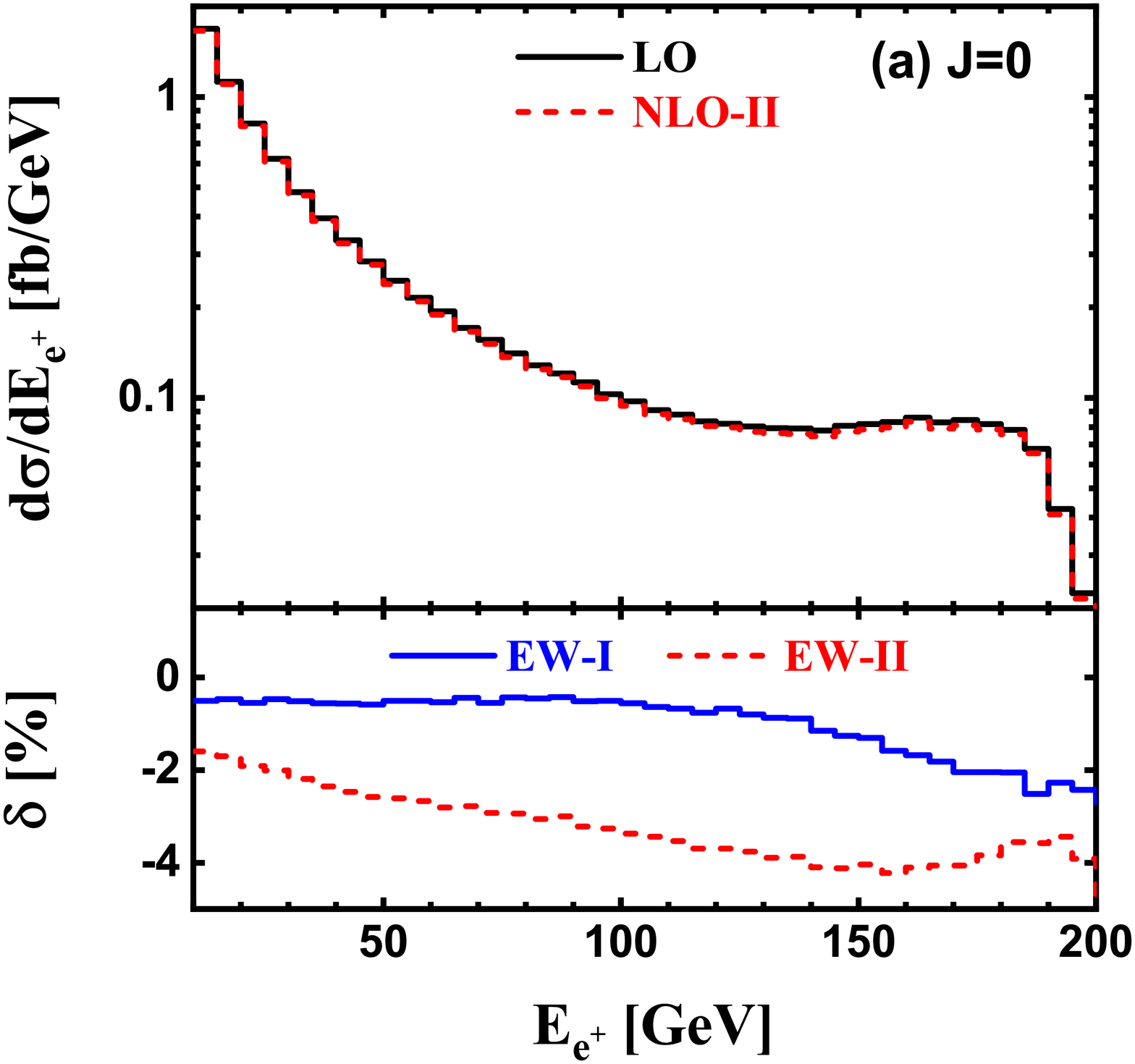}
\includegraphics[width=0.45\textwidth]{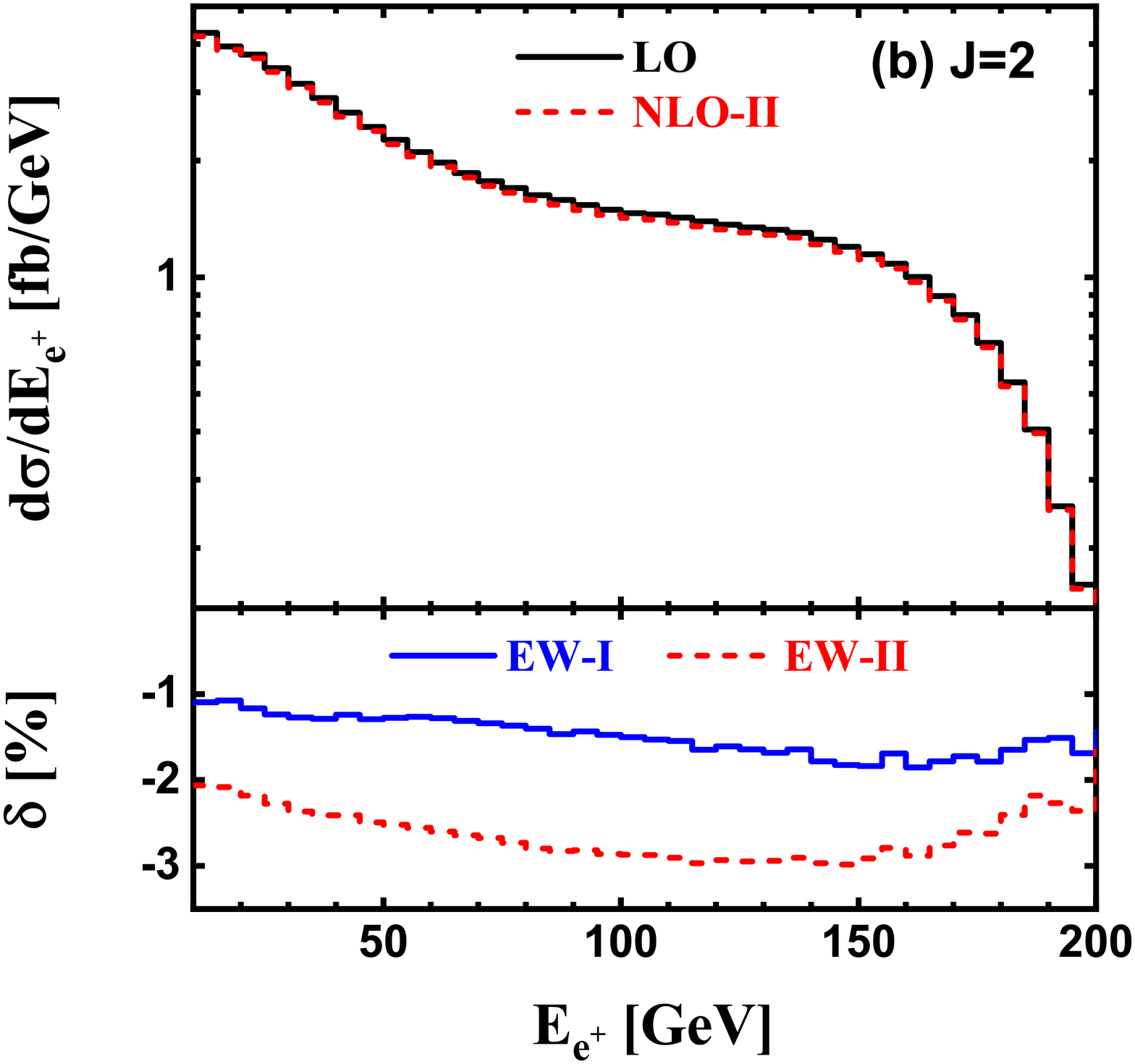}
\includegraphics[width=0.45\textwidth]{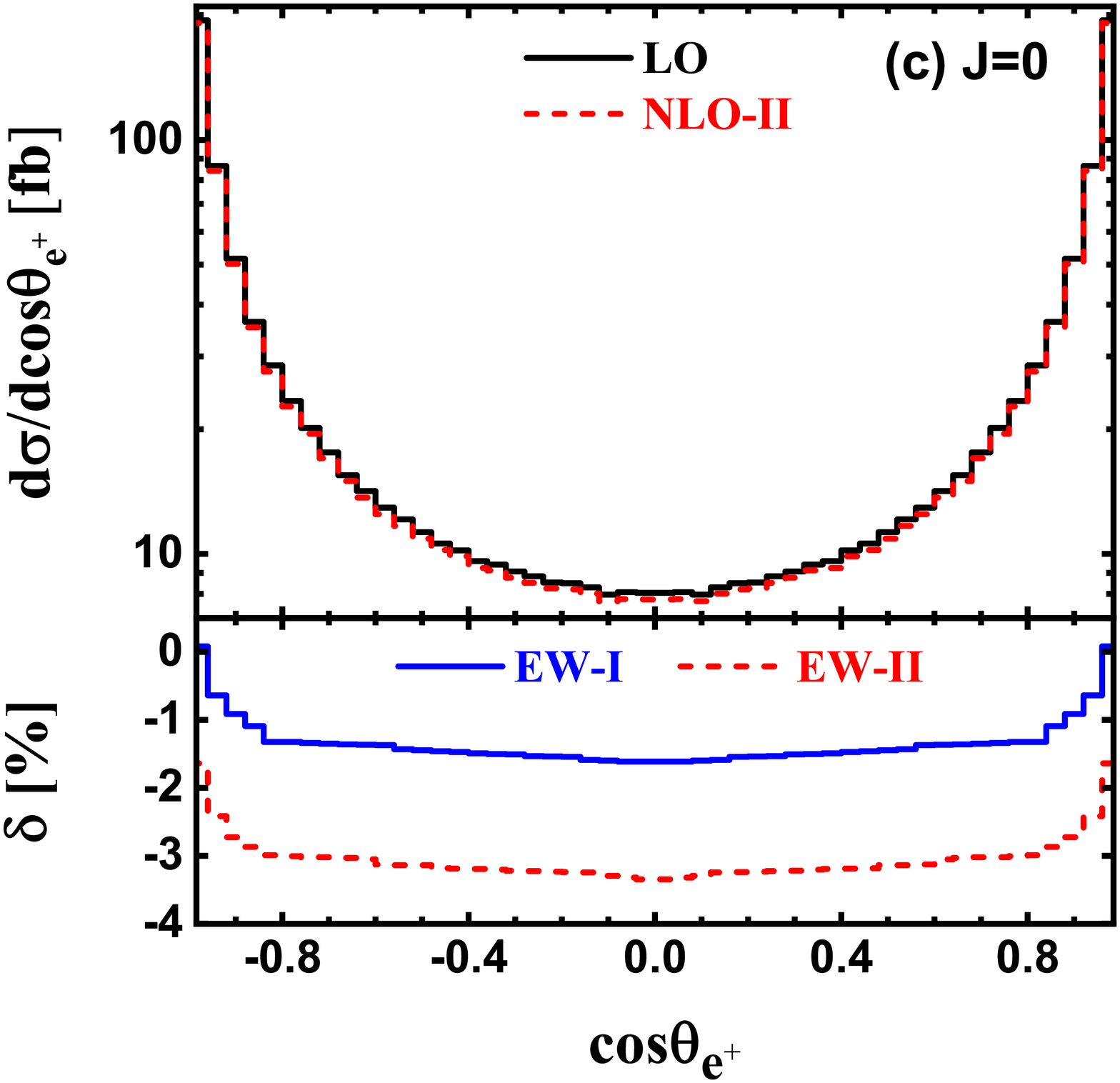}
\includegraphics[width=0.45\textwidth]{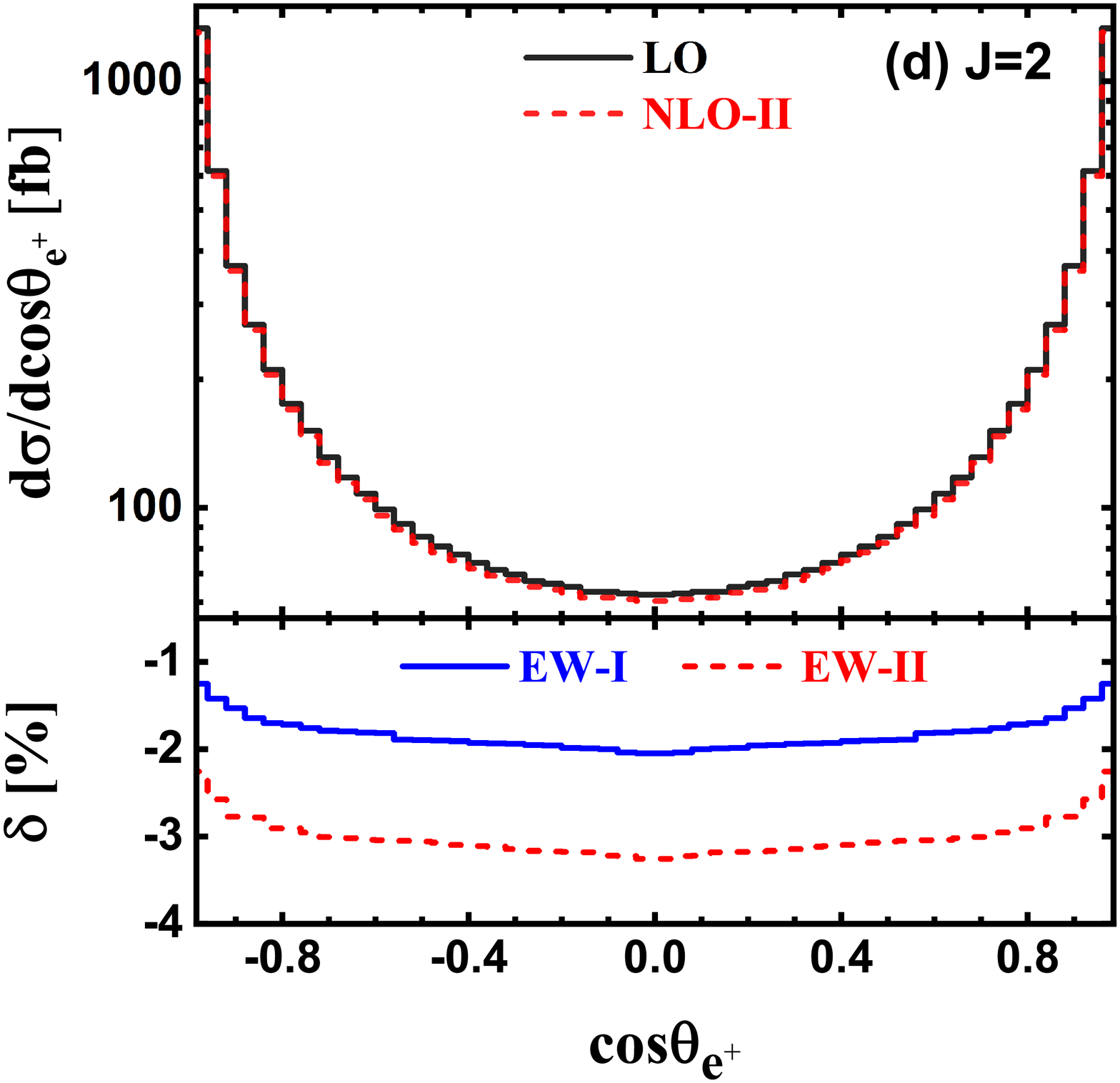}
\caption{LO, NLO EW corrected energy and angular distributions of the final-state positron and the corresponding EW relative corrections for $ee \rightarrow \gamma\gamma \rightarrow e^+e^-\gamma$ via $\text{J} = 0$ and $\text{J} = 2$ collisions of Compton back-scattered photons at $\sqrt{s} = 500~ \text{GeV}$.}
\label{fig12}
\end{center}
\end{figure}
\begin{figure}[!htbp]
\begin{center}
\includegraphics[width=0.45\textwidth]{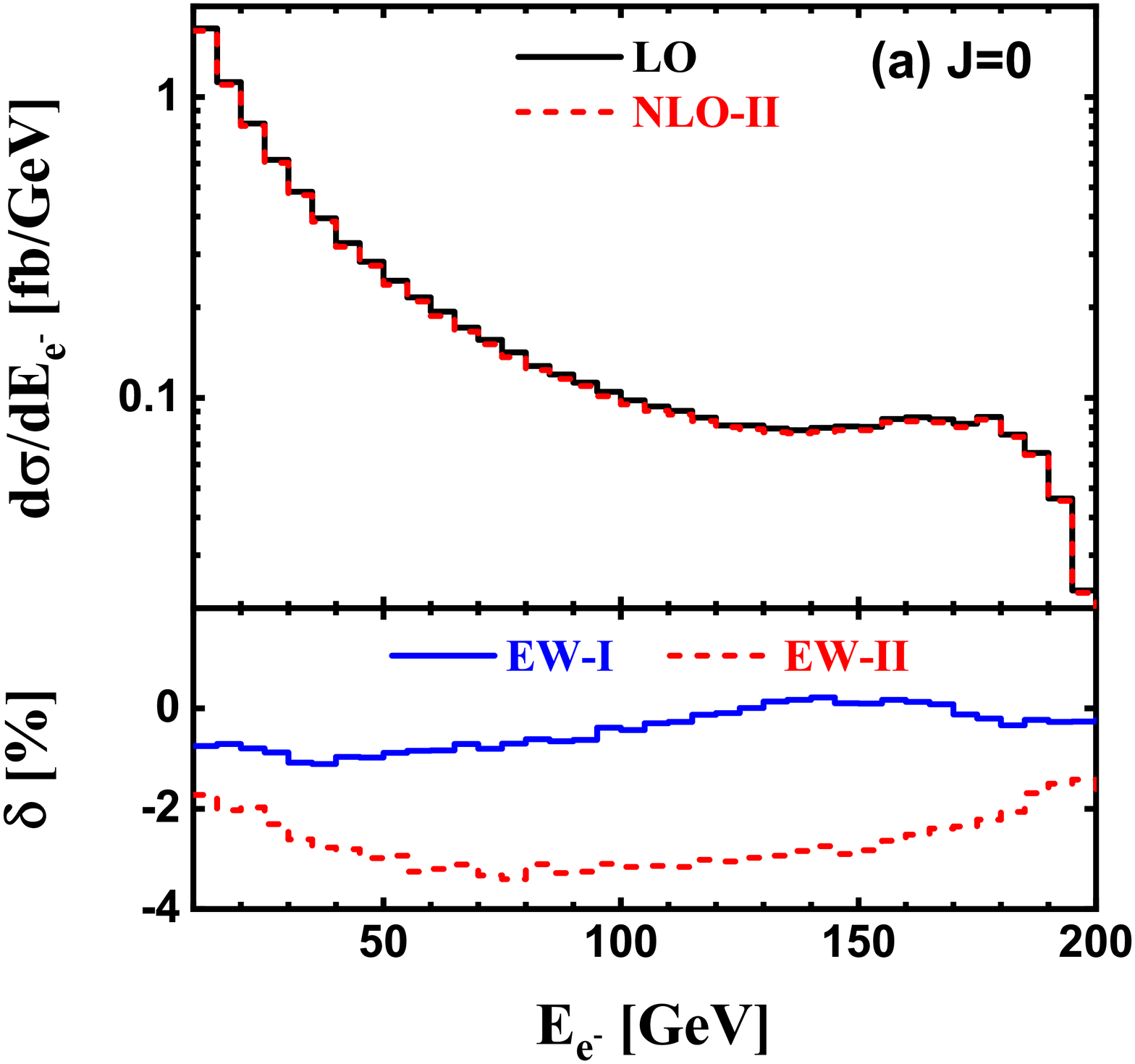}
\includegraphics[width=0.45\textwidth]{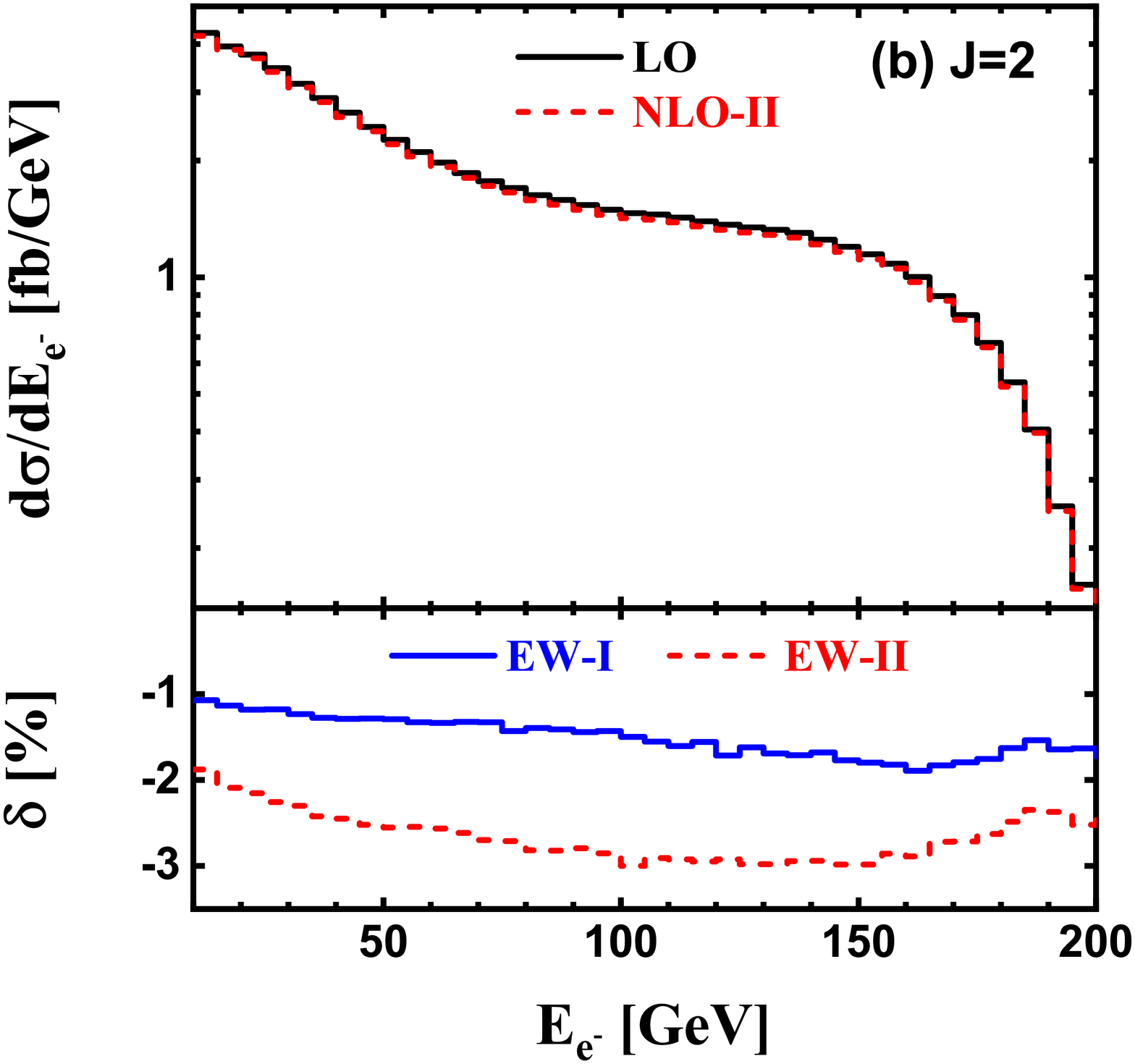}
\includegraphics[width=0.45\textwidth]{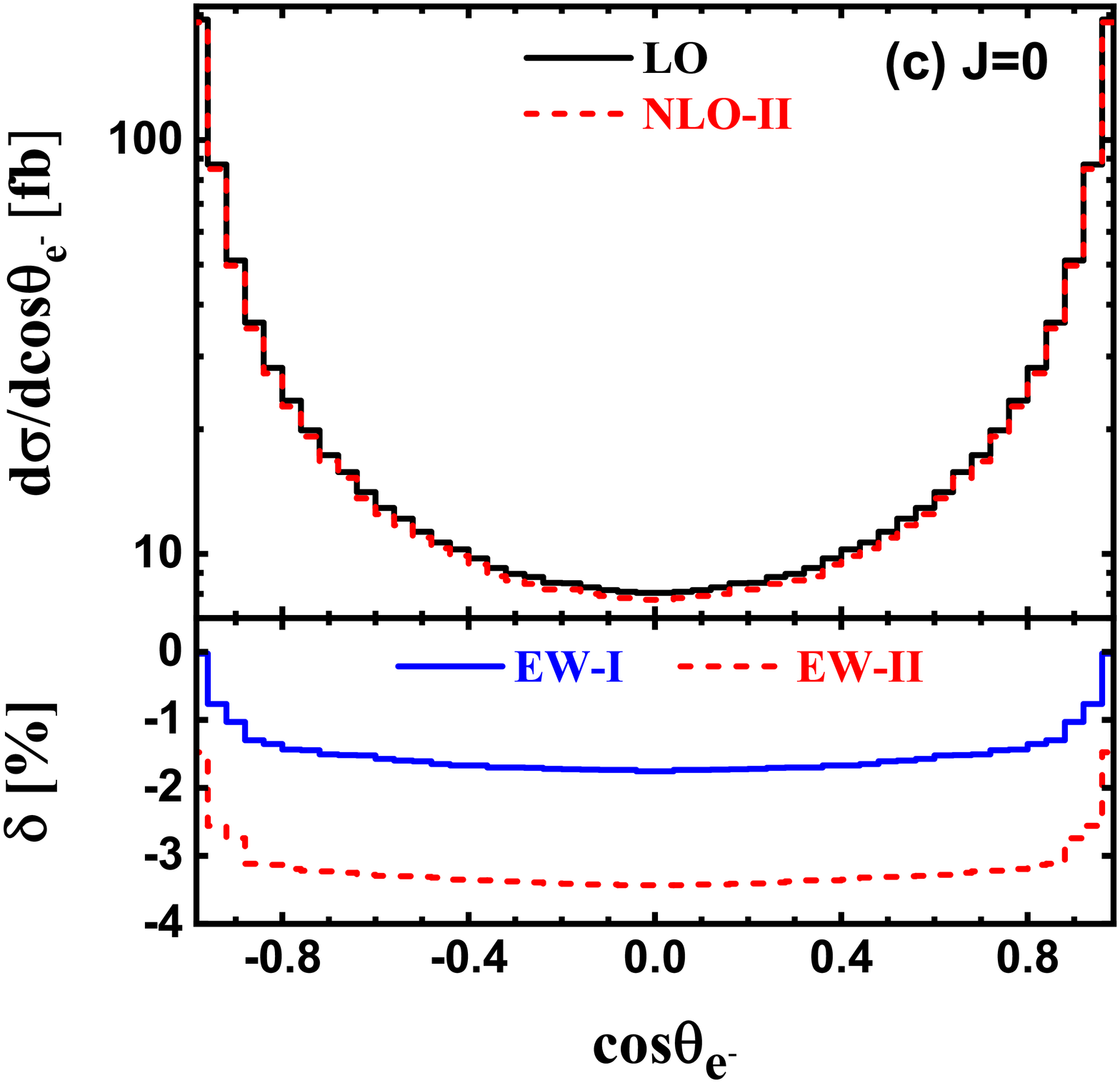}
\includegraphics[width=0.45\textwidth]{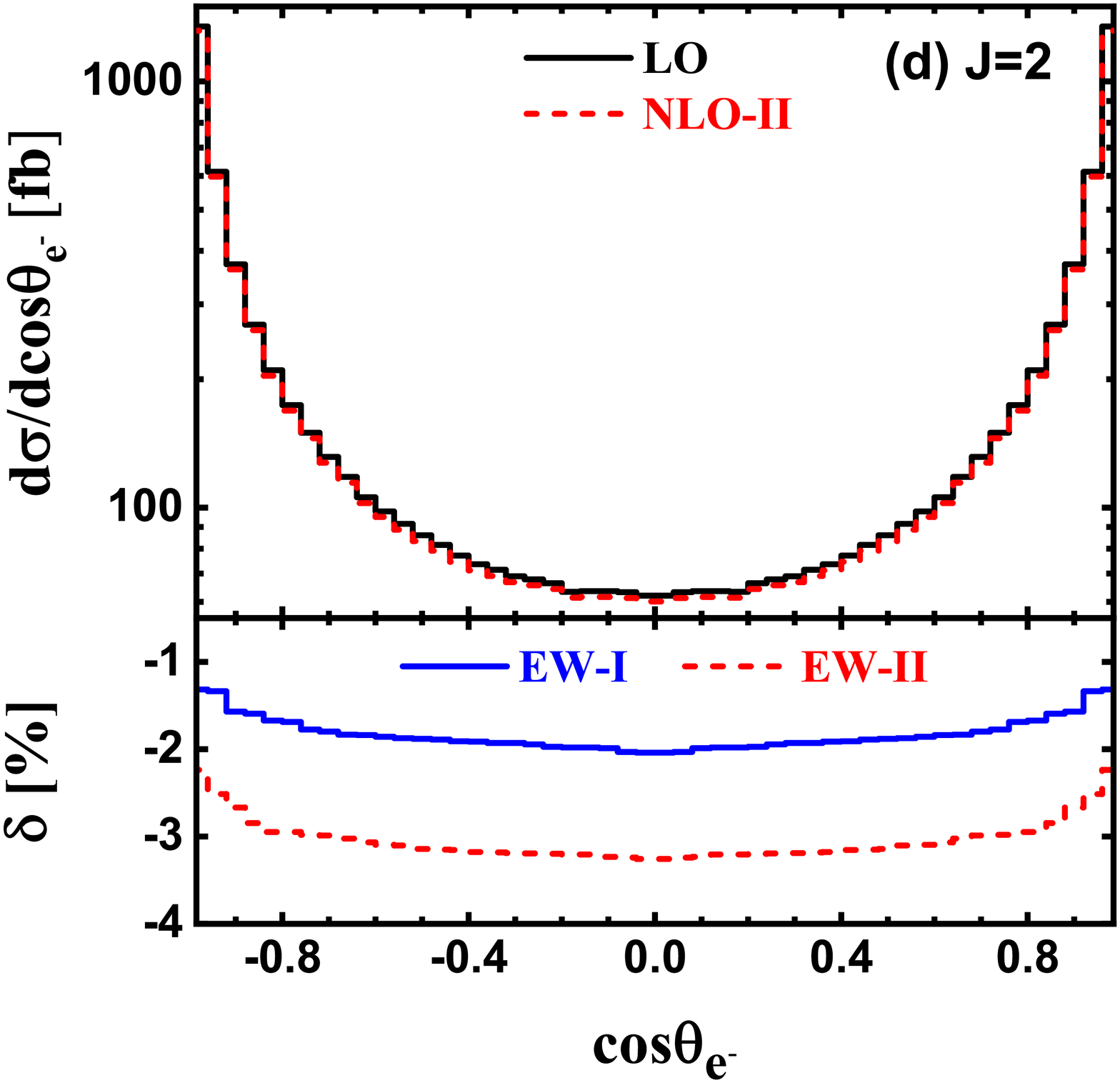}
\caption{Same as Fig.\ref{fig12}, but for the final-state electron.}
\label{fig13}
\end{center}
\end{figure}
\begin{figure}[!htbp]
\begin{center}
\includegraphics[width=0.45\textwidth]{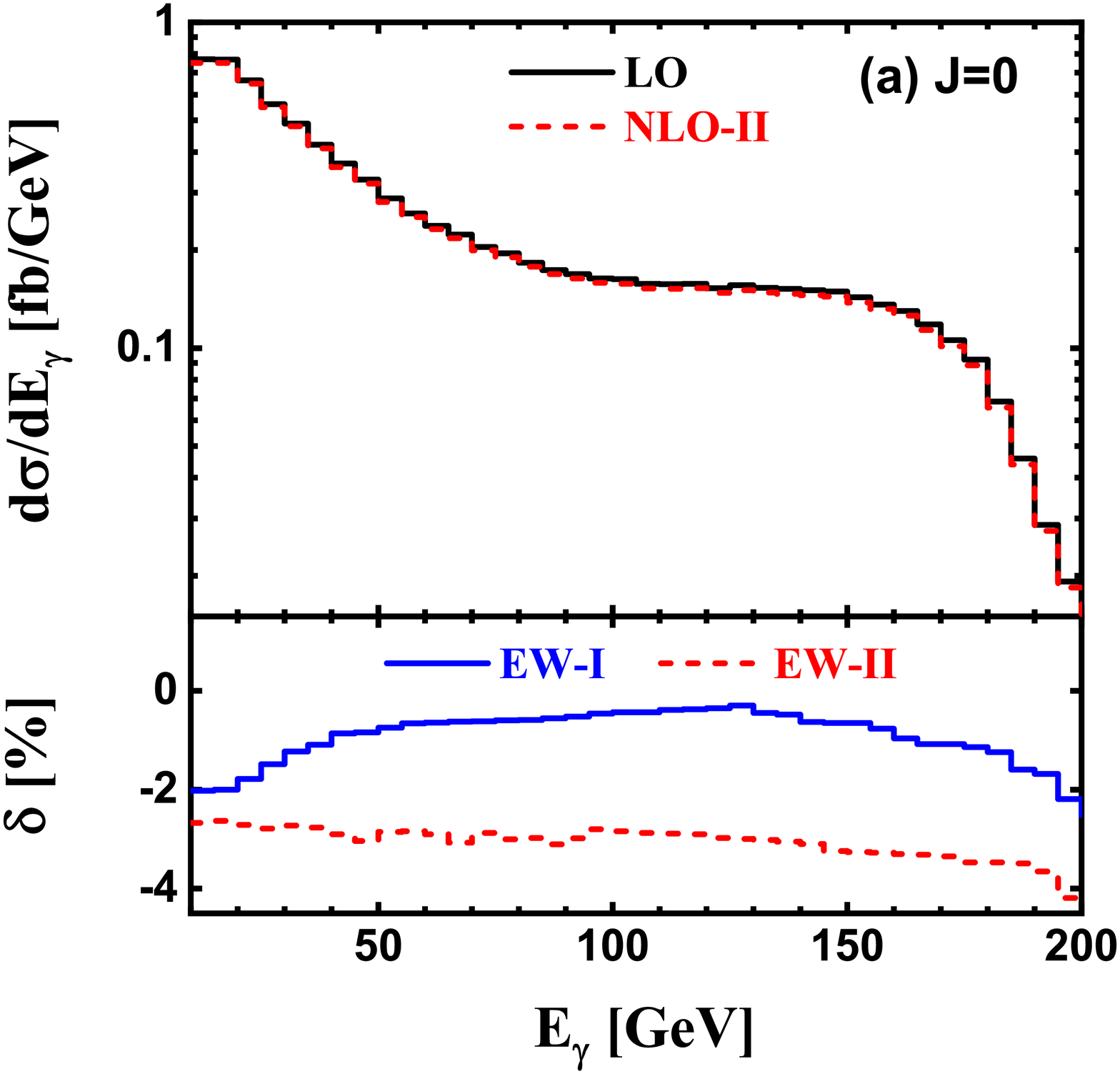}
\includegraphics[width=0.45\textwidth]{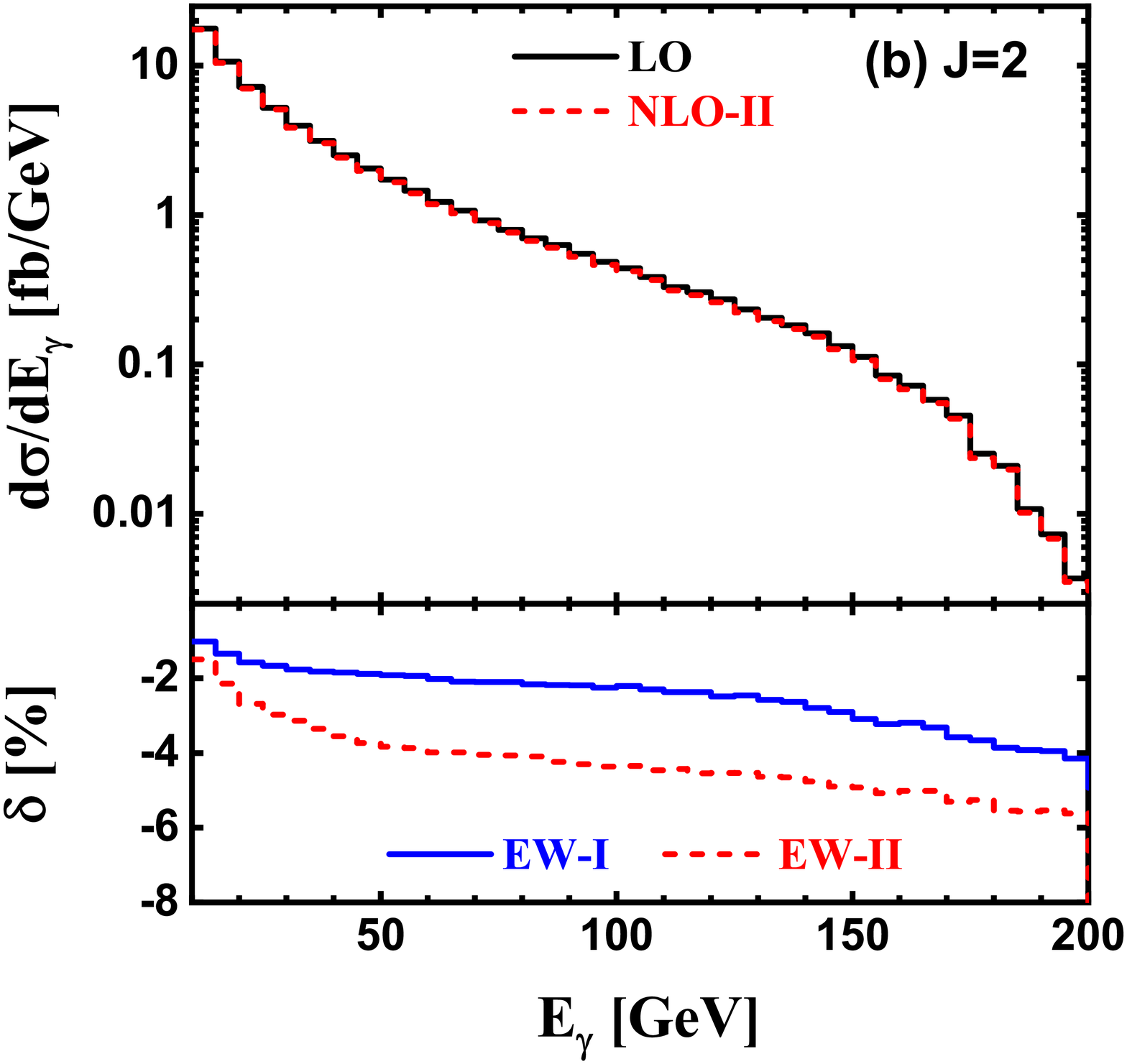}
\includegraphics[width=0.45\textwidth]{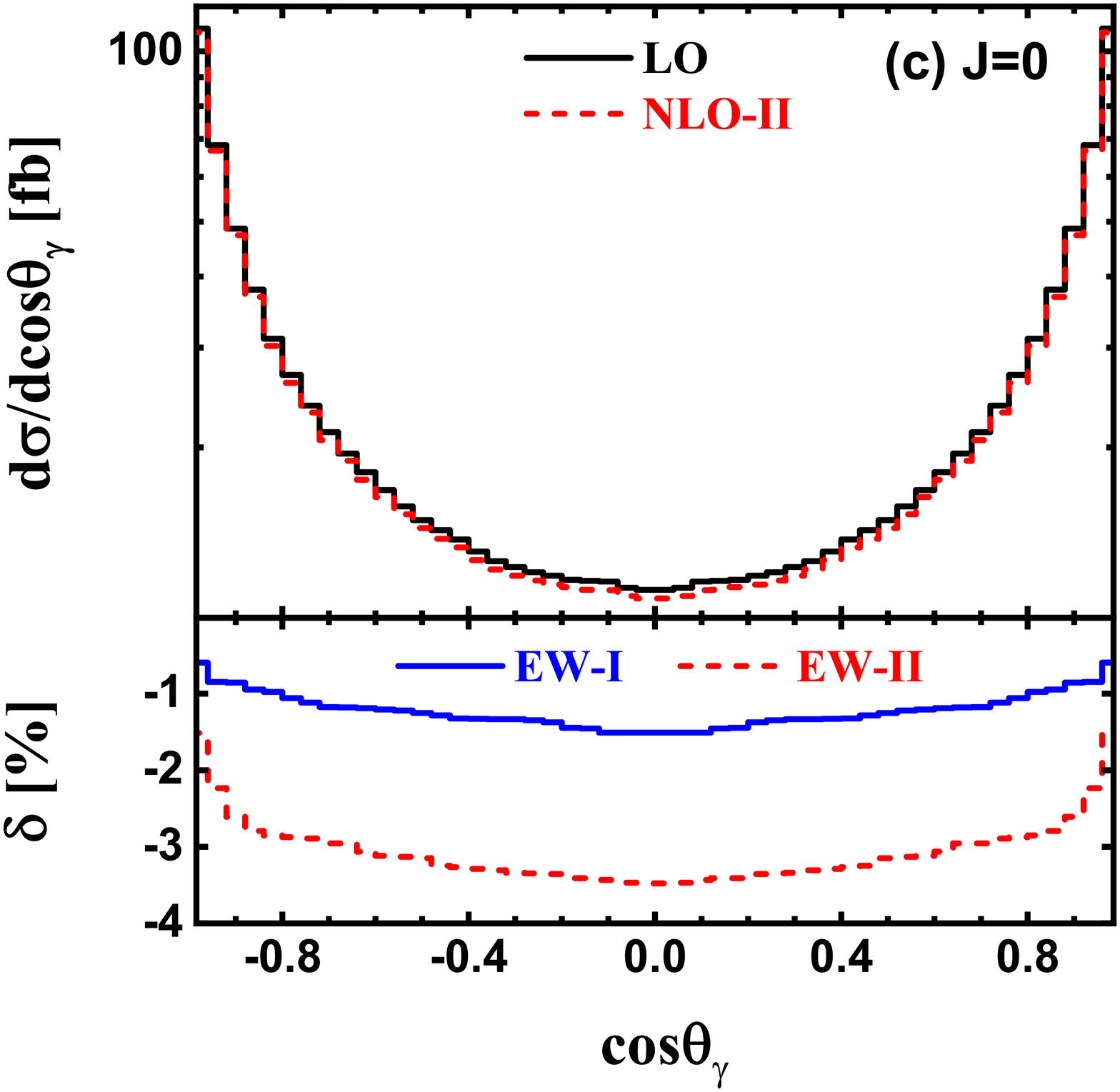}
\includegraphics[width=0.45\textwidth]{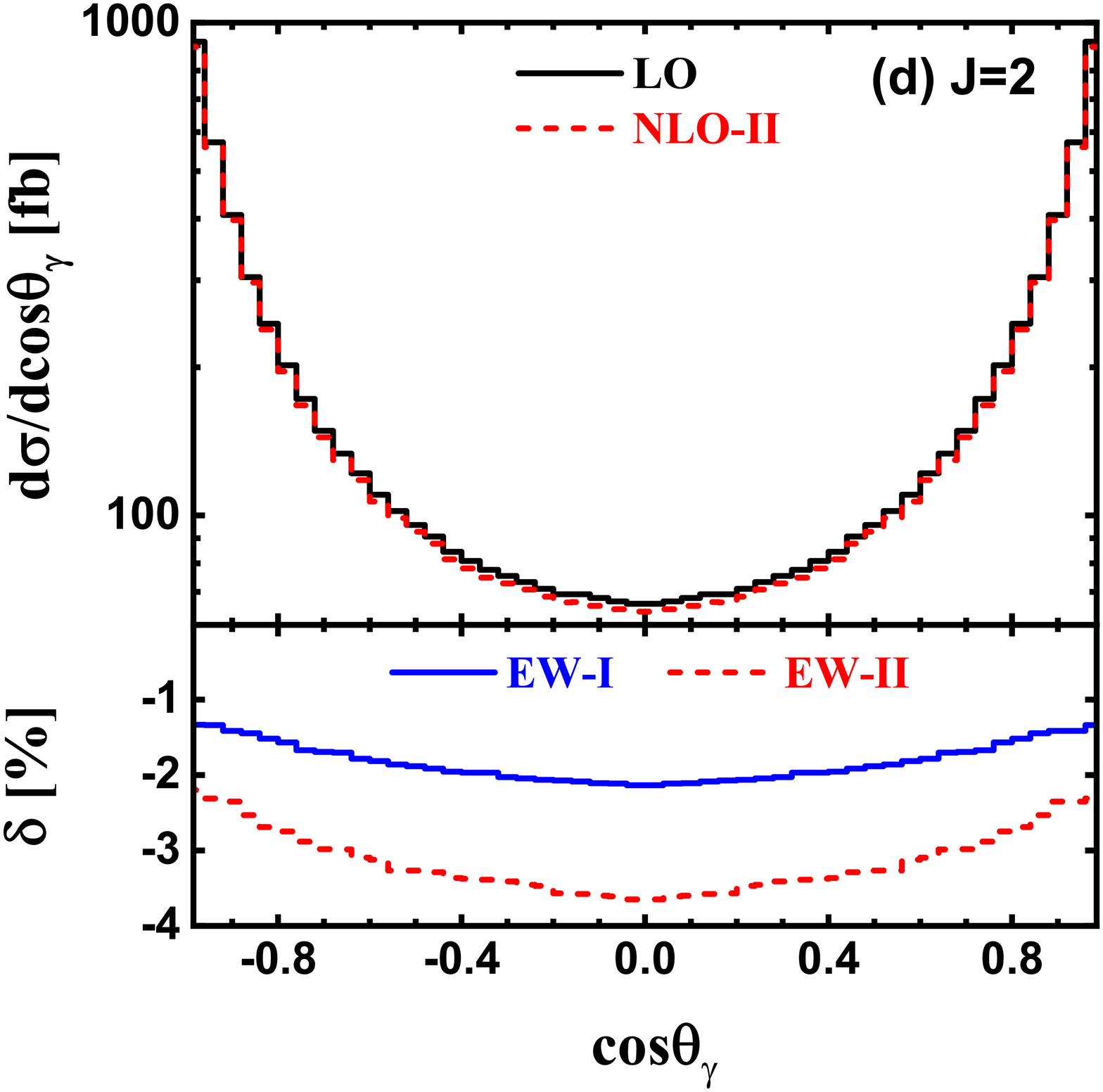}
\caption{Same as Fig.\ref{fig12}, but for the final-state leading photon.}
\label{fig14}
\end{center}
\end{figure}
\begin{figure}[!htbp]
\begin{center}
\includegraphics[width=0.45\textwidth]{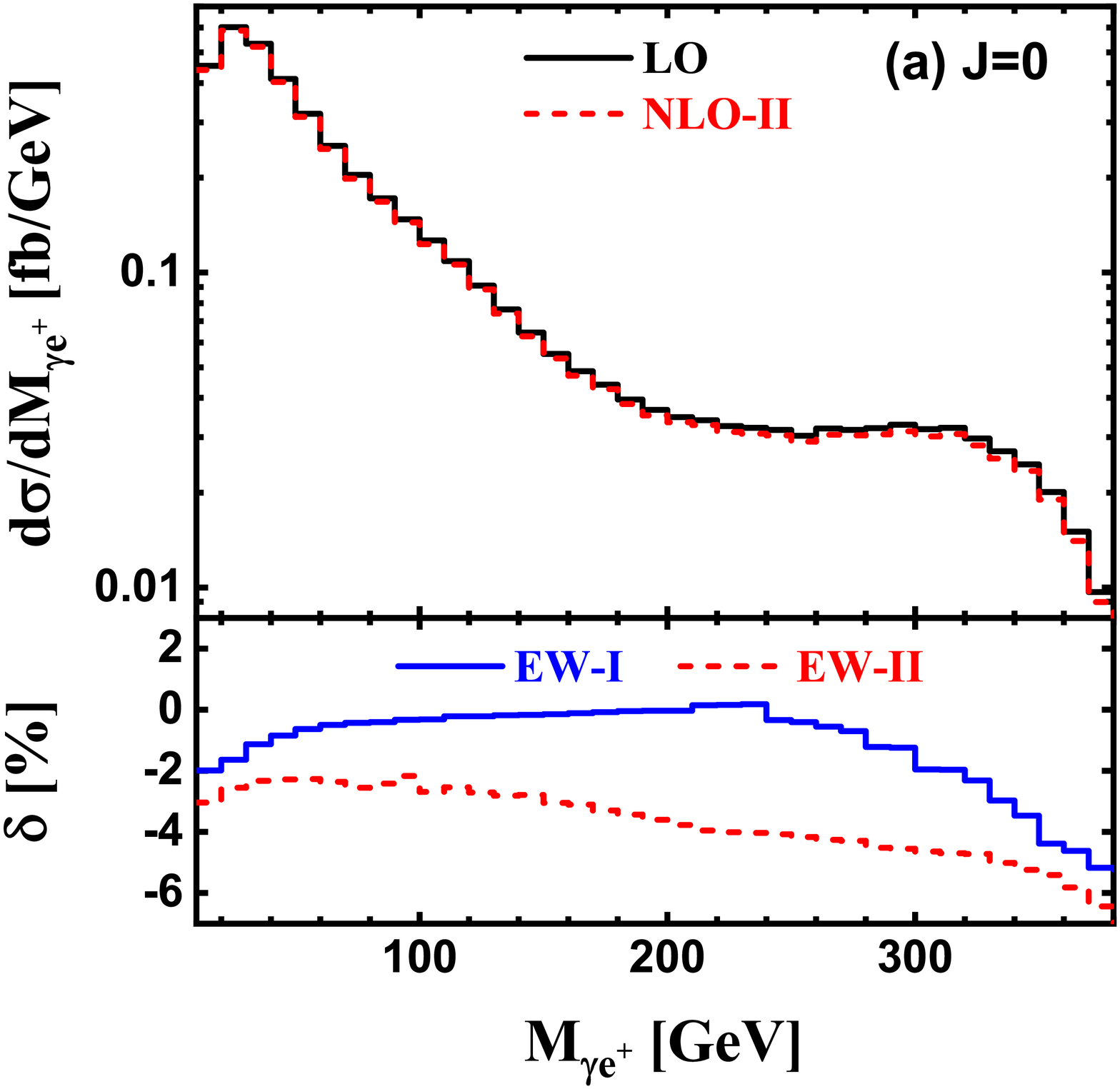}
\includegraphics[width=0.45\textwidth]{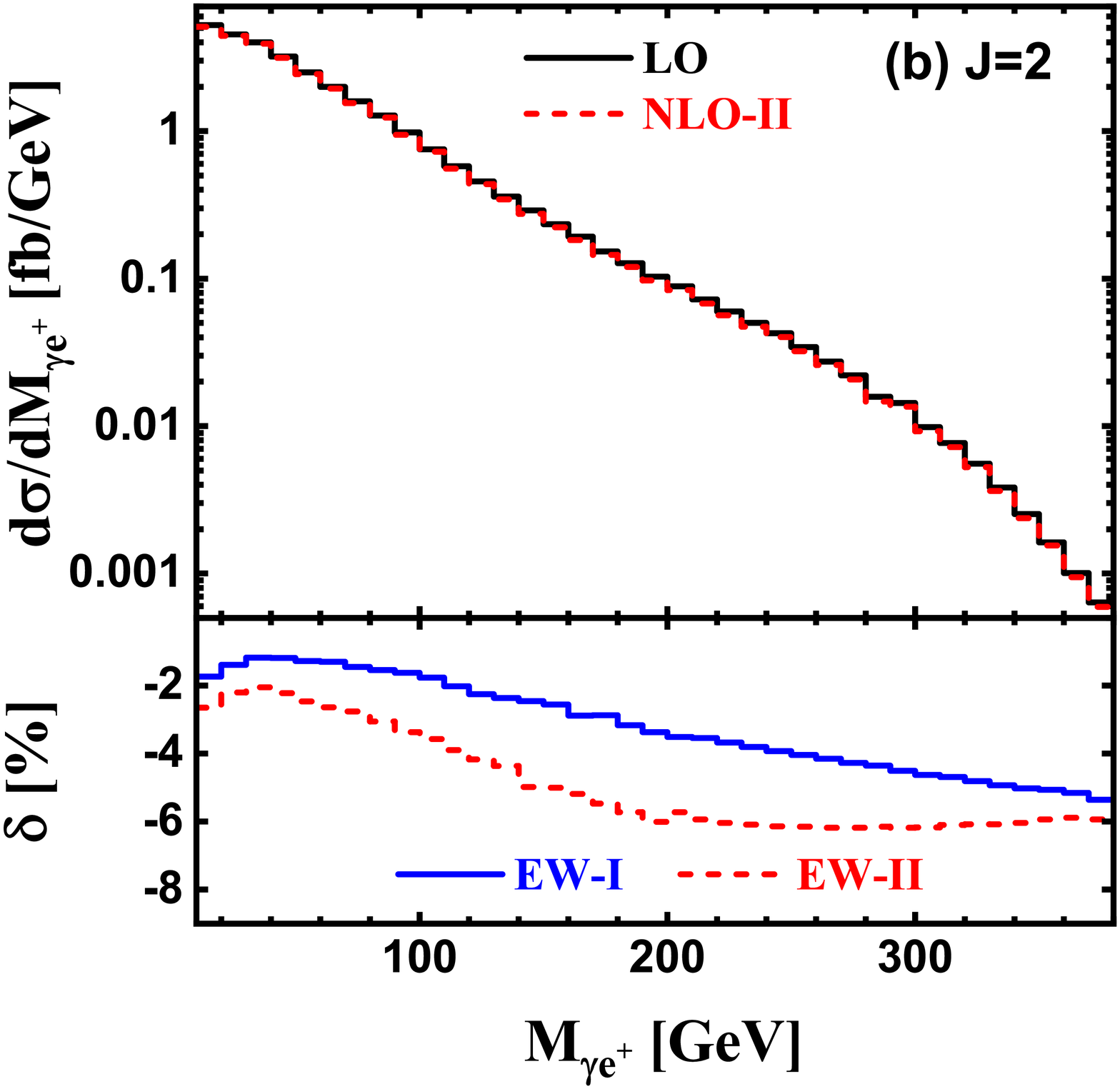}
\includegraphics[width=0.45\textwidth]{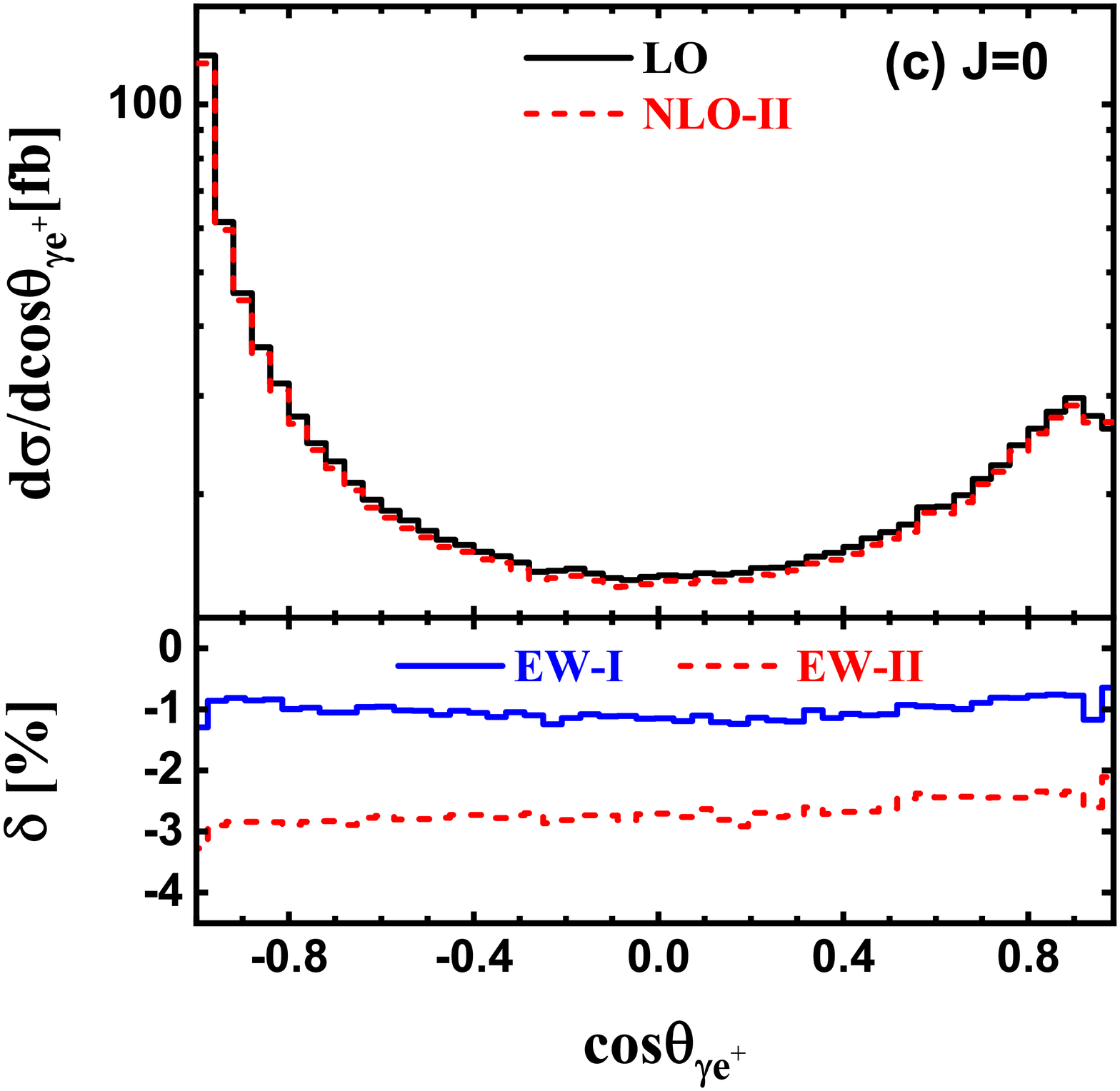}
\includegraphics[width=0.45\textwidth]{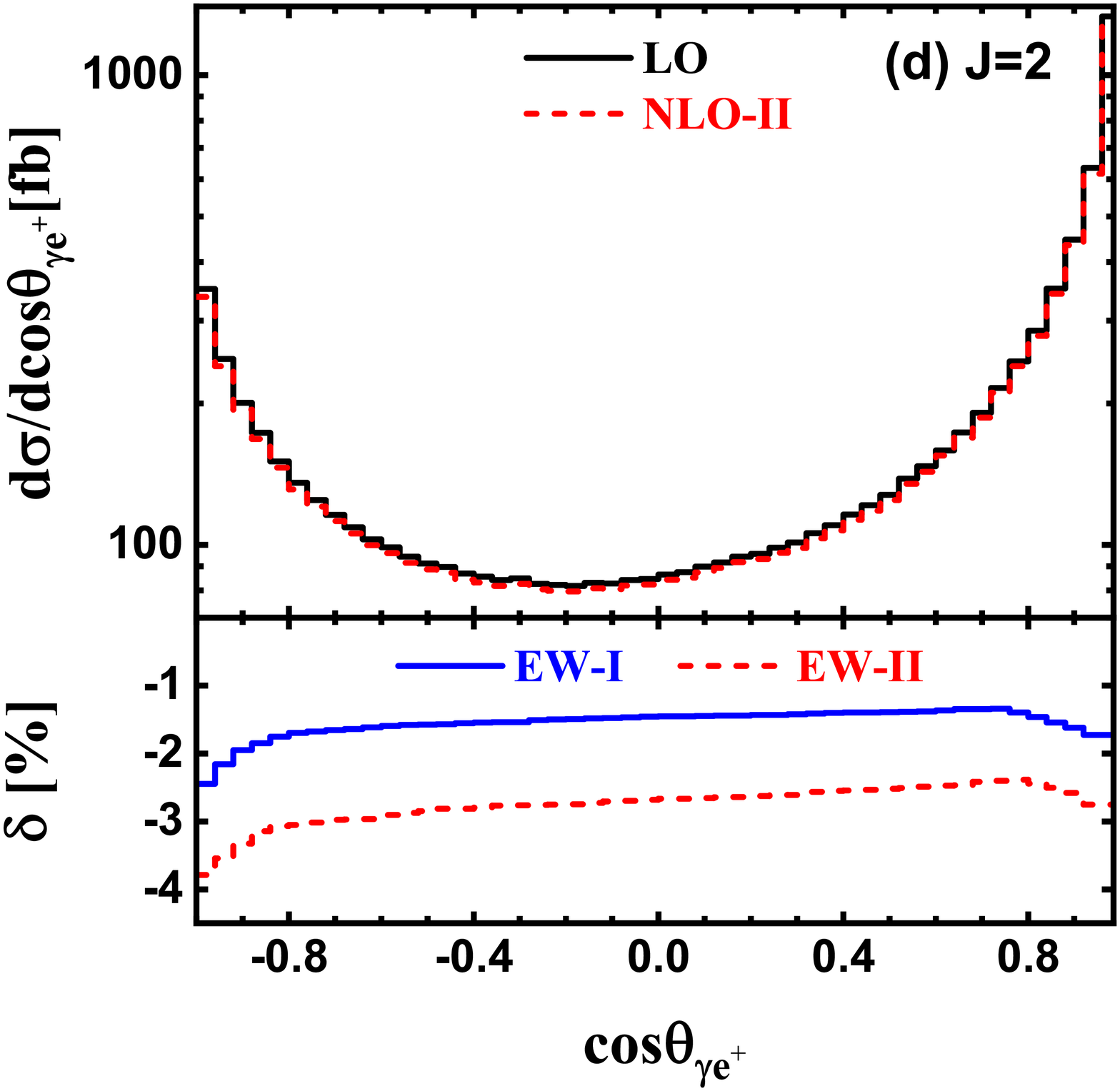}
\caption{LO, NLO EW corrected invariant mass and opening angle distributions of the final-state $\gamma e^+$ system and the corresponding EW relative corrections for $ee \rightarrow \gamma\gamma \rightarrow e^+e^-\gamma$ via $\text{J} = 0$ and $\text{J} = 2$ collisions of Compton back-scattered photons at $\sqrt{s} = 500~ \text{GeV}$.}
\label{fig15}
\end{center}
\end{figure}
\begin{figure}[!htbp]
\begin{center}
\includegraphics[width=0.45\textwidth]{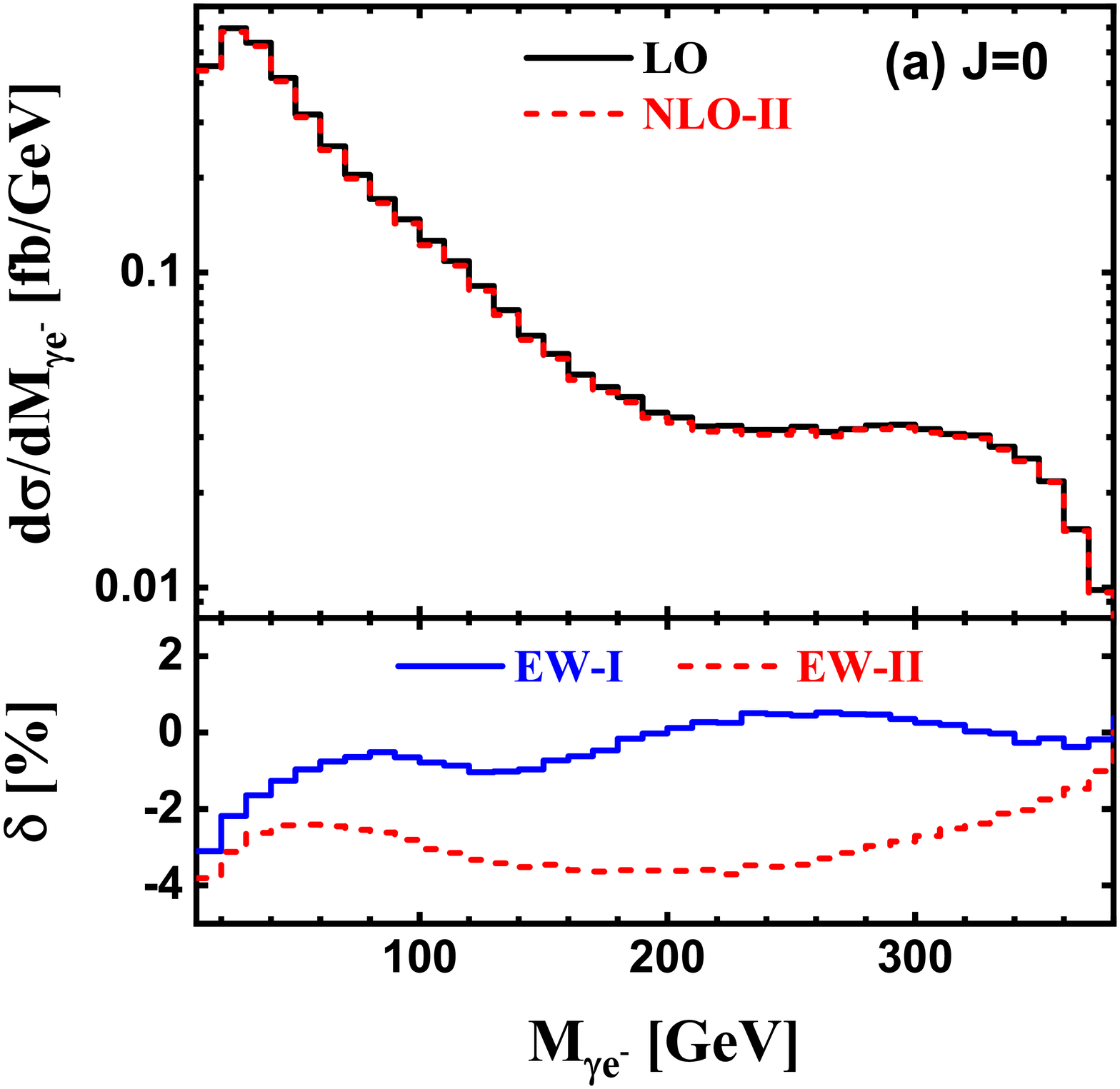}
\includegraphics[width=0.45\textwidth]{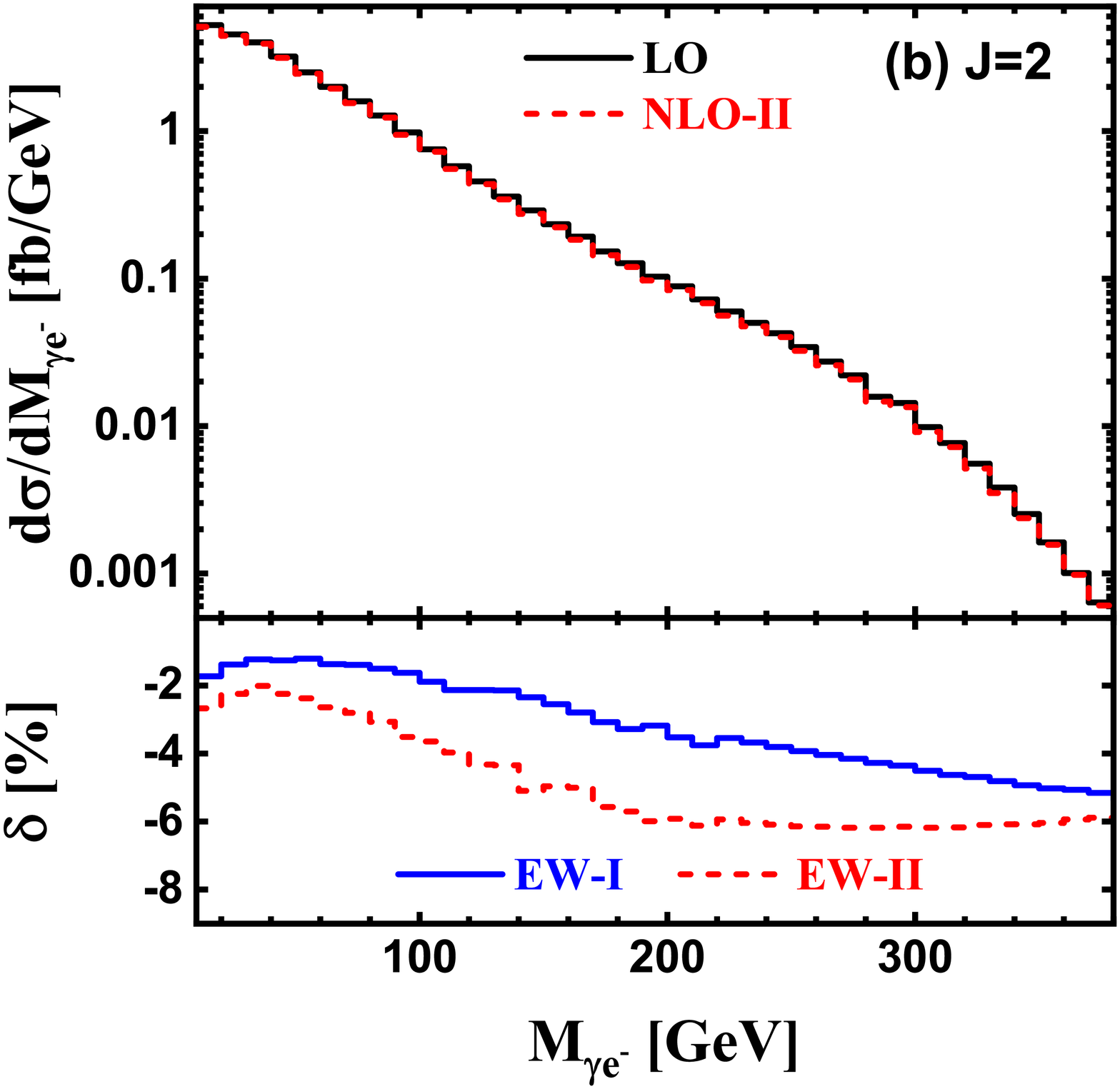}
\includegraphics[width=0.45\textwidth]{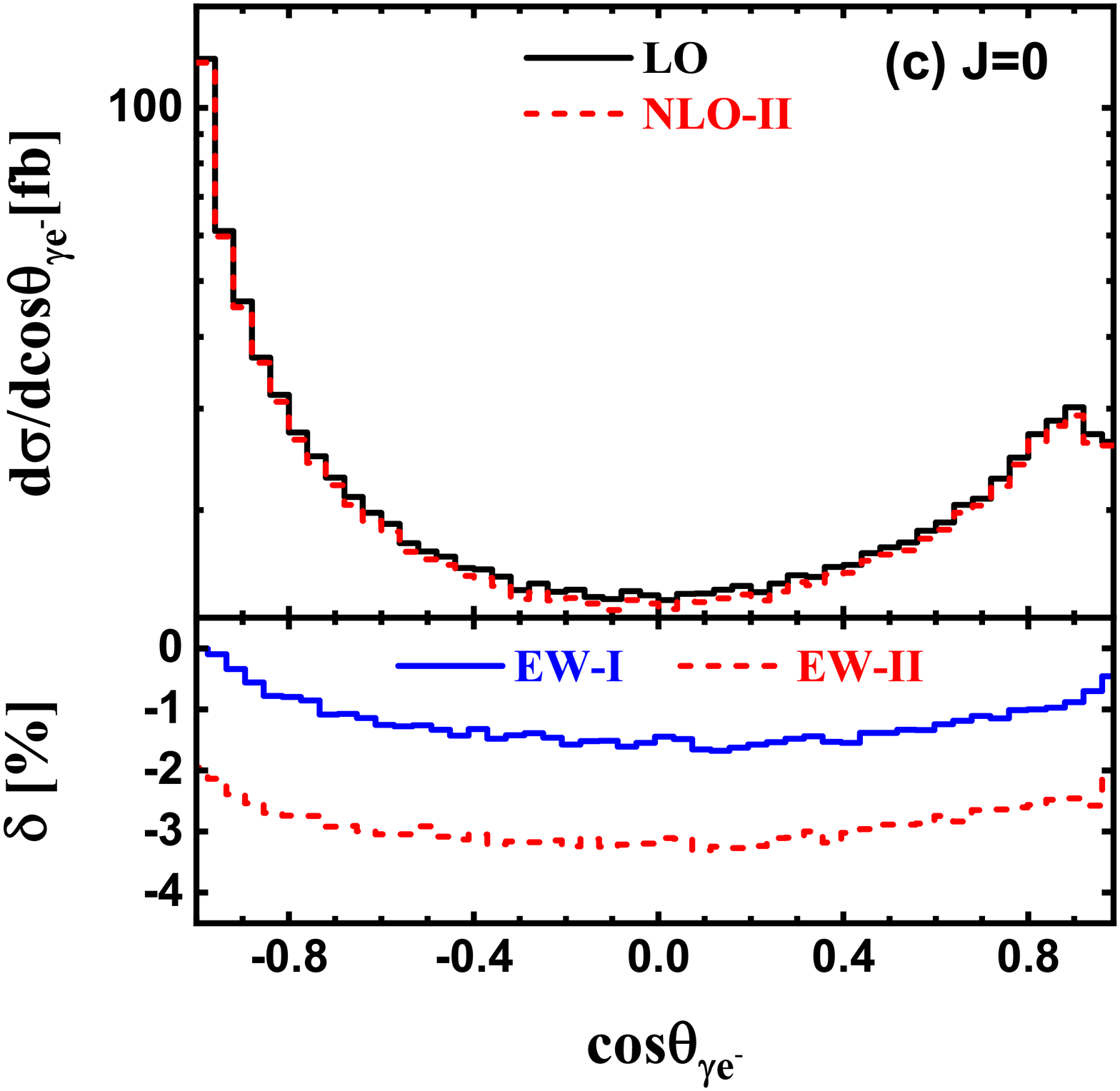}
\includegraphics[width=0.45\textwidth]{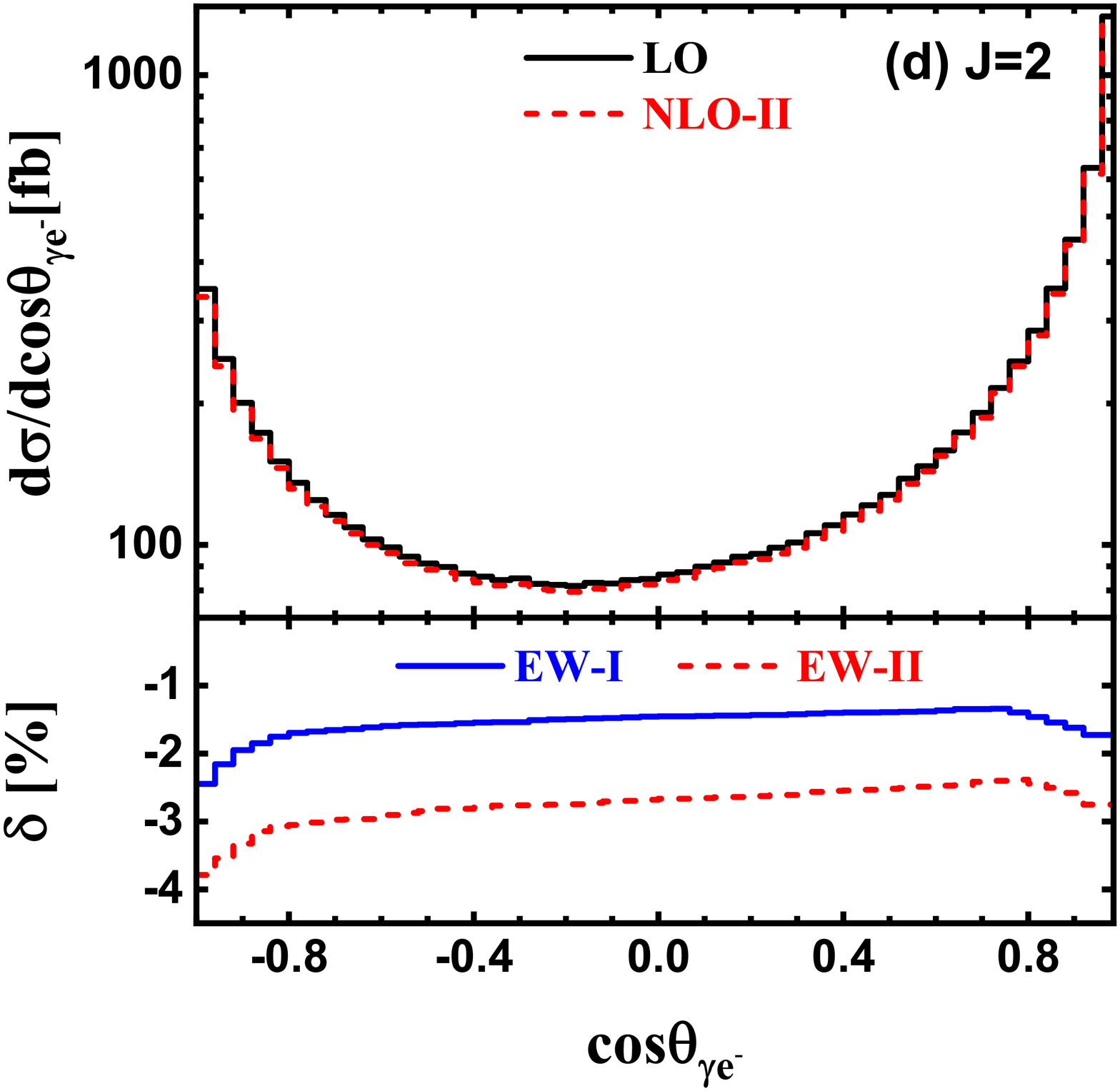}
\caption{Same as Fig.\ref{fig15}, but for the final-state $\gamma e^-$ system.}
\label{fig16}
\end{center}
\end{figure}

In Figs.\ref{fig12} - \ref{fig16} we present the LO, NLO EW corrected kinematic distributions of the final-state positron, electron and leading photon as well as the corresponding EW relative corrections for $ee \rightarrow \gamma\gamma \rightarrow e^+e^-\gamma$ via $\text{J} = 0$ and $\text{J} = 2$ collisions of Compton back-scattered photons at $\sqrt{s} = 500~ \text{GeV}$ separately. As expected, the kinematic distributions of $e^-$ and $\gamma e^-$ system are the same as the corresponding ones of $e^+$ and $\gamma e^+$ system at both LO and EW NLO within the calculation errors for the $\text{J} = 2$ polarization configuration of back-scattered photons.

\end{document}